\documentclass[%
 aip,
 amsmath,amssymb,
]{revtex4-1}

\usepackage{comment}
\usepackage{graphicx}
\usepackage{dcolumn}
\usepackage{bm}
\usepackage{float} 

\usepackage[utf8]{inputenc}
\usepackage[T1]{fontenc}
\usepackage{mathptmx}

\draft 

\usepackage[dvipsnames]{xcolor}
\usepackage{xspace}

\newcommand{\ddt}[1]{ \frac{\partial {#1}}{\partial t} }
\newcommand{\ol}{ \overline }
\newcommand{\p}[1]{ \acute{#1} }
\newcommand{\pp}[1]{ \acute{\acute{#1}} }
\usepackage[colorinlistoftodos]{todonotes}
\usepackage{hyperref}
\hypersetup{%
  colorlinks=false,
  linkbordercolor=red,
  pdfborderstyle={/S/U/W 1}
}

\def\red{\textcolor{red}}

\newif\ifcomments 
\commentsfalse
\commentstrue 
\newcommand{\daniel}[1]{{\ifcomments \color{Magenta} Daniel says: #1\xspace \fi}}

\begin{document}

\preprint{LA-UR-20-29879} 

\title[Filtering, averaging and scale dependency in HVDT]
{Filtering, averaging and scale dependency in homogeneous variable density turbulence}
\thanks{Invited for Physics of Fluids special issue in memory of Edward E. (Ted) O'Brien.}

\author{J. A. Saenz}
 \email{juan.saenz@lanl.gov}
\author{D. Aslangil}%
\author{D. Livescu}
\affiliation{%
Los Alamos National Laboratory
}%

\date{\today}

\begin{abstract}

We investigate relationships between statistics obtained from filtering and from ensemble or Reynolds-averaging turbulence flow fields as a function of length scale. Generalized central moments in the filtering approach are expressed as inner products of generalized fluctuating quantities, $q'(\xi,x)=q(\xi)-\overline q(x)$, representing fluctuations of a field $q(\xi)$, at any point $\xi$, with respect to its filtered value at $x$. For positive-definite filter kernels, these expressions provide a {\it scale-resolving} framework, with statistics and realizability conditions at any length scale. In the small-scale limit, scale-resolving statistics become zero. In the large-scale limit, scale-resolving statistics and realizability conditions are the same as in the Reynolds-averaged description. Using direct numerical simulations (DNS) of homogeneous variable density turbulence, we diagnose Reynolds stresses, $\mathcal{T}_{ij}$, resolved kinetic energy, $k_r$, turbulent mass-flux velocity, $a_i$, and density-specific volume covariance, $b$, defined in the scale-resolving framework. These variables, and terms in their governing equations, vary smoothly between zero and their Reynolds-averaged definitions at the small and large scale limits, respectively. At intermediate scales, the governing equations exhibit interactions between terms that are not active in the Reynolds-averaged limit. For example, in the Reynolds-averaged limit, $b$ follows a decaying process driven by a destruction term; at intermediate length scales it is a balance between production, redistribution, destruction, and transport, where $b$ grows as the density spectrum develops, and then decays when mixing becomes strong enough. This work supports the notion of a generalized, length-scale adaptive model that converges to DNS at high resolutions, and to Reynolds-averaged statistics at coarse resolutions.

\end{abstract}

\maketitle


\section{Introduction}

Turbulent flows are often characterized using statistics obtained from averaging or from filtering fields of interest. 
Averaging is used to obtain a statistical description of the flow, whereby a notional ensemble of realizations is averaged at a given point in space-time $(x,y,z,t)$. 
By averaging the Navier-Stokes equations, in the statistical description one obtains a set of governing equations for primitive variables.
These Reynolds-averaged Navier-Stokes (RANS) equations have additional terms that lead to a closure problem, and unclosed terms are modeled with varying levels of complexity.
When using the filtering approach, on the other hand, the Navier-Stokes equations are filtered, leading also to a closure problem, where models are used to represent the effects of small, unresolved scales on the resolved fields.
When the filter size is such that the resolved fields represent the largest eddies of the flow, this approach to modeling turbulence is referred to as large eddy simulations (LES), in contrast to RANS modeling where all scales of the flow are modeled.
For more details on RANS and LES modeling, refer to [\onlinecite{pope_2001}].
Further, several modeling approaches exist in which varying levels of resolution of scales, between LES and RANS, are addressed. 
These approaches are sometimes broadly referred to as scale resolving simulations (e.g. see [\onlinecite{menter_etal_2012}] and references therein), and are sometimes also called hybrid models, as many of them use a combination of LES and RANS ideas and subclosures. For more detailed information on hybrid models, refer to the recent reviews in [\onlinecite{chaouat_2017}] and [\onlinecite{heinz_2020}].

Most often, each choice of LES \cite{meneveau_katz_2000, pope_2001}, hybrid \cite{fasel_etal_2006,grinstein_etal_2020} or RANS \cite{pope_2001} model is closely associated to a level of resolution of the length scales that are relevant to the flow.
However, the idea that different characterizations (namely either filtering or averaging), and different models are appropriate for discrete, distinct levels of resolution of the flow can seem somewhat arbitrary.
Indeed, in practice it is often times difficult to draw a clear line that delimits where different models are valid and others are not.
For a given flow, and perhaps for a given type of flows, it is reasonable to expect that it would be possible to have a characterization that is valid at any scale relevant to the flow, and a model of the flow dynamics that can be applied at an arbitrary level of resolution.
An example of this can be found in [\onlinecite{perot_gadebusch_2007}] and [\onlinecite{perot_gadebusch_2009}], where a {\it self-adapting} model was presented and used to simulate decaying homogeneous isotropic turbulence; the model was able to calculate total kinetic energy with the same fidelity regardless of whether the resolution corresponded to DNS, RANS, or something in between.

In this work, we investigate the formal relationships between filtering and averaging, and present a generalized, scale-resolving (SR) statistical description for homogeneous turbulent flows that is valid at an arbitrary length scale, filter width, or resolution, i.e. any resolution from the smallest to the largest length scales of the flow.
Traditionally, variables are defined using central moments in the RANS statistical approach, and generalized central moments \cite{germano_1992} when considering filtering quantities, which are constructed by expressing the central moments as residuals.
We express both (RANS and filtered quantities) as central moments of generalized fluctuating quantities, thus expressing both approaches in the same general form. 
Generalized central moments in the filtering approach are expressed as inner products of generalized fluctuating quantities, $q'(\xi,x)=q(\xi)-\overline q(x)$, which represent fluctuations of a field variable $q$ at points $\xi$ with respect to its filtered value at a point $x$. 
The SR statistical description of the flow is consistent with the Navier-Stokes (NS) equations.
We derive realizability conditions for SR statistics, and we show that, for filter kernels that are positive definite, these realizability conditions are equivalent to the realizability conditions of their RANS counterparts, and that the latter constitute a special case of the former in the limit of large filter widths or length scales.

We illustrate these concepts by deriving SR statistics equivalent to those often used for RANS characterization and modeling of variable density turbulence using Favre averaging \cite{besnard_etal_1992}, namely the density-specific volume covariance, $b$, the turbulence mass flux velocity, $a_i$, the Favre averaged Reynolds stress tensor, $\mathcal{T}_{ij}$, and the large scale kinetic energy, $k_r$, along with governing equations for each of these variables, and investigate variable density effects using this formulation.
We diagnose these variables, and the terms in the governing equations for $b, a_1, k_r$ using data from recent DNS of homogeneous variable density turbulence \cite{aslangil_etal_2020}.
We discuss how some terms in the governing equations are related to length scales.
For example: for length scales that are larger than the vertical integral length scale, the variables in the SR statistics, as well as their governing equations, are fully represented by the RANS description;
the rate of transfer of energy between resolved and unresolved kinetic energy, in a volume integrated sense, peaks at length scales of the order of the horizontal Taylor micro-scale.

This work is consistent with some aspects and concepts of hybrid modeling and can be used to further the development of models for scale resolving simulations.
The results strongly suggest that the dynamics and processes relevant to the turbulence physics in HVDT transition smoothly, as a function of length scale, from the NS limit to the RANS limit.
The dynamical processes represented by the terms in the balance equations of the SR variables that we diagnose for HVDT, $b, a_1, k_r$, are all trivially zero in the NS limit, correspond to the RANS balance for this flow in the RANS limit, where some are active and some are not, and are all active at intermediate length scales.
For example, in HVDT, only destruction is active in the RANS governing equation for $b$ at scales approximately equal to or larger than the integral length scale, while at intermediate length scales and scales below the integral length scale for this flow, there is a balance between production, redistribution, destruction, and transport.
From the perspective of modeling, this work supports the notion of a generalized, length-scale adaptive model in terms of the SR variables, that converges to DNS at high resolutions, and to classical RANS statistics at coarse resolutions.
A model that relies on computing SR variables in terms of RANS statistics alone, for example by scaling the RANS statistics \cite{fasel_etal_2006}, would not be able to capture the full SR dynamics.

We begin by recalling the equations used for filtering and averaging in section \ref{sec:filtering_and_RANSaveraging}, followed by a derivation of the realizability conditions for the filtered variables in section \ref{sec:innerprod_realizability}.
The flow that we use for diagnostics is described in section \ref{section:hvdt_flow_description}, where we present the governing equations for HVDT, and we derive the equations for the SR variables.
Then, in section \ref{sec:diagnostics}, we present these diagnostics. Finally, in section \ref{sec:duscussion_conclusions}, we provide a summary and discussion on the merits of this work for investigating physics underlying turbulence at different scales, and for model development.


\section{Filtering and statistical averaging}
\label{sec:filtering_and_RANSaveraging}

The Reynolds-averaged Navier-Stokes (RANS) method, also commonly referred to as the statistical approach, is used to compute flow statistics for diagnostic analysis of DNS and experiments of turbulent flows and for turbulence modeling.
In this approach, ensemble averages are used to calculate central moments in terms of fluctuating quantities, where the latter are defined as departures from means, e.g. $q = \langle q \rangle + q'$ and $q = \langle \rho q \rangle / \langle \rho \rangle + q''$.
Invoking ergodicity, such averages can be calculated along space-time directions of homogeneity in the flow.
Here, we consider spatial homogeneity only and indicate the volume average of a spatially and temporally varying quantity $q(x,y,z,t)$ by angle brackets, such that
\begin{equation}
\langle q \rangle = \frac{1}{V} \int_V q \, dV . 
\end{equation}
with $\langle q\rangle$ retaining time dependency.
When it is necessary to make the distinction between RANS quantities and filtered quantities, we use non-italicized roman subscript `$\mathrm{e}$' to denote RANS statistical quantities defined using central moments of fluctuating quantities. 
Using the statistical approach for VDT flows, several statistical quantities have been identified for playing an important role in the dynamics of the mean flow \cite{besnard_etal_1992}, and subsequently used for RANS modeling of VDT flows \cite{stalsberg-zarling_gore_2011, schwarzkopf_etal_2011, schwarzkopf_etal_2015}. We will now introduce these quantities, and we will discuss them throughout this paper. 

The Reynolds stress tensor 
\begin{equation}
\mathcal R_{ij} = \frac{ \langle \rho u''_i u''_j \rangle }{\langle \rho \rangle}
 = \frac{\langle \rho u_i u_j \rangle}{\langle  \rho \rangle } 
- \frac{\langle \rho  u_i \rangle}{\langle  \rho \rangle}
\frac{\langle  \rho  u_j \rangle}{\langle  \rho \rangle},
\label{eq:rans_Rij}
\end{equation}
plays an important role in modulating the exchange of momentum and kinetic energy between the mean and fluctuating portions of the flow~\cite{livescu_etal_2009}.
The turbulent mass flux velocity
\begin{equation}
a_{\mathrm{e}i} = \frac{ \langle \rho' u_i' \rangle }{\langle \rho \rangle},
\end{equation}
has the effect of moderating the production of Reynolds stresses $\mathcal R_{ij}$ and turbulent kinetic energy~\cite{livescu_etal_2009}.
The density-specific volume covariance
\begin{equation}
b_{\mathrm{e}} = - \langle \rho' v' \rangle = \langle \rho \rangle \langle v \rangle - 1
\end{equation}
is a measure of mixing in VDT, and affects the production of $a_{\mathrm{e}i}$~\cite{livescu_etal_2009}.

In the filtering approach to modeling turbulence, variables are filtered in space using
\begin{equation}
\ol{q}(x) = \int G(\xi,x; \sigma) q(\xi) d\xi,
\label{eq:def_filter}
\end{equation}
where $G(\xi,x; \sigma)$ is a prescribed filtering kernel with parametric dependence on a filter width $w=\sigma$. 
In general, the filtering kernel can be a function of space-time, but here we restrict it to spatial filters.
Favre filtered quantities, commonly used in variable density turbulence, are defined as
\begin{equation}
\widetilde{q} = \frac{\ol{\rho q}}{\ol \rho}.
\label{eq:favre_filter}
\end{equation}
Dynamical variables obtained through filtering are often times defined by analogy to the residual form of the RANS quantities\cite{germano_1992}, e.g. as in (\ref{eq:rans_Rij}).
This way, in the context of scale resolving filtered flows, we define the density-specific volume covariance, the turbulent mass-flux velocity, and the turbulent stress tensor as
\begin{equation}
b = \ol{\rho} \,  \ol{\mathrm{v}} - 1,
\label{eq:def_b}
\end{equation}
\begin{equation}
a_i = \widetilde{u}_i - \ol{u}_i ,
\label{eq:def_ai}
\end{equation}
and
\begin{equation}
\mathcal{T}_{ij} = \widetilde{u_i u_j} - \widetilde{u}_i \widetilde{u}_j ,
\label{eq:def_turb_stress}
\end{equation}
respectively.

Generalized central moments $\varphi(u_i, u_j) = \ol{u_i u_j} - \ol{u_i}\ \ol{u_j}$ of the flow field $u_i$ can be used to  obtain transport equations for filtered quantities. 
This way, the equations that result from filtering the NS equations are mathematically identical to their RANS counterparts -- a property referred to as the averaging invariance of the filtered NS equations \cite{germano_1992} -- with the variables in the equations having different interpretations.
These developments have since been extensively used in developing methods used in large eddy simulations (LES). 
%
%

When employing RANS statistics, the flow is decomposed into a mean component and a fluctuating component, while, as remarked in [\onlinecite{germano_1992}], when using filtering and generalized central moments, representations of the flow field at different levels of filtering are compared.
In this paper, we will write the above definitions for $b$, $a_i$ and $\mathcal{T}_{ij}$ in equations (\ref{eq:def_b}-\ref{eq:def_turb_stress}) in terms of generalized moments of \emph{fluctuating quantities with respect to filtered fields}.

\subsection{Inner products and realizability}
\label{sec:innerprod_realizability}

The two approaches of employing RANS statistics and using filtering and the associated generalized central moments to represent turbulent flows are closely related.
This becomes clear when the moments from both approaches are expressed in terms of inner products.

In [\onlinecite{vreman_etal_1994}], RANS statistical moments are expressed in terms of inner products to derive realizability conditions for turbulent stresses in incompressible flows. 
Similarly, in the filtering approach, generalized central moments can be expressed in terms of inner products of generalized fluctuating quantities, which in turn represent moments of generalized fluctuating quantities, and which, under special conditions, at large filtering length scales become the RANS central moments.
Thus, it is important to note the connections between generalized central moments and inner products, between filtering and averaging, and the notion of generalized fluctuations. 
This will be the subject of this section, and in section  \ref{sec:diagnostics} we will illustrate these concepts by diagnosing some quantities of interest and their budgets, from DNS of HVDT flow. This flow is described in section \ref{section:hvdt_flow_description}.


To obtain the realizability conditions for the turbulent stress tensor, we express $\mathcal{T}_{ij}$ as an inner product. 
Starting from (\ref{eq:def_turb_stress}) and using (\ref{eq:def_filter}-\ref{eq:favre_filter}), with the notation $\widetilde{(\cdot)}(x)$, e.g. as in $\widetilde{u_i u_j}(x)$ and $\widetilde{u}_i(x)$, to indicate functions of space $x= (x_1, x_2, x_3)$, we write
\begin{eqnarray}
\mathcal{T}_{ij} &=& \widetilde{u_i u_j}(x) - \widetilde{u}_i(x) \widetilde{u}_j(x) \nonumber \\
&=& \widetilde{u_i u_j}(x) - \widetilde{u}_i(x) \widetilde{u}_j(x) - \widetilde{u}_i(x) \widetilde{u}_j(x) + \widetilde{u}_i(x) \widetilde{u}_j(x) \nonumber \\
&=& \frac{1}{\ol \rho} \left [ \int G(\xi, x) \rho (\xi) u_i(\xi) u_j(\xi) d\xi - \widetilde{u}_i(x) \int G(\xi,x)  \rho (\xi) u_j(\xi) d\xi \right . \nonumber \\
& & - \widetilde{u}_j(x) \int G(\xi,x)  \rho (\xi) u_i(\xi) d\xi + \left . \widetilde{u}_i(x) \widetilde{u}_j(x) \int G(\xi,x) \rho (\xi) d\xi \right] \nonumber \\
&=& \frac{1}{\ol \rho} \int G(\xi,x) \rho (\xi) \left [ u_i(\xi) -\widetilde{u}_i(x) \right ] \left [ u_j(\xi) -\widetilde{u}_j(x) \right ] d\xi \nonumber \\
\mathcal{T}_{ij} &=& \left (\pp u_i(\xi,x), \pp u_j(\xi,x) \right ) ^\rho_x,
\label{eq:Tij_inner_product}
\end{eqnarray}
where 
\begin{equation}
\left ( f, g \right)^\rho_x =  \frac{1}{\ol \rho} \int \rho(\xi) G(\xi,x) f(\xi,x) g(\xi,x)d\xi.
\label{eq:dens_innerprod}
\end{equation}
For positive $\rho$ and $G$, it can be shown that (\ref{eq:dens_innerprod}) is an inner product \cite{vreman_etal_1994}, or more specifically a density-weighted inner product, which is positive semi-definite. It can also be interpreted as a density weighted convolution.
We have used the definition of a generalized fluctuating quantity, namely
\begin{equation}
\pp u_i(\xi,x) \equiv u_i(\xi) - \widetilde{u}_i(x),
\label{eq:gen_fluctuation}
\end{equation}
%
which represents fluctuations of a field variable $u_i(\xi)$ at points $\xi$, with respect to its filtered value $\tilde{u}_i(x)$ at a point $x$.
Note that the generalized fluctuating quantity $\pp u_i(\xi,x)$ depends on a separation distance from $x$, namely $\zeta = \xi - x$.
In terms of $\zeta$ 
\begin{equation}
\pp u_i(\zeta, x) = u_i(x+\zeta) - \widetilde{u}_i(x),
\label{eq:gen_fluctuation2}
\end{equation}
which is an alternative expression for the generalized fluctuations.

Similar to the discussion in [\onlinecite{vreman_etal_1994}], $\pp u_i(\xi, x) \ne u''_i(x)$, since $\pp u_i(\xi,x) = u_i(\xi) - \widetilde{u}_i(x)$ in general is a two-point quantity that depends on $\xi$ and $x$, while $u''_i(x) = u_i(x) - \left\langle \rho(x) {u}_i(x) \right \rangle/\left\langle \rho(x) \right\rangle$ is a single point quantity that depends only on $x$. 
However, as the filtering length scale becomes large and approaches or exceeds some dynamically relevant integral length scale $\mathcal{L}$, the x dependency can be dropped so that the generalized fluctuating quantity becomes the fluctuating quantity defined in the statistical approach, 
\begin{equation}
\pp u_i \rightarrow u''_i,
\end{equation}
and, in this limit, it becomes a single-point quantity.

Similar to $\pp u_i(\xi,x)$, we define generalized fluctuations based on non-density weighted filtered quantities:

\begin{equation}
\p \rho(\xi,x) \equiv \rho(\xi) - \ol \rho(x)   ,
\label{eq:b_gen_fluctuation_one_prime}
\end{equation}
and 
\begin{equation}
\p u_i(\xi,x) \equiv u(\xi) - \ol u_i(x).
\label{eq:u_gen_fluctuation_one_prime}
\end{equation}

Applying the integral Schwarz inequality to the product of the filtered density and filtered specific volume, using the definition of the filter, we can obtain a realizability condition for $b = \ol \rho \, \ol{\mathrm{v}}-1$, as follows
\begin{eqnarray}
\int f g dV &\le & \left ( \int f^2 dV \right)^{1/2}  \left ( \int g^2 dV \right)^{1/2} \nonumber \\
f^2 &=& G\rho \nonumber \\
g^2 &=& G \frac{1}{\rho} \nonumber \\
\int \left ( G\rho \right )^{1/2} \left (G \frac{1}{\rho}\right)^{1/2} dV &\le & \left ( \int G\rho dV \right)^{1/2}  \left ( \int G \frac{1}{\rho} dV \right)^{1/2} \nonumber \\
1 &\le & ( \ol \rho )^{1/2} \; (\ol{\mathrm{v}})^{1/2} \nonumber \\
1 &\le & \ol \rho \; \ol{\mathrm{v}} \nonumber \\
\ol \rho \; \ol{\mathrm{v}} - 1 &\ge& 0
\label{eq:inner_realizability_b}
\end{eqnarray}
A similar derivation can be used in the RANS limit to show that $b_e\ge 0$ \cite{livescu_ristorcelli_2007}. Using a similar approach as for the stress tensor, it can be shown that
\begin{equation}
b = \left ( \frac{1}{\rho \ol \rho} \p \rho, \p \rho \right)_x. 
\end{equation}

Now, we note that $a_i = \widetilde{u}_i - \ol u_i$ can also be written as an inner product,
\begin{eqnarray}
\ol \rho(x) a_i(x) &=& \ol \rho(x) \left [ \widetilde{u}_i(x) - \ol u_i(x) \right ]  \nonumber \\
&=& \ol \rho \widetilde{u}_i - \ol u_i \ol \rho - \ol \rho \ol u_i + \ol \rho \ol u_i \nonumber \\
&=& \int G(\xi, x) \rho(\xi) u_i(\xi) d\xi - \ol u(x) \int G(\xi, x)  \rho(\xi) d\xi  -  \ol \rho(x) \int G(\xi, x)  u_i(\xi) d\xi \nonumber \\
&& + \ol \rho(x) \ol u_i(x) \int G(\xi,x) d\xi \nonumber \\
&=& \int G \left ( \rho(\xi) - \ol \rho(x) \right ) \left ( u_i(\xi) - \ol u_i(x) \right ) d\xi \nonumber \\
&=& \left ( r_x, \ol v_i^x \right )_x \nonumber \\
a_i &=& \frac{\left (  \p \rho, \p u_i \right )_x}{\ol \rho},
\label{eq:a_inner_product_a}
\end{eqnarray}
Alternatively,
\begin{eqnarray}
\ol \rho(x) a_i(x) &=& \ol \rho(x) (\widetilde{u}_i(x) - \ol u_i(x)) \nonumber \\
&=& \ol \rho \widetilde{u}_i - \widetilde u_i \ol \rho + \ol \rho \widetilde u_i - \ol \rho \ol u_i \nonumber \\
&=& \int G(\xi, x) \rho(\xi) u_i(\xi) d\xi - \widetilde u_i(x) \int G(\xi, x) \rho(\xi)  d\xi  + \ol \rho(x) \widetilde u_i(x) \int G(\xi, x)  d\xi  \nonumber \\
& & - \ol \rho(x) \int G(\xi,x) u_i(\xi) d\xi  \nonumber \\
&=& \int G(\xi, x) \left [ \rho(\xi) - \ol \rho(x) \right] \left[ u_i(\xi) - \widetilde u_i(x) \right ] d\xi \nonumber \\
&=& (\p \rho , \pp u_i)_x \nonumber \\
a_i &=& \frac{(\p \rho , \pp u_i)_x}{\ol \rho}
\label{eq:a_inner_product_b}
\end{eqnarray}
As before, when the filtering length scale is large, similar to a dominating length scale, $w \sim \mathcal{L}$, the generalized density fluctuation becomes the RANS density fluctuation $\p \rho \rightarrow \rho - \langle \rho \rangle = \rho'$, and $a_i$ becomes the RANS quantity used in the statistical approach \cite{besnard_etal_1992}, $a_i \rightarrow a_{\mathrm{e}i} = \langle \rho' u_i' \rangle / \rho = \langle \rho' u_i'' \rangle / \rho$.

With this, we can write a realizability condition for $a_\alpha$ in terms of $b$ and $\mathcal{T}_{\alpha \alpha}$ (where double Greek letter indices imply no summation), using the Schwartz integral inequality as follows,
\begin{eqnarray}
\ol \rho a_\alpha &=& \left (\p \rho, \pp u_\alpha \right)_x \nonumber \\
&=& \left (\frac{\p \rho}{\rho^{1/2}}, \pp u_\alpha \rho^{1/2} \right)_x \nonumber \\
&\le& \left ( \frac{1}{\rho} \p \rho, \p \rho \right)_x^{1/2} \left( \pp u_\alpha, \pp u_\alpha \rho \right)_x^{1/2} \nonumber \\ 
&=& \ol \rho b^{1/2} \mathcal{T}_{\alpha \alpha}^{1/2},
\end{eqnarray}
which leads to
\begin{eqnarray}
a_\alpha ^2 \le b \mathcal{T}_{\alpha \alpha}.
\end{eqnarray}
Again, as the filter length scale increases and becomes comparable to a relevant integral length scale, $w \sim \mathcal{L}$, the filtering operation converges to a volume average; in this limit, the relationship above becomes $a_{\mathrm{e}\alpha}^2 \le b_\mathrm{e} \mathcal{R}_{ij}$, the realizability condition in the statistical approach \cite{besnard_etal_1992}.

The above realizability conditions are true only for positive kernels, such as the Gaussian filter kernel or the box filter kernel. 
Negative kernels, while mathematically adequate, and while sometimes desirable for validation of LES (e.g. [\onlinecite{buzzicotti_etal_2018}]), will yield different realizability conditions, and do not preserve scalar bounds. 
For example, the sharp spectral filter is non-local and its kernel oscillates around zero in physical space. 
As a result, a sharp spectral filter can lead to $\ol{\rho} \le \rho_1$ in VDT flows in which two pure fluids with different densities $\rho_1 \le \rho_2$ mix, leading to values of $b$ that violate realizability conditions. Further, in flows with moderate to large Atwood numbers, this can lead to $\ol{\rho} \le 0$, and to numerical issues and artifacts in the definition of Favre filtered quantities using (\ref{eq:favre_filter}) when $\ol \rho$ is small. 
The realizability conditions that we present above are most attractive here due to (\emph{i}) their physical interpretations, e.g. positive kinetic energy $\mathcal{T}_{ii}/2$, positive densities $\ol \rho$ and positive $b$, and (\emph{ii}) because they converge to the realizability conditions for their counterparts in the RANS statistical description of the flow. 
For these reasons, we limit ourselves to filtering with positive kernels, and we use a Gaussian filter, with filter kernels in physical and spectral spaces given by
\begin{equation}
G(\xi,x) = \frac{1}{\sigma \sqrt{2\pi}} \exp{ \left [ -\frac{1}{2} \left ( \frac{x-\xi}{\sigma} \right )^2 \right ] },
\label{eq:filter_gaussian_kernel}
\end{equation}
and
\begin{equation}
\widehat{G}(x; w) = \exp{ \left [ - \frac{\sigma^2}{2} \mathbf{k}^2 \right] },
\end{equation}
respectively.
Note that the variance of the Gaussian filter commonly used in LES ($\sigma_L^2$) is 24 $\times$ the variance of the Gaussian filter we use here, $\sigma_L^2 = 24 \, \sigma^2$.
For this filter, the filtered density remains bounded, i.e. $0 \le \rho_1 \le \ol \rho \le \rho_2$.

\subsection{Additional properties and summary}
\label{sec:sr_stats_prop_summary}

Inner products have several defining properties for real, two-point quantities $f$ and $g$ (e.g. [\onlinecite{boyd_vandenberghe_2018}]).
They are commutative in $f,g$
\begin{equation}
\left ( f, g \right)_x =  \left ( g, f \right)_x
\end{equation}
distributive, or linear in the first argument,
\begin{equation}
\left ( a f, g \right)_x =  a \left ( g, f \right)_x ,
\end{equation}
\begin{equation}
\left ( h+ f, g \right)_x =  \left ( h, g \right)_x + \left ( f, g \right)_x
\end{equation}
and positive definite
\begin{equation}
\left ( f, f \right)_x   \ge 0.
\end{equation}
%
%
%
This way, the SR variables in (\ref{eq:inner_realizability_b}, \ref{eq:a_inner_product_a}, \ref{eq:a_inner_product_b}, \ref{eq:Tij_inner_product}), can be expressed as
\begin{equation}
  b 
  = \left ( \frac{1}{\rho \ol \rho} \p \rho, \p \rho \right)_x 
  \label{eq:b_inner_expectedvalue}
\end{equation}
\begin{equation}
  a_i 
  = \frac{\left (  \p \rho, \p u_i \right )_x}{\ol \rho}
  = \frac{(\p \rho , \pp u_i)_x}{\ol \rho}
  \label{eq:a1_inner_expectedvalue}
\end{equation}
\begin{eqnarray}
\mathcal{T}_{ij} 
= \left (\pp u_i, \pp u_j \right ) ^\rho_x 
\label{eq:Tij_inner_expectedvalue}
\end{eqnarray}
%
in terms of the generalized fluctuations defined in (\ref{eq:gen_fluctuation}), (\ref{eq:b_gen_fluctuation_one_prime}), (\ref{eq:u_gen_fluctuation_one_prime}).
%

Realizability conditions are given by the properties of the inner product \cite{boyd_vandenberghe_2018}, namely
\begin{equation}
\mathcal{T}_{\alpha \alpha} \ge 0, \quad \mathcal{T}_{\alpha \alpha} \mathcal{T}_{\beta \beta} - (\mathcal{T}_{\alpha \beta})^2  \ge 0, \quad \det \mathcal{T}_{ij} \ge 0, \\
\label{eq:realizability_Tij}
\end{equation}
\begin{equation}
  b \ge 0,
  \label{eq:realizability_b}
\end{equation}
\begin{equation}
a_{\alpha}^2 \le b \mathcal{T}_{\alpha\alpha}.
\label{eq:realizability_a}
\end{equation}

In the limit as the filter width becomes small, $w \rightarrow 0$, the filter approaches a delta function and filtered quantities become close to the pointwise values of the underlying quantities
so that the instantaneous, or Navier-Stokes flow fields are recovered, e.g.
\begin{eqnarray}
\ol \rho \rightarrow \rho, \quad
\tilde u_i \rightarrow u_i. \quad
\label{eq:filtered_quantities_dns_limit}
\end{eqnarray}

Since integration limits much larger than the filter width do not change the integral, the effective parts (i.e. the parts that contribute to inner products) of the generalized fluctuations,
$\pp u_i(\xi,x)$, $\p u_i(\xi,x)$, $\p \rho(\xi,x)$
and the generalized filtered quantities $\mathcal{T}_{ij}$, $a_i$,
and $b$ all approach zero.
Thus, we will refer to this limit as the {\it Navier-Stokes limit} or the {\it Navier-Stokes description} of the flow.
We will refer to the statistical description corresponding to intermediate length scales, or filter widths, where the generalized filtering statistical description $\mathcal{T}_{ij}$, $a_i$ and $b$ is nontrivial, as the {\it scale-resolving} (SR) description of the flow.



We can expect the behavior of the SR description of the flow to vary smoothly between the NS and RANS descriptions.
At the largest scales, $\mathcal{T}_{ij}$, $a_i$, and $b$ become equal to the RANS statistical description and they are non-zero, while in the NS limit they are zero.
At the smallest dissipative and diffusive scales, the flow field is smooth and $\mathcal{T}_{ij}$, $a_i$, and $b$ in the SR description can therefore be expected to transition smoothly from zero to their RANS values as the length-scales increase.
Similarly, the processes affecting the balances of $\mathcal{T}_{ij}$, $a_i$, and $b$, e.g. production, dissipation, destruction and transport of the quantities in the SR description of the flow, can be expected to vary smoothly between the two limits. 

In what follows, we will systematically investigate the SR description, and the transitions between the NS and RANS descriptions of the flow, by diagnosing variables in the SR statistical description, $\mathcal{T}_{ij}$, $a_i$ and $b$, using DNS of homogeneous variable density turbulence.


\section{Flow description}
\label{section:hvdt_flow_description}

\begin{figure}[tb]
    \centering
    \includegraphics[trim=0 0 32 0, clip, width=0.45\textwidth]{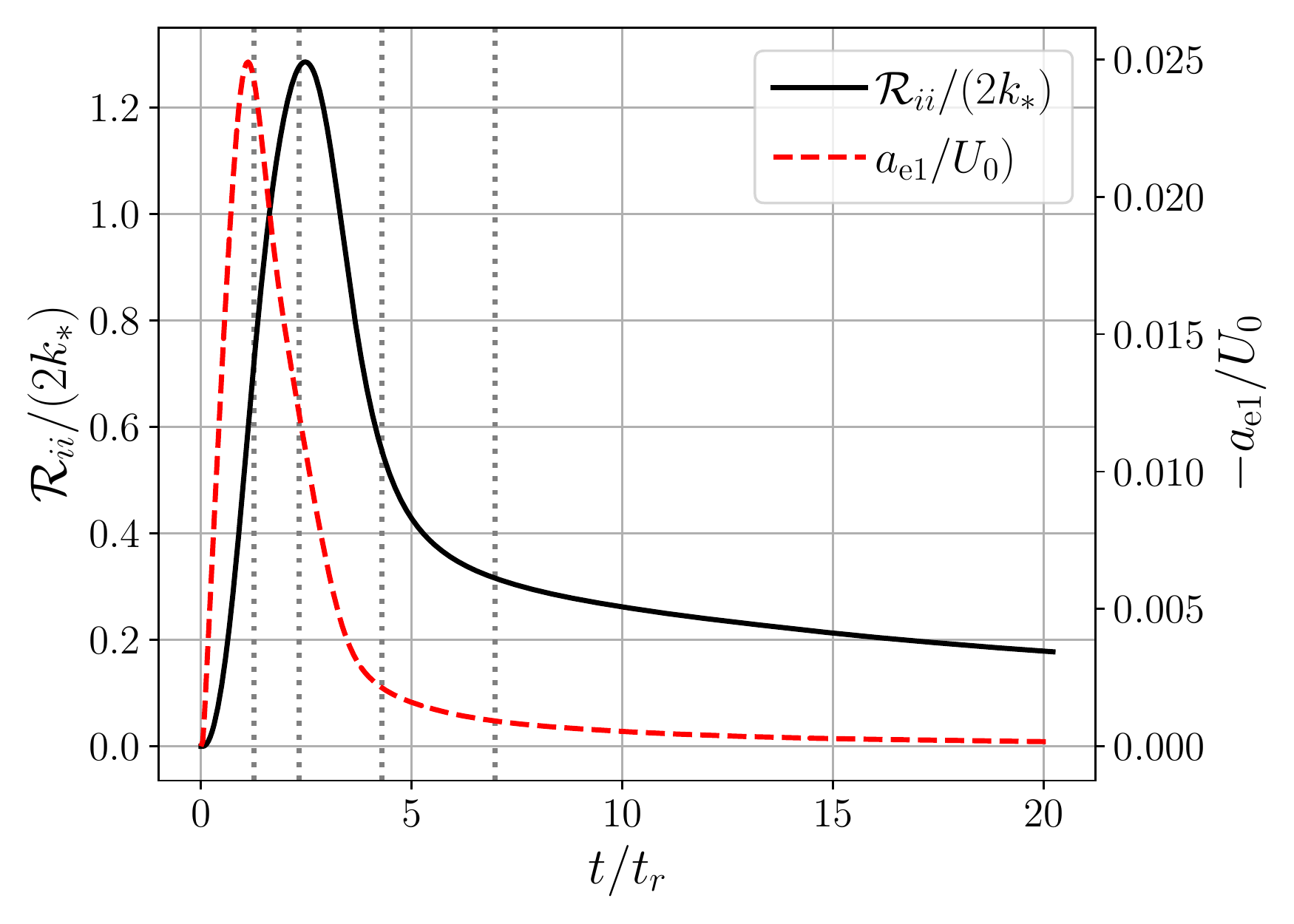}
    \includegraphics[trim=32 0 0 0, clip, width=0.45\textwidth]{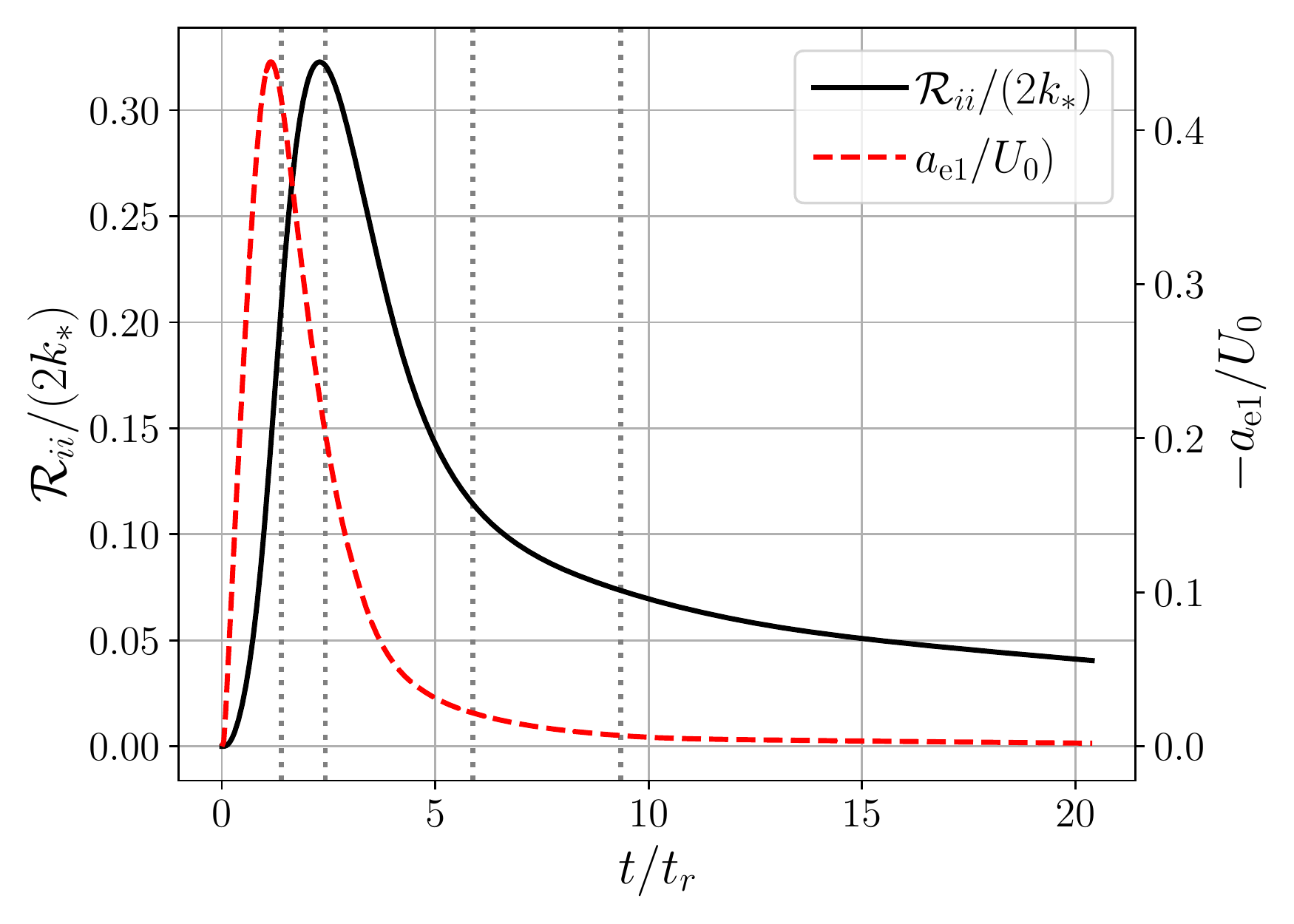}
    \caption{Normalized turbulent kinetic energy (solid curve) and normalized rate of conversion of potential energy to turbulent kinetic energy (dashed curve) as a function of $t/t_r$ ($t_r = \sqrt{Fr^2/A}$) for $A=0.05$ (left) and $A=0.75$ (right). The flows considered here have $F_r=1$. Times at which diagnostics are calculated for each flow, listed in Table \ref{tab:regime_times}, are indicated by the vertical dashed lines, of which the first three also depict the separation of the four flow regimes discussed in the text.}
    \label{fig:flow_regimes}
\end{figure}
Buoyancy driven homogeneous variable density turbulence (HVDT) was first introduced in [\onlinecite{batchelor_etal_1992}] and [\onlinecite{sandoval_etal_1997}], and further developed and discussed in [\onlinecite{ livescu_ristorcelli_2007, livescu_ristorcelli_2008, aslangil_etal_2020,aslangil_etal_2020b}].
HVDT is a canonical flow configuration in which two incompressible, compositionally different fluids with densities $\rho_1$ and $\rho_2$, with $\rho_2 > \rho_1$, are randomly distributed within an accelerated, triply periodic cube with side dimension $L=2 \pi$.
Here, we consider flows in which the initial probability density function (PDF) of density is initially symmetrical, with $\langle \rho \rangle = (\rho_1 + \rho_2)/2$, resembling a double-delta distribution with two peaks at densities $\rho = \rho_1$ and $\rho = \rho_2$.
Aslangil et al. \cite{Aslangil_book_ch_2019, aslangil_etal_2020} divide the highly non-linear evolution of the flow into four distinct regimes based on the sign of the time derivatives of the turbulent kinetic energy, as depicted in Figure \ref{fig:flow_regimes}. 
The flow is initially quiescent, and at time $t=0$, due to instability to buoyancy forces, the flow starts moving as available potential energy is converted to kinetic energy, which, at first, rapidly increases with time \cite{livescu_ristorcelli_2007, aslangil_etal_2020}. This first regime is dubbed the explosive growth regime. 
As time advances and the flow develops, it transitions to a turbulent state and, as a result, turbulent kinetic energy and the rates of mixing and dissipation increase. 
Following this stage, as a result of increasing turbulence and mixing, the density PDF is populated at intermediate densities ($\rho_1 \le \rho \le \rho_2$). 
The lighter fluid has less inertia, so it is stirred more and mixes faster than the heavy fluid \cite{livescu_ristorcelli_2007,livescu_ristorcelli_2008,aslangil_etal_2020,aslangil_etal_2020b}, and the density PDF quickly becomes asymmetrical as the lighter fluid densities ($\rho \le \langle \rho \rangle$) are populated more than the heavier fluid densities ($\rho \le \langle \rho \rangle$) - these effects become more pronounced as $A$ increases.
As a result, buoyancy forces decrease and the rate of conversion from potential energy to kinetic energy peaks, as dissipation increases.
The peak of kinetic energy production by buoyancy forces marks the beginning of the saturated growth regime, during which kinetic energy continues to grow at an increasingly slower rate, as turbulence develops, leading to more mixing and dissipation, and consequently less production.
Eventually, the rate of dissipation overcomes the rate of production of kinetic energy, and the kinetic energy peaks. After this point, kinetic energy starts decaying rapidly during the fast decay regime, and then more slowly during the gradual decay regime.

Aslangil et al. [\onlinecite{Aslangil_book_ch_2019}], [\onlinecite{aslangil_etal_2020}] discuss how the identification of these regimes makes it possible to draw parallels between HVDT and more complex VDT flows, such as Rayleigh-Taylor Instability (RTI) \cite{rayleigh_1882, taylor_1949}, RTI under variable-acceleration \cite{ramaprabhu_etal_2013,aslangil_etal_2016,aslangil_etal_2020c,livescu_2020}, Richtmyer–Meshkov Instability (RMI) \cite{richtmyer_1960,meshkov_1969}, VD mixing layers \cite{baltzer_livescu_2020}, and VD jets \cite{charonko_prestridge_2017}, among others.
Recent reviews of some of these flows and the relevant VD processes involved may be found in [\onlinecite{zhou_2017a, zhou_2017b, livescu_2020}].
For this reason, and since HVDT reaches much larger Reynolds numbers and is relatively simple to post-process and analyze thanks to spatial homogeneity, we will use it here to diagnose and investigate scale dependence of the dynamical processes in VD flows.


\subsection{Governing equations}

The flow is governed by the equations for conservation of mass and momentum, 
\begin{equation}
\ddt{\rho} + \frac{\partial}{\partial x_j} (\rho u_j) = 0, 
\label{eq:mass}
\end{equation}
\begin{equation}
\ddt{\rho u_i} + \frac{\partial}{\partial x_j} (\rho u_i u_j) = 
-\frac{\partial p}{\partial x_i} 
+ \frac{\partial \tau_{ij}}{\partial x_j}
+ \frac{1}{Fr^2} \rho g_i,
\label{eq:mom}
\end{equation}
where the viscous stress tensor for the Newtonian fluids and strain rate tensor are given by 
\begin{equation}
\tau_{ij} 
= \frac{\rho}{Re_0} \left( 
\frac{\partial u_i}{\partial x_j} 
+ \frac{\partial u_j}{\partial x_i} 
- \frac{2}{3} \frac{\partial u_k}{\partial x_k} \delta_{ij} 
\right).
\label{eq:tau}
\end{equation}
%
%
%
The divergence of the velocity field is not zero because of the effect that mixing of VD fluids has on the specific volume \cite{livescu_2020},
\begin{equation}
\frac{\partial u_{j}}{\partial x_j} = - \frac{1}{Re_0 Sc} \frac{\partial^2}{\partial x_j \partial x_j} (\ln \rho).
\label{eq:div_and_rho}
\end{equation}
There is a degree of freedom in the mean pressure gradient due to the triply periodic boundary conditions, which is constrained by requiring that the mean pressure gradient maximizes the production of the total kinetic energy from conversion of the available potential energy \cite{livescu_ristorcelli_2007,aslangil_etal_2020}. As a result, the mean pressure gradient is given by
\begin{equation}
\frac{\partial \langle p \rangle}{\partial x_i} = 
\frac{1}{V} \left (  
\frac{1}{Fr^2} g_i - \langle v' p'_{,i} \rangle + \langle u'_{i} u'_{j,j} \rangle + \langle v' \tau'_{ij,j} \rangle 
\right ).
\label{eq:mean_press_grad}
\end{equation}
Note that due to homogeneity, the mean pressure gradient is constant in space, varying only in time. The definition of the mean pressure gradient also leads to $\overline{u_i}=0$ at all times during the flow evolution.

Relevant non-dimensional numbers are Atwood number $A$, Froude number $Fr$, computational Reynolds number $Re_0$ and Schmidt number $Sc$, defined as
\begin{equation}
A = \frac{\rho_2 - \rho_1}{\rho_2 + \rho_1},
\end{equation}
\begin{equation}
Fr^2 = \frac{U_0^2}{gL_0},
\end{equation}
\begin{equation}
Re_0 = \frac{\rho_0 L_0 U_0}{\mu_0},
\end{equation}
\begin{equation}
Sc= \frac{\mu_0}{\rho_0 D_0},
\end{equation}
where $\mu_0$ is the reference dynamic viscosity, $\rho_0=(\rho_2+\rho_1)/2$ for the symmetric initial conditions in this study is the reference density, $L_0$, $U_0$, $t_0$ are the reference length, velocity, and time scales, respectively. All cases considered here have unity $Fr$ and $Sc$ numbers. The simulations analyzed here represent a subset of the cases presented in [\onlinecite{aslangil_etal_2020}]: the computational Reynolds numbers are $Re_0=10^4$ and $Re_0=1563$ for the low and high Atwood number cases, $A=0.05$ and $A=0.75$, respectively.


\subsection{Governing equations for scale-resolving variables}
\label{sec:generalized_gov_eqns}


We obtain the governing, or transport, equations for the generalized statistics in the scale resolving description of the flow in the same way as for the RANS statistical description of the flow.
Applying the filtering operations defined in (\ref{eq:def_filter}) and (\ref{eq:favre_filter}) on the NS equations (\ref{eq:mass}-\ref{eq:mom}) results in
\begin{equation}
\ddt{\overline \rho} + \frac{\partial}{\partial x_j} (\overline \rho \widetilde u_j) = 0 ,
\label{eq:filtered_cont}
\end{equation}
\begin{equation}
\ddt{\overline \rho \widetilde u_i} + \frac{\partial}{\partial x_j} (\overline \rho \widetilde u_i \widetilde u_j) = 
-\frac{\partial \overline p}{\partial x_i} 
+ \frac{\partial \overline \tau_{ij}}{\partial x_j}
+ \frac{1}{Fr^2} \overline \rho g_i
- \frac{\partial}{\partial x_j} \left ( \overline \rho \mathcal{T}_{ij} \right ) .
\label{eq:filtered_mom}
\end{equation}

Transport equations for the filtered density-specific volume covariance $b$, turbulent mass-flux velocity $a_i$ and turbulent stresses $\mathcal{T}_{ij}$ can be obtained following the same procedure used to derive the transport equations for the RANS BHR statistics \cite{besnard_etal_1992}, by applying the filtering operations (\ref{eq:def_filter}), (\ref{eq:favre_filter}) on the NS equations (\ref{eq:mass}-\ref{eq:mom}), and using the definitions (\ref{eq:def_b}),  (\ref{eq:def_ai}), (\ref{eq:def_turb_stress}).
Alternatively, replacing central moments with their corresponding generalized central moments \cite{germano_1992}, the transport equations for the unclosed RANS variables \cite{besnard_etal_1992, stalsberg-zarling_gore_2011, schwarzkopf_etal_2011} can be converted to equations for the filtered quantities, using the following rules.
The generalized central moments of relevance here are 
\begin{eqnarray}
\varphi(f,g) &=& \ol{fg} - \ol{f}\ol{g} \rightarrow \ol{f' g'} ,\\
\varphi(f,g,h) &=& \ol{fgh} - \ol{f} \varphi(g,h) - \ol{g} \varphi(f,h) - \ol{h} \varphi(f,g) - \ol{f}\ol{g}\ol{h}  \rightarrow \ol{f' g' h'} , \\
%
\tilde \varphi(f,g) &=& \widetilde{f g} - \tilde f \tilde g \rightarrow \ol{\rho f''g''} / \ol \rho , \\
\tilde \varphi(f,g,h) &=& \widetilde{fgh} - \tilde \varphi(f g) \tilde h - \tilde \varphi(f h) \tilde g  - \tilde \varphi(gh) \tilde f  - \tilde f \tilde g \tilde h \rightarrow \ol{\rho f''g''h''} / \ol \rho ,
\end{eqnarray}
%
%
%
%
where the arrows indicate the corresponding central moments in the RANS statistical formulation.
Either way, the resulting transport equations are

\begin{eqnarray}
\ddt{\ol \rho b }  + \left( \ol \rho \tilde u_k b \right)_{,k} 
= - 2 (b+1) a_k \ol \rho_{,k} 
+ 2 \ol \rho a_k b_{,k} 
+ \ol \rho^2 \left( \frac{ \varphi(\rho, v, u_k)}{\ol \rho} \right)_{,k} 
  +  2 \ol{\rho}^2 \varphi(v, u_{k,k}),
\label{eq:trans_b}
\end{eqnarray}
\begin{eqnarray}
\ddt{\ol \rho a_i }  + (\ol \rho \tilde u_k a_i)_{,k} 
&=& b ( \ol p_{,i} -\ol \tau_{ki,k}) - \mathcal{T}_{ik} \ol \rho_{,k} - \ol \rho a_k (\tilde u_i - a_i)_{,k} + \ol \rho (a_i a_k)_{,k} \nonumber \\
&-& \ol \rho \left( \frac{\varphi(\rho, u_i, u_k)}{\ol \rho} \right)_{,k} +  \ol \rho \, \varphi (v, p_{,i}) - \ol \rho \, \varphi(v,\tau_{ki,k}) )
- \ol{\rho} \varphi(u_i, u_{k,k}),
\label{eq:trans_ai}
\end{eqnarray}
and 
%
%
\begin{eqnarray}
\ddt{\ol \rho \mathcal{T}_{ij}}  + \left( \ol \rho \tilde u_k \mathcal{T}_{ij} \right)_{,k} 
&=& a_i \ol P_{,j} + a_j \ol P_{,i} 
- \ol\rho \mathcal{T}_{ik} \tilde u_{j,k} - \ol\rho \mathcal{T}_{jk} \tilde u_{i,k} 
- a_i \tilde \varphi_{jk,k} - a_j \tilde \varphi_{ik,k}  \nonumber \\
&-& \left[ \ol \rho \, \tilde \varphi(u_i, u_j, u_k) \right]_{,k}  \nonumber \\
&+& \left[ \hat\varphi(u_i, \tau_{jk}) + \hat \varphi(u_j, \tau_{ik}) \right]_{,k}  
- \left[ \hat \varphi(u_i, p) \right]_{,j} - \left[ \hat \varphi(u_j, p) \right ] _{,i}  \nonumber \\
&+&  \hat \varphi(u_{i,j}, p) +  \hat \varphi(u_{j,i}, p) 
-  \hat \varphi(\tau_{jk}, u_{i,k}) - \hat \varphi(\tau_{ik}, u_{j,k}).
\label{eq:trans_Tij}
\end{eqnarray}

The sub-scale kinetic energy is related to the turbulence stress tensor by
\begin{equation}
{k}_s = \frac{1}{2} \mathcal{T}_{kk}.
\label{eq:subscale_ke}
\end{equation}
Here, we use the term sub-scale kinetic energy for $k_s$, but in the context of filtering it can be referred to as sub-filter kinetic energy, and in the context of LES it can also be referred to as unresolved kinetic energy.
Complementing the sub-scale kinetic energy is the scale-resolved, or resolved, kinetic energy,
\begin{equation}
{k}_r =  \frac{1}{2} \widetilde u_i \widetilde u_i,
\label{eq:resolved_ke}
\end{equation}
such that the sum of the two is the total kinetic energy
\begin{equation}
{k} =  k_s + k_r.
\end{equation}
%
%
%
It is useful to investigate the sub-scale kinetic energy equation, for its physical significance, its ties to the stress tensor, and also for its role in modeling.
The transport equation for the sub-scale kinetic energy is given by, following the form used in [\onlinecite{stalsberg-zarling_gore_2011}],
\begin{eqnarray}
\ddt{\ol \rho k_s}  + \left( \ol \rho \tilde u_k k_s \right)_{,k} 
&=& a_k (\ol p_{,k} - \ol \tau_{ik,k} )
- \ol\rho \mathcal{T}_{ik} \tilde u_{i,k}
- \frac{1}{2} \left[ \ol \rho \, \tilde \varphi(u_i, u_i, u_k) \right]_{,k}  \nonumber \\
&-& \left[ \hat \varphi(u_k, p) \right]_{,k}  +  \hat \varphi(u_{k,k}, p) 
- \hat \varphi(\tau_{ki}, u_{i,k}) + \left[ \hat\varphi(u_i, \tau_{ki}) \right]_{,k}  .
\label{eq:BHR_subscale_ke}
\end{eqnarray}
%
%
%
For completeness, the Favre averaged scale-resolved kinetic energy is
\begin{equation}
\ddt{\ol \rho k_r} + \frac{\partial}{\partial x_j} (\ol \rho \widetilde u_j k_r) 
= \frac{\partial}{\partial x_j} \left(  \widetilde u_i \ol \tau_{ij} - \ol \rho \widetilde u_i \mathcal{T}_{ij}  - \widetilde u_j \ol{p'} \right) 
+ \frac{1}{Fr^2} \tilde u_i \overline \rho g_i
+ \ol{p'} \frac{\partial \widetilde u_i}{\partial x_i}
- \widetilde u_i \frac{\partial \langle p \rangle}{\partial x_i}
- \epsilon_{s}
- \epsilon.
\label{eq:scaleresolvedke}
\end{equation}
We decompose the pressure gradient using the RANS definition, as it is used in the DNS, as described above.
As a result, here we use
\begin{equation}
\frac{\partial \overline p}{\partial x_i} = \frac{\partial \langle p \rangle}{\partial x_i} 
+ \frac{\partial \overline{p'}} {\partial x_i} .
\label{eq:filtered_p}
\end{equation}
The average molecular viscous stress tensor is given by
%
%
%
\begin{equation}
\overline \tau_{ij} 
=  \overline { \mu \left( 
\frac{\partial  u_i}{\partial x_j} 
+ \frac{\partial  u_j}{\partial x_i} 
- \frac{2}{3} \frac{\partial  u_k}{\partial x_k} \delta_{ij} 
\right) } .
\label{eq:filtered_tau}
\end{equation}
and the molecular dissipation is
\begin{equation}
\epsilon = \ol \tau_{ij} \widetilde S_{ij},
\end{equation}
where the resolved strain is given by
\begin{equation}
\widetilde{S}_{ij} = \frac{1}{2} \left( \frac{\partial \widetilde u_i}{\partial x_j} + \frac{\partial \widetilde u_j}{\partial x_i} \right).
\end{equation}
We define the kinetic energy transfer between resolved and sub-filter scale kinetic energies as
\begin{equation}
\epsilon_{s} = -\overline{\rho} \, \mathcal{T}_{ij} \widetilde{S}_{ij} .
\end{equation}

From equation (\ref{eq:subscale_ke}), the sub-scale kinetic energy $k_s$ transitions to 0 in the NS limit, and to the finite RANS quantity in the RANS limit, as $\mathcal{T}_{ij}$ does.
In the NS limit, the Favre averaged velocity $\tilde{u}_i$ converges to the instantaneous velocity $\tilde u_i \rightarrow u_i$, and $\widetilde{u_i u_j} \rightarrow u_i u_j$, so the scale-resolved kinetic energy $k_r$ transitions to the total kinetic energy in the NS limit, $k_r \rightarrow u_i u_j / 2$.
In the RANS limit, $k_r$ converges to the RANS mean kinetic energy.

As in section \ref{sec:sr_stats_prop_summary}, for smooth flows, we can expect the governing equations for the scale resolved statistics $b$, $a_i$, $\mathcal{T}_{ij}$ and $k_s$ in equations 
(\ref{eq:trans_b} - \ref{eq:trans_Tij}), (\ref{eq:BHR_subscale_ke}) to have similar transitions to NS and RANS limits as the variables themselves. As a result, we expect the governing equations for the SR variables to transition to zero in the NS limit, and to the governing equations for their RANS counterparts in the RANS limit. The governing equation for the scale-resolved kinetic energy $k_r$ transitions to the governing equation for the total kinetic energy in the NS limit, and to the governing equation for the RANS mean kinetic energy in the RANS limit.


\section{Diagnostics for homogeneous variable density turbulence}
\label{sec:diagnostics}

We now investigate the scale-resolving generalized statistics using direct numerical simulations of homogeneous variable density turbulence from  [\onlinecite{aslangil_etal_2020}]. 
The HVDT DNS at the times indicated in Table \ref{tab:regime_times} is filtered using a Gaussian kernel as defined in equation (\ref{eq:filter_gaussian_kernel}). 
The filter width $w$ is given by,
\begin{equation}
w = f_0 \left ( \frac{f_1}{f_0}\right ) ^{ \left ( \frac{i-1}{N_w-1} \right ) } \Delta x
\label{eq:w_of_i}
\end{equation}
where $f_0=1/\pi$ and $f_1=512$ are parameters used to control the lower and upper bounds of $w$.
We perform diagnostics at $N_w = 15$ filter widths normalized by the box size $L=2 \pi$, varying between a fraction of the grid size, where $i=1$ and $w/L=(\Delta x/\pi)/L = 3.1\times10^{-4}$, and half the box size, where $i=N_w=15$ and $w=1/2$, such that $(\Delta x/\pi)/L \le w/L \le 1/2$, as listed in Table \ref{tab:filter_width_list}, as this range is observed to be large enough to contain the transition of the SR quantities between the NS and the RANS limits.

The DNS for this flow have many degrees of freedom, as the spatial resolution is $N^3=1024^3$, and there are several dynamical variables of interest. 
To simplify the analysis, a number of diagnostics are used to investigate scale dependence of the VDT statistics presented in section \ref{sec:filtering_and_RANSaveraging} and of the budgets in the governing equations for some of these statistics.
Since the flow is homogeneous, RANS statistics and, in general, volume averages have no spatial variability.
For this reason, we will first investigate volume-averaged SR statistics.
However, scale resolving statistics do vary in space, and this will be investigated too by looking at probability density function distributions as a function of filter width.

The evolution of the kinetic energy $\mathcal{R}_{ii}/2$ and kinetic energy production are shown in Figure \ref{fig:flow_regimes}. Throughout this section, we will be looking at diagnostics from HVDT DNS with $A=0.05$ and $A=0.75$ at the four times $t/t_r$ ($t_r = \sqrt{Fr^2/A}$) listed in Table \ref{tab:regime_times}, corresponding to the four regimes in HVDT, plotted in Figure \ref{fig:flow_regimes}. The simulations considered here have $Fr=1$.
These four times correspond to {\it i)} when the kinetic energy production peaks, at the end of the explosive growth regime, {\it ii)} when the kinetic energy peaks, at the end of the saturated growth regime, {\it iii)} the end of the fast decay regime when the net rate of decay of kinetic energy starts decreasing, and {\it iv)} at a time during the gradual decay regime.

\begin{table}[tb]
  \caption{\label{tab:regime_times}
  Normalized times $t/t_r$ ($t_r = \sqrt{Fr^2/A}$) at which diagnostics are computed at each Atwood number. }
  \begin{ruledtabular}
  \begin{tabular}{rcc}
  Time instance & $t/t_r$ for $A=0.05$ case& $t/t_r$ for $A=0.75$ case\\
  \hline
  End of explosive growth & 1.3 & 1.4 \\
  End of saturated growth & 2.3 &  2.4 \\
  End of fast decay & 4.3 & 5.9 \\
  Within gradual decay & 7.0 & 9.4 \\
  \end{tabular}
  \end{ruledtabular}
\end{table}

\begin{table}[tb]
  \caption{
  \label{tab:filter_width_list}
  List of filter number $i$ and their corresponding normalized filter widths $w/L$, obtained from (\ref{eq:w_of_i}). }
  \begin{ruledtabular}
  \begin{tabular}{rccc}
  $i$ & $w/L$ & $i$ & $w/L$\\
  \hline
    1  &   $(\Delta x/\pi)/L = 3.1\times10^{-4}$ &    9  &   $2.1\times10^{-2}$\\
    2  &   $5.3\times10^{-4}$                   &    10 &   $3.6\times10^{-2}$\\
    3  &   $8.9\times10^{-4}$                   &    11 &   $6.1\times10^{-2}$\\
    4  &   $1.5\times10^{-3}$                   &    12 &   $1.0\times10^{-1}$\\
    5  &   $2.6\times10^{-3}$                   &    13 &   $1.7\times10^{-1}$\\
    6  &   $4.3\times10^{-3}$                   &    14 &   $3.0\times10^{-1}$\\
    7  &   $7.4\times10^{-3}$                   &    15 &   $1/2 = 5.0\times10^{-1}$\\
    8  &   $1.2\times10^{-2}$                   &       &
    \end{tabular}
  \end{ruledtabular}
\end{table}


\subsection{Scale-resolving variables}

\begin{figure}[htb]
  \centering
  \includegraphics[trim=0 32 0 0, clip, width=0.3\textwidth]{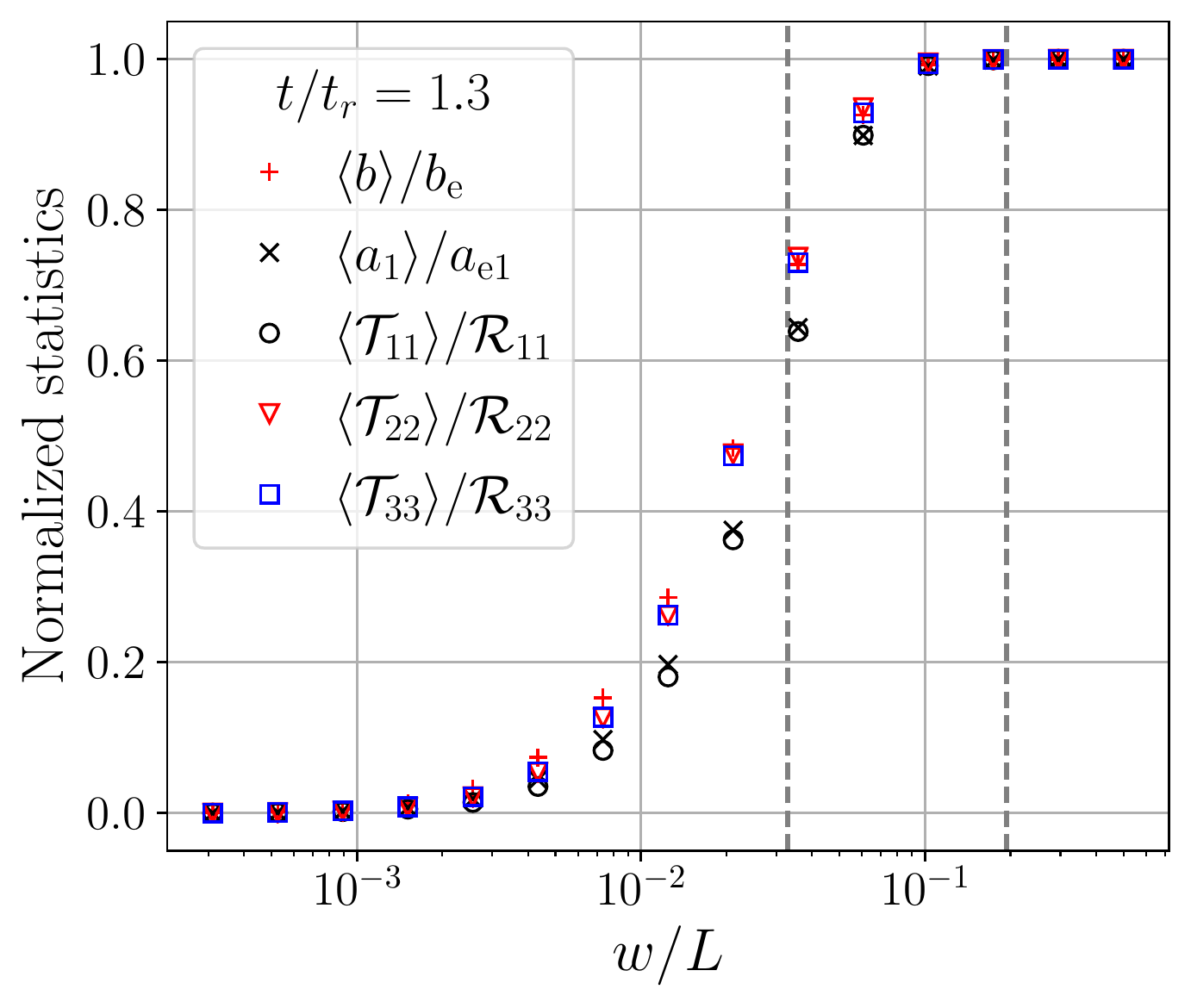}
  \includegraphics[trim=28 32 0 0, clip, width=0.28\textwidth]{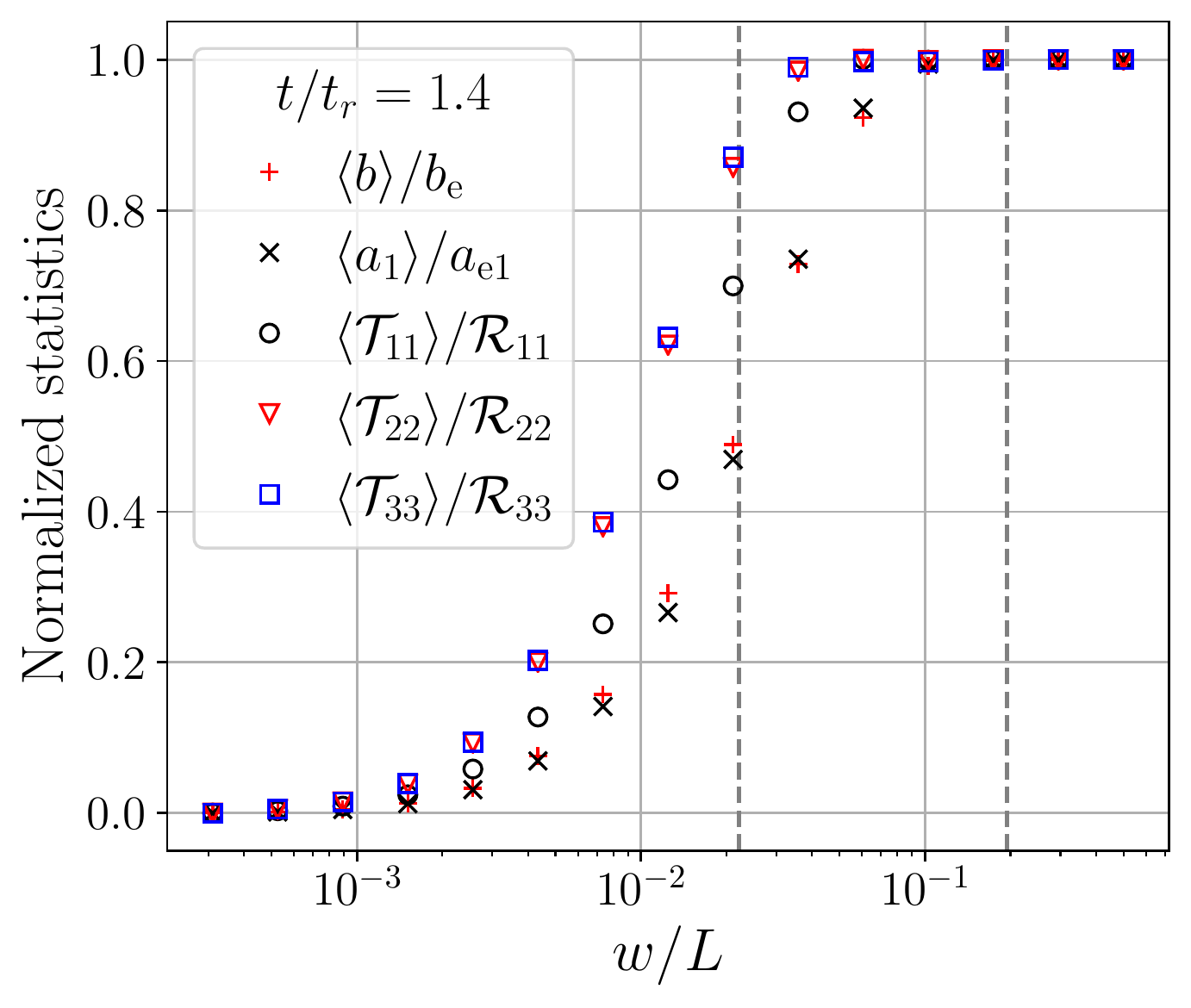}\\
  \includegraphics[trim=0 32 0 0, clip, width=0.3\textwidth]{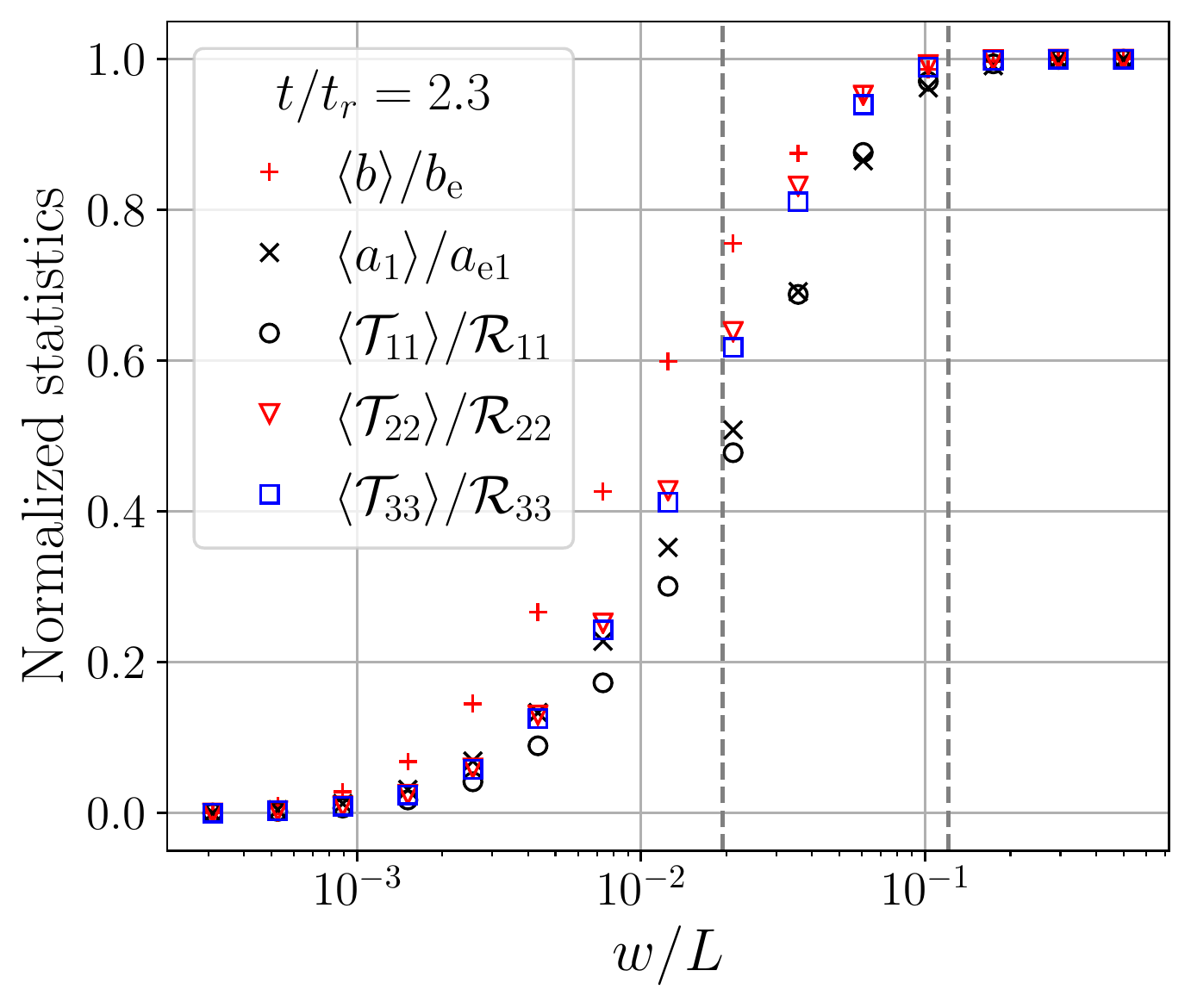}
  \includegraphics[trim=28 32 0 0, clip, width=0.28\textwidth]{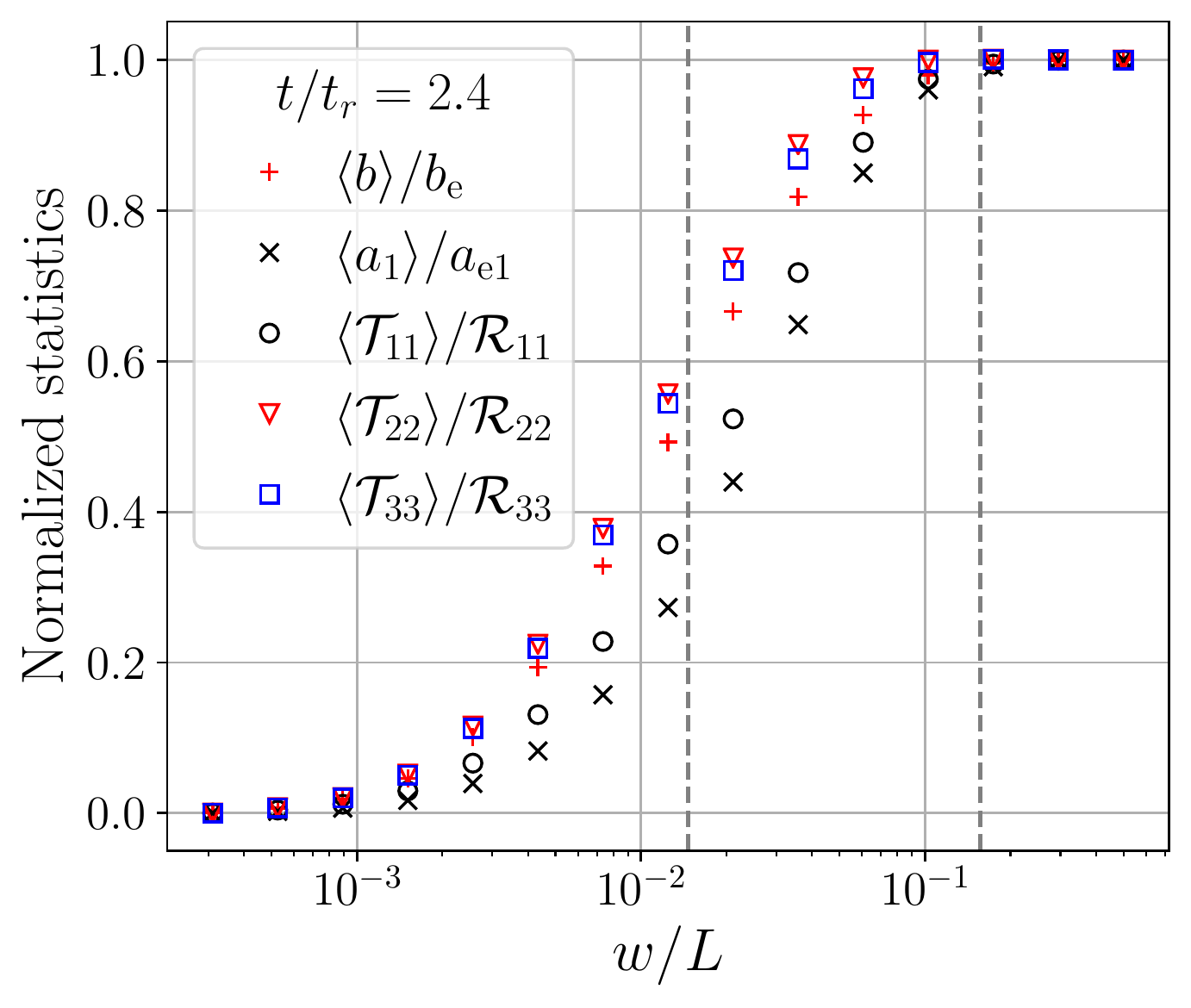}\\
  \includegraphics[trim=0 32 0 0, clip, width=0.3\textwidth]{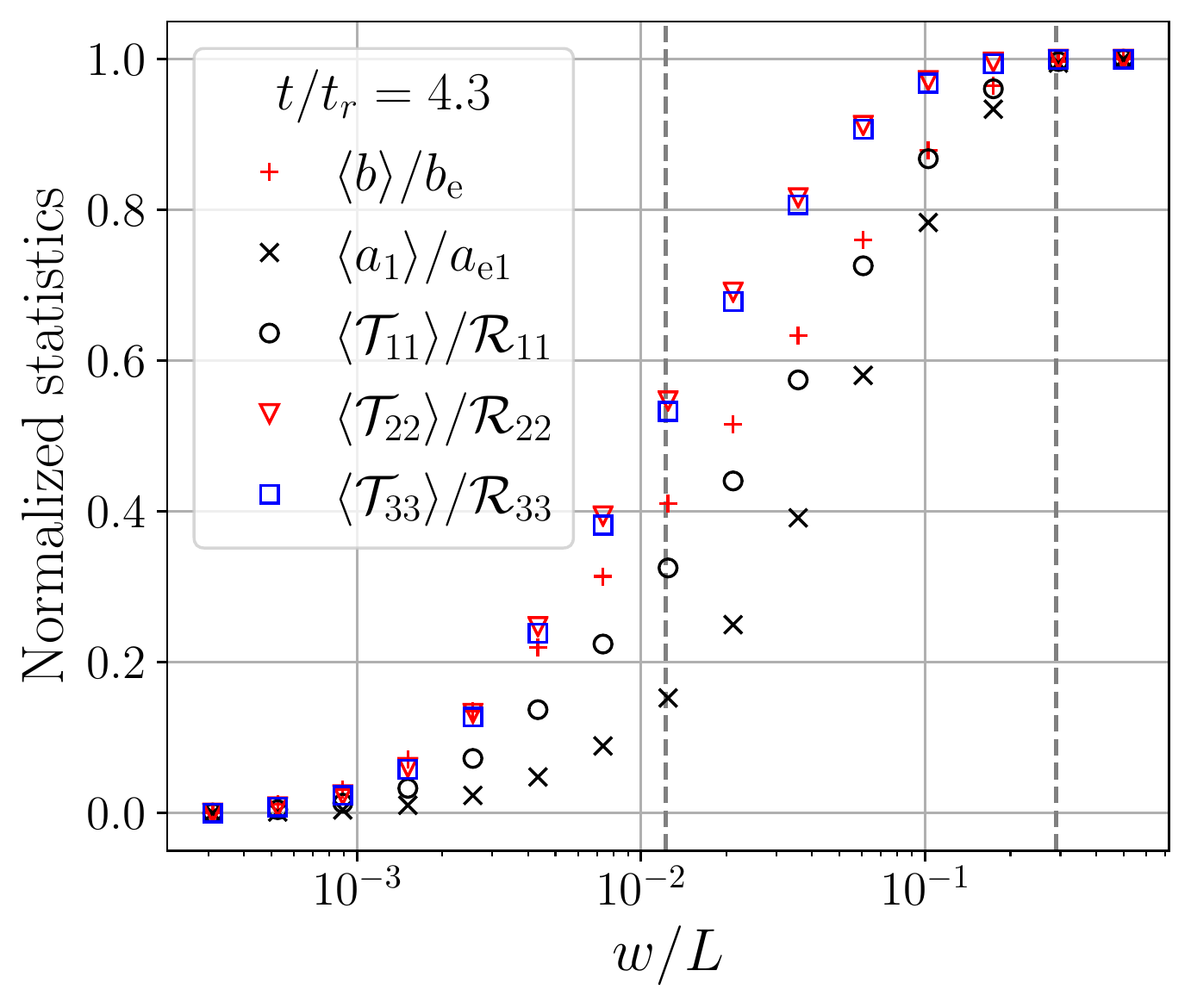}
  \includegraphics[trim=28 32 0 0, clip, width=0.28\textwidth]{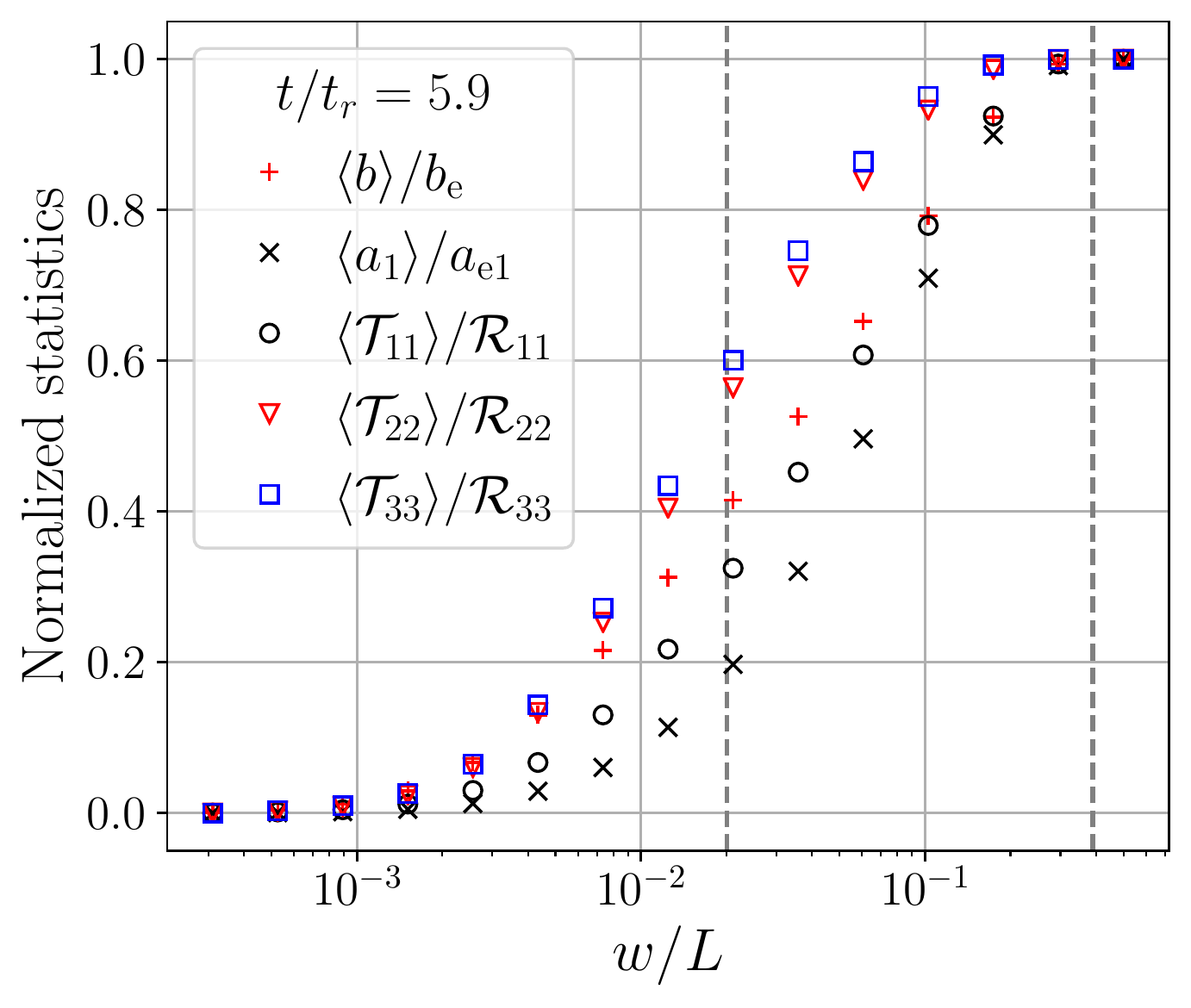}\\
  \includegraphics[trim=0 0 0 0, clip, width=0.3\textwidth]{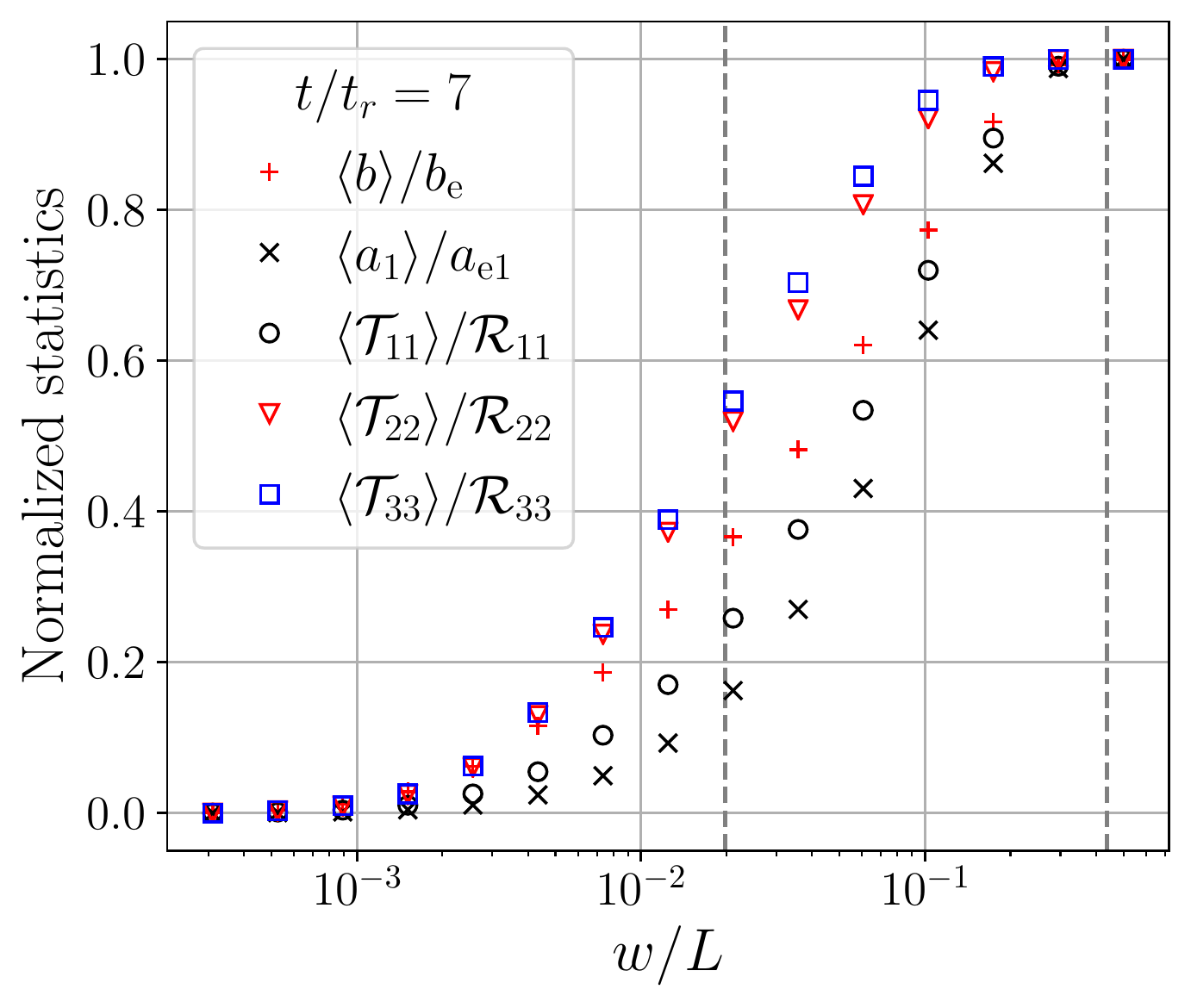}
  \includegraphics[trim=28 0 0 0, clip, width=0.28\textwidth]{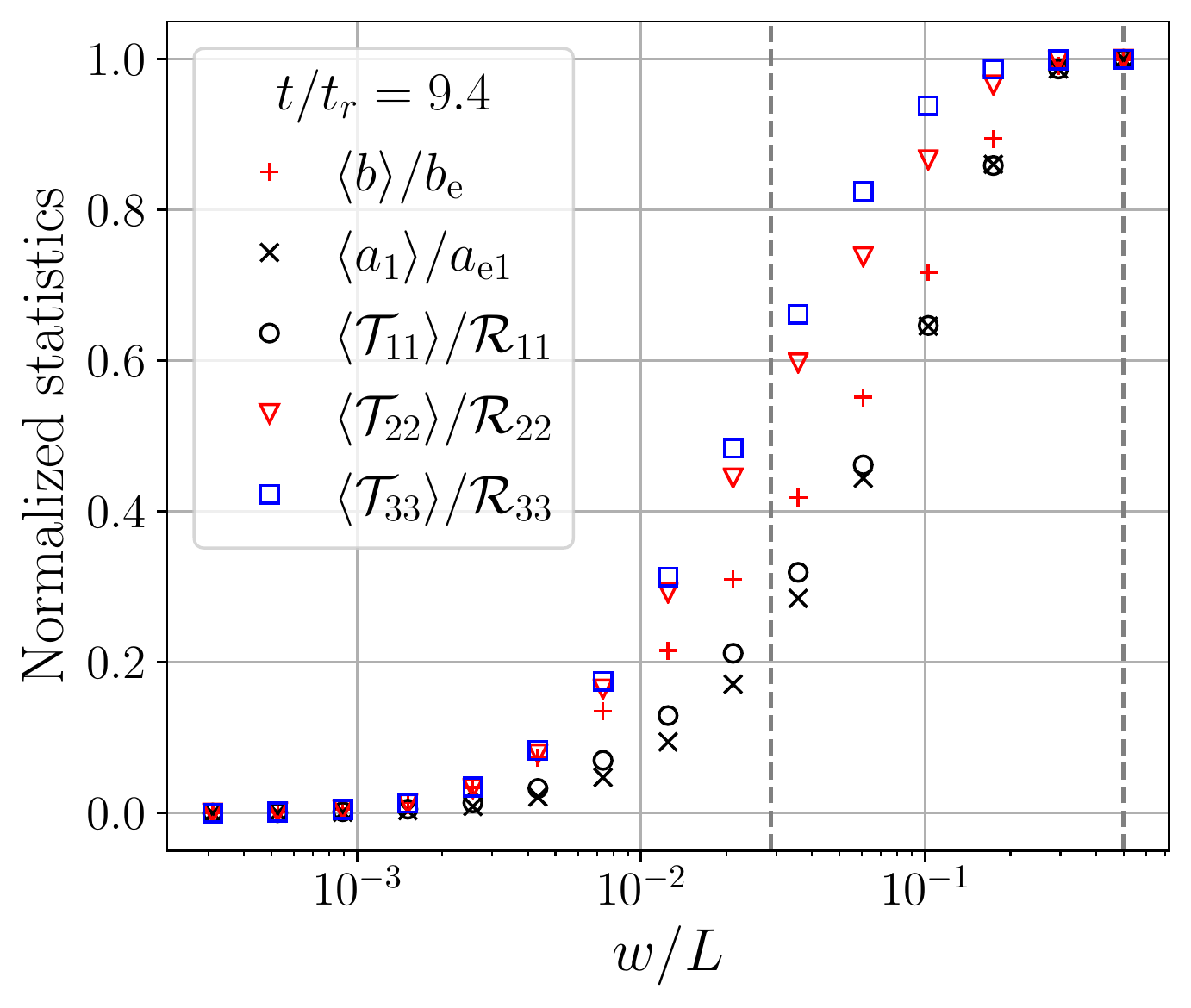}
  \caption{Volume averaged filtered quantities, normalized by their corresponding RANS counterparts, as a function of filter width normalized by the box size, $w/L$. The left and right columns correspond to $A=0.05$ and $A=0.75$, respectively. Dashed vertical lines indicate the horizontal Taylor micro-scale $\lambda_h$ and the integral length scale calculated based on the vertical velocity $\mathcal{L}_v$, with $\lambda_h < \mathcal{L}_v$.}
  \label{fig:vol_avg_stat_vars}
\end{figure}

Volume integrated SR variables are plotted, normalized by their RANS statistics counterparts, in Figure \ref{fig:vol_avg_stat_vars}. 
For the small Atwood number $A=0.05$ snapshot at the end of the explosive growth period, the first time shown, the statistics collapse reasonably well to a single curve.
However, at later times this is no longer the case for either $A=0.05$ or $A=0.75$ cases, and the spread between the curves increases until the rapid decay regime, after which the spread seems to either stabilize or slightly decrease.
The horizontal components of the turbulent stresses tensor $\mathcal{T}_{22}$ and $\mathcal{T}_{33}$ have the same scaling.
However, the vertical component $\mathcal{T}_{11}$ and the horizontal components $\mathcal{T}_{22}, \mathcal{T}_{33}$ of the stress tensor, the vertical component $a_1$, and $b$, each have different length scalings, and these scalings vary in time. 

We compare the scaling of the SR variables to smallest Taylor micro-scale, which for this flow corresponds to the horizontal Taylor micro-scale,
\begin{equation}
  \lambda_h = \frac{1}{2} \sum_{\beta=2}^3 \sqrt{ \frac{\langle u'^2_\beta \rangle}{ \left\langle \left(\frac{\partial u'_\beta}{\partial x_\beta}\right)^2\right\rangle}},
  \end{equation}
and to the largest integral length scale, namely, based on the vertical velocity,
%
%
\begin{equation}
  \mathcal{L}_v = \frac{2\pi\int_0^\infty k^{-1}E_{u_1}(k) dk}{\int_0^\infty E_{u_1}(k) dk}, \quad
\end{equation}
which are shown as vertical lines in Figure \ref{fig:vol_avg_stat_vars}.
The volume integrated SR variables converge to their RANS values when the filter width is comparable to the vertical integral length scale, $w\approx\mathcal{L}_v$, which is about an order of magnitude smaller than the box size for these DNS.
The NS values are reached at length scales that are at least one order of magnitude smaller than $\lambda_h$.
In the low Atwood number simulation, the NS limit is reached at larger length scales at early times, and at smaller length scales at later times, reflecting the population of smaller length scales as the turbulence develops in time.
However, in the large Atwood number simulation, the NS limit is reached at more or less the same length scale for the first three times, and at slightly larger length scales for the last snapshot.
Note that the volume integrated density $\langle \ol \rho \rangle$ is not shown in Figure \ref{fig:vol_avg_stat_vars}, as $\langle \ol \rho \rangle = \langle \rho \rangle = (\rho_1 + \rho_2)/2$ is constant throughout the flow.

\begin{figure}[htb]
  \centering
  \includegraphics[width=0.321\textwidth]{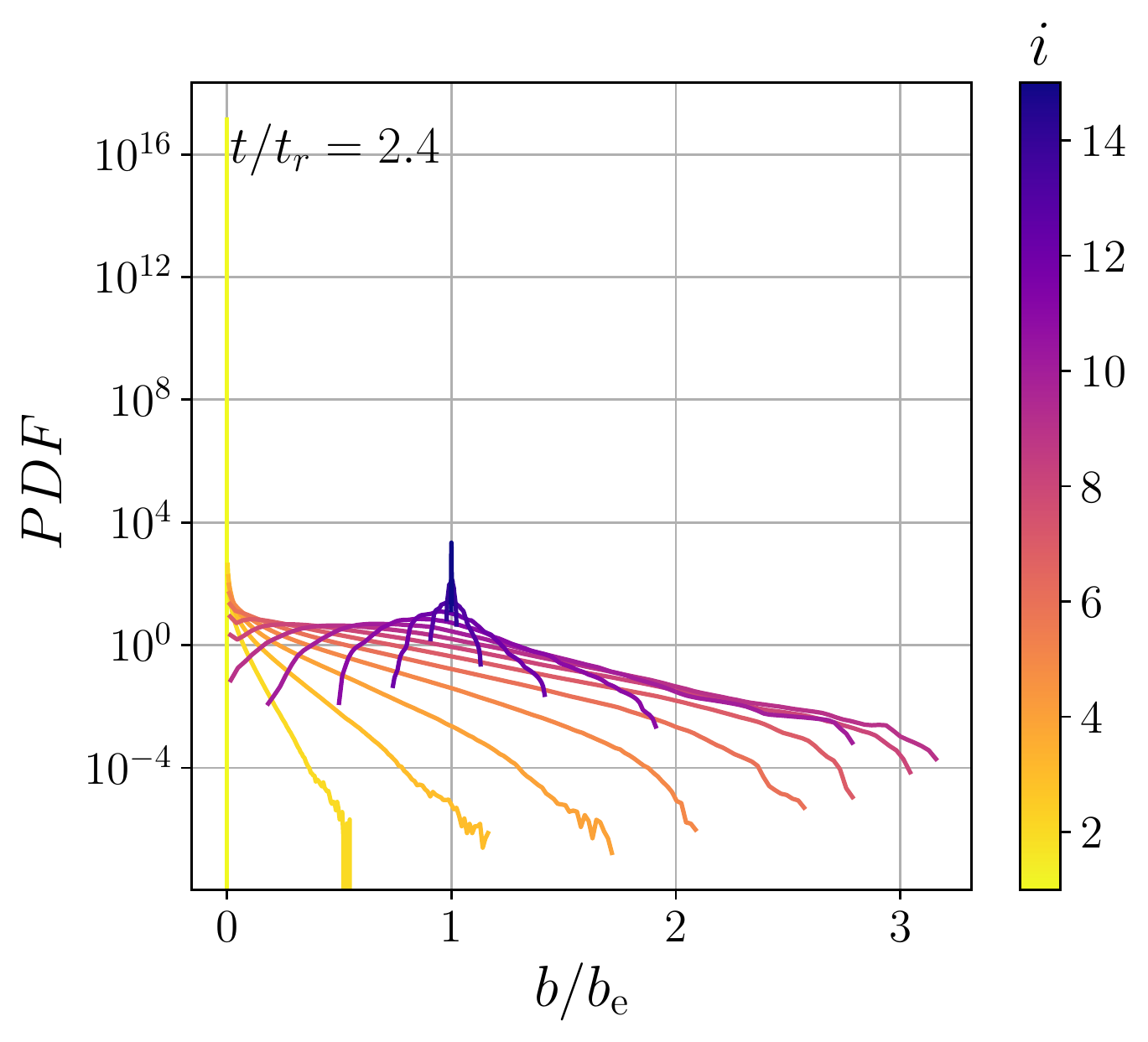}
  \includegraphics[trim=28 0 0 0, clip, width=0.3\textwidth]{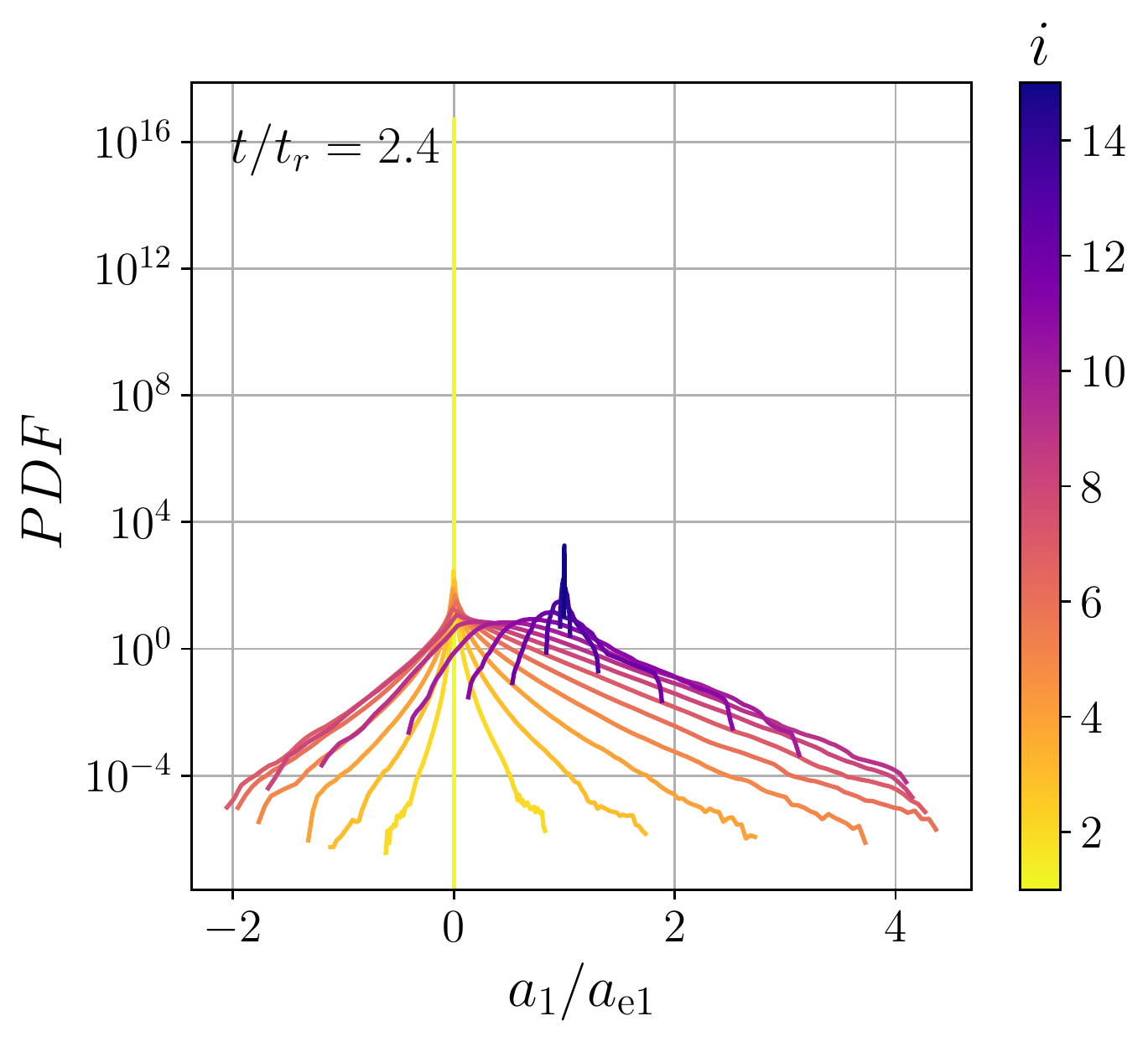}\\
  \includegraphics[width=0.321\textwidth]{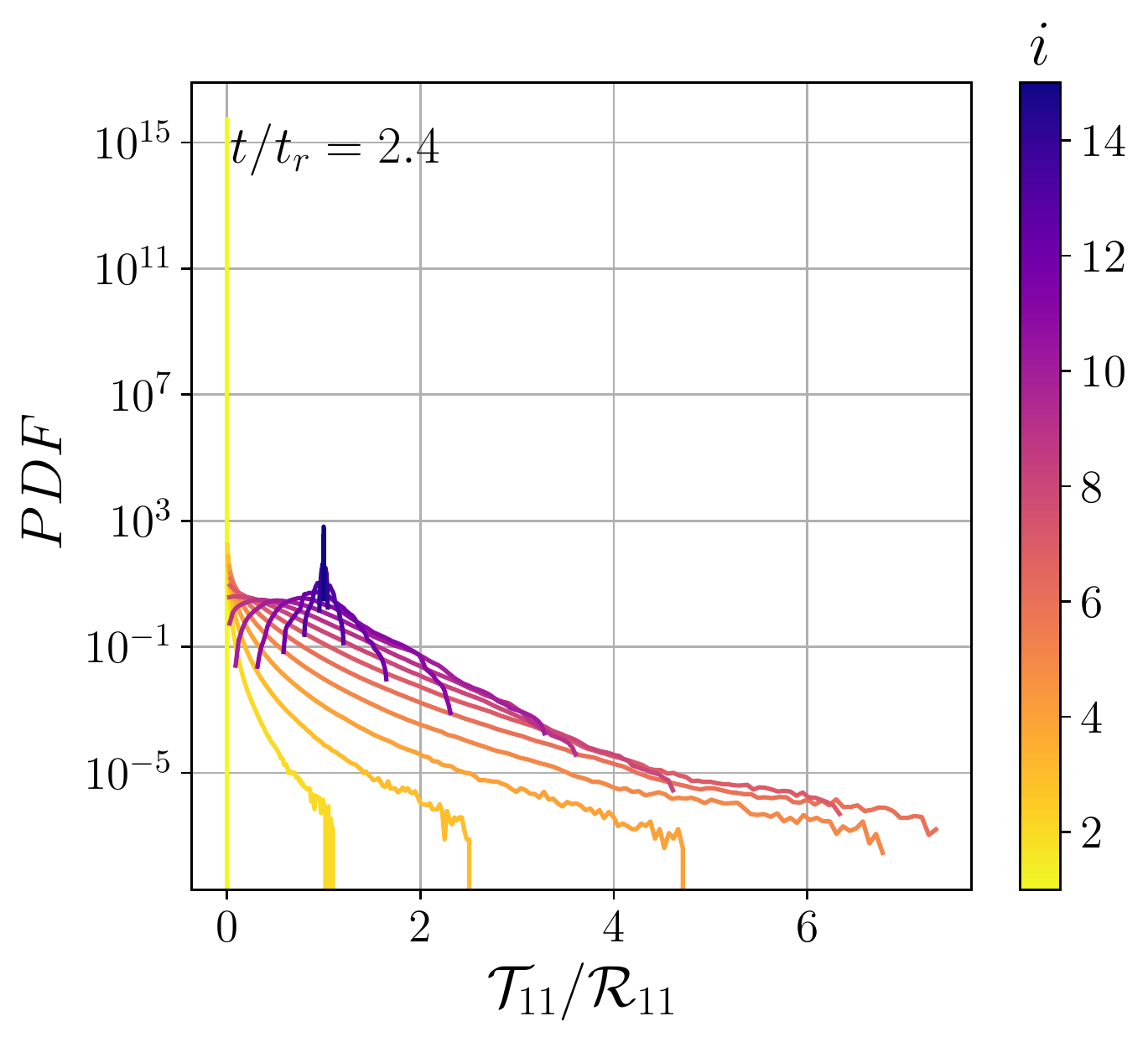}
  \includegraphics[trim=28 0 0 0, clip, width=0.3\textwidth]{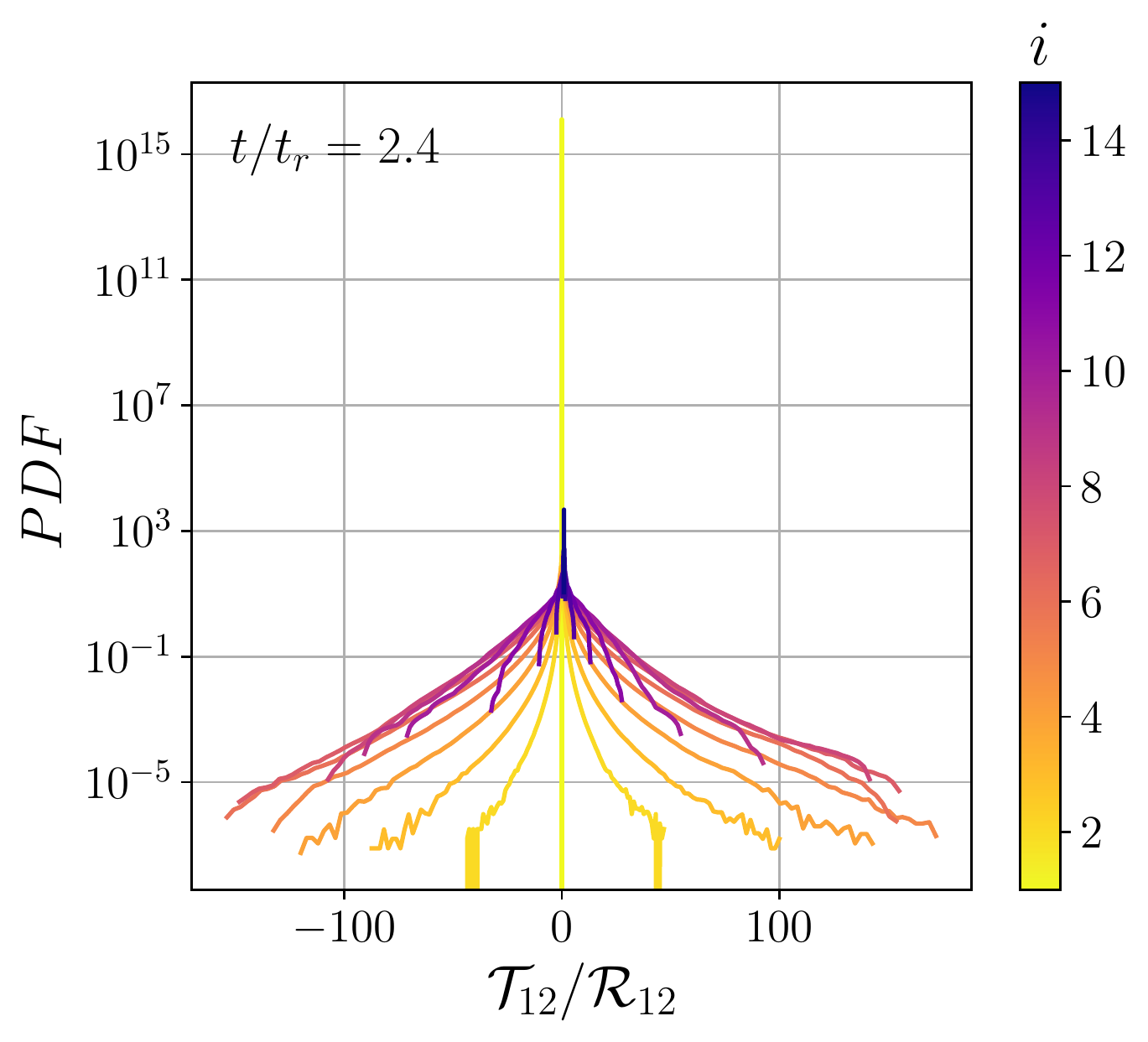}
  \caption{ Probability density functions for $b$, $a_1$, $\mathcal{T}_{11}$ and $\mathcal{T}_{12}$, normalized by their RANS counterparts, at $t/t_r = 2.4$ when $\mathcal{R}_{ii}$ peaks, for $A=0.75$. The colorbar indicates the filter width used, according to (\ref{eq:w_of_i}), Table \ref{tab:filter_width_list}, where $i=1$ corresponds to the smallest filter width $w/L = (\Delta x/\pi)/L = 3.1\times10^{-4}$ and $i=15$ corresponds to the largest filter width $w/L = 1/2$.}
  \label{fig:pdf_b_a1_T11_T12}
\end{figure}

At any given time, at intermediate filter widths, the SR statistical quantities have large variability in space. To quantify this, we compute the probability density function of $b$, $a_1$, $\mathcal{T}_{11}$ and $\mathcal{T}_{12}$, normalized by their RANS counterparts, shown in Figure \ref{fig:pdf_b_a1_T11_T12} for $A=0.75$ at the time when $\mathcal{R}_{ii}$ peaks. 
The realizability conditions (\ref{eq:realizability_b}) and (\ref{eq:realizability_Tij}) for $b$ and $\mathcal{T}_{ij}$, respectively, are satisfied by the Gaussian filter we use here.
The yellow line corresponding to $i=1$ shows that at the NS limit, where the filter size is a fraction of a grid cell, $w/L = (\Delta x/\pi)/L = 3.1\times10^{-4}$, all quantities are zero everywhere in the domain, and the PDFs correspond to a delta function centered at 0.
As the filter width increases, the range of values in the PDFs broaden until the filter width at which the widest PDF is observed is reached, namely $i=9$, $w/L=2.1\times10^{-2}$, for $b$, where the filter width is just around the horizontal Taylor micro-scale $\lambda_h$ (Figure \ref{fig:vol_avg_stat_vars}), $i=6$, $w/L=4.3\times10^{-3}$ for $a_1$ and $\mathcal{T}_{11}$, and $i=5$, $w/L=2.6\times10^{-3}$ for $\mathcal{T}_{12}$.
At the broadest shape of the PDFs, the variability is such that values of about 3, 4 and 6 times the RANS values are observed in $b$, $a_1$ and $\mathcal{T}_{11}/\mathcal{R}_{11}$, respectively, while for $\mathcal{T}_{12}/\mathcal{R}_{12}$, this variability is much larger, and values of up to 100 times the RANS values are observed. 
Subsequently, as $w/L$ continues to increase, the PDFs become narrower until eventually they again become a delta function at the RANS limit, $w/L=1/2$.
In the RANS limit, illustrated by the dark blue lines at $i=15$ corresponding to $w/L=1/2$, the SR quantities have no spatial variability as they reach their RANS values, where $b/b_\mathrm{e} = 1$, $a_1/a_{\mathrm{e}1} = 1$, $\mathcal{T}_{11}/\mathcal{R}_{11} = 1$ and $\mathcal{T}_{12}/\mathcal{R}_{12} = 1$. Similar behavior is observed at other times as well as for $A=0.05$.

\begin{figure}[t]
  \centering
  \includegraphics[width=0.18\textwidth]{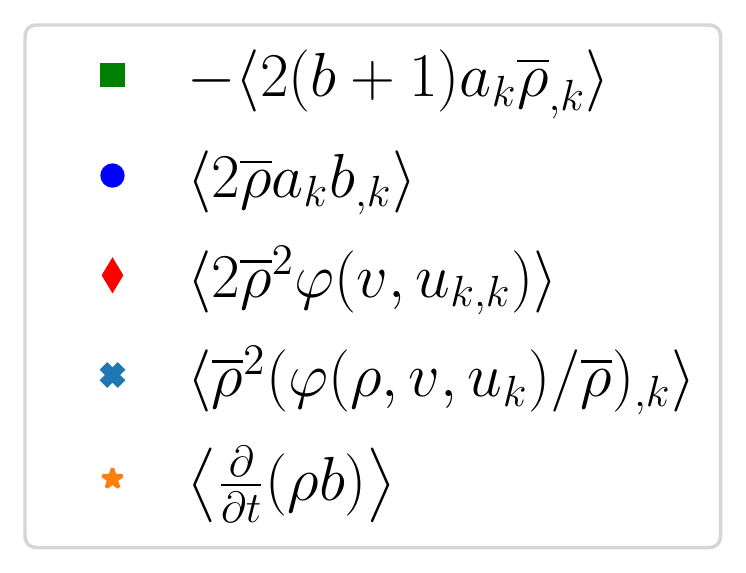}
  \includegraphics[trim=0 32 0 0, clip, width=0.3\textwidth]{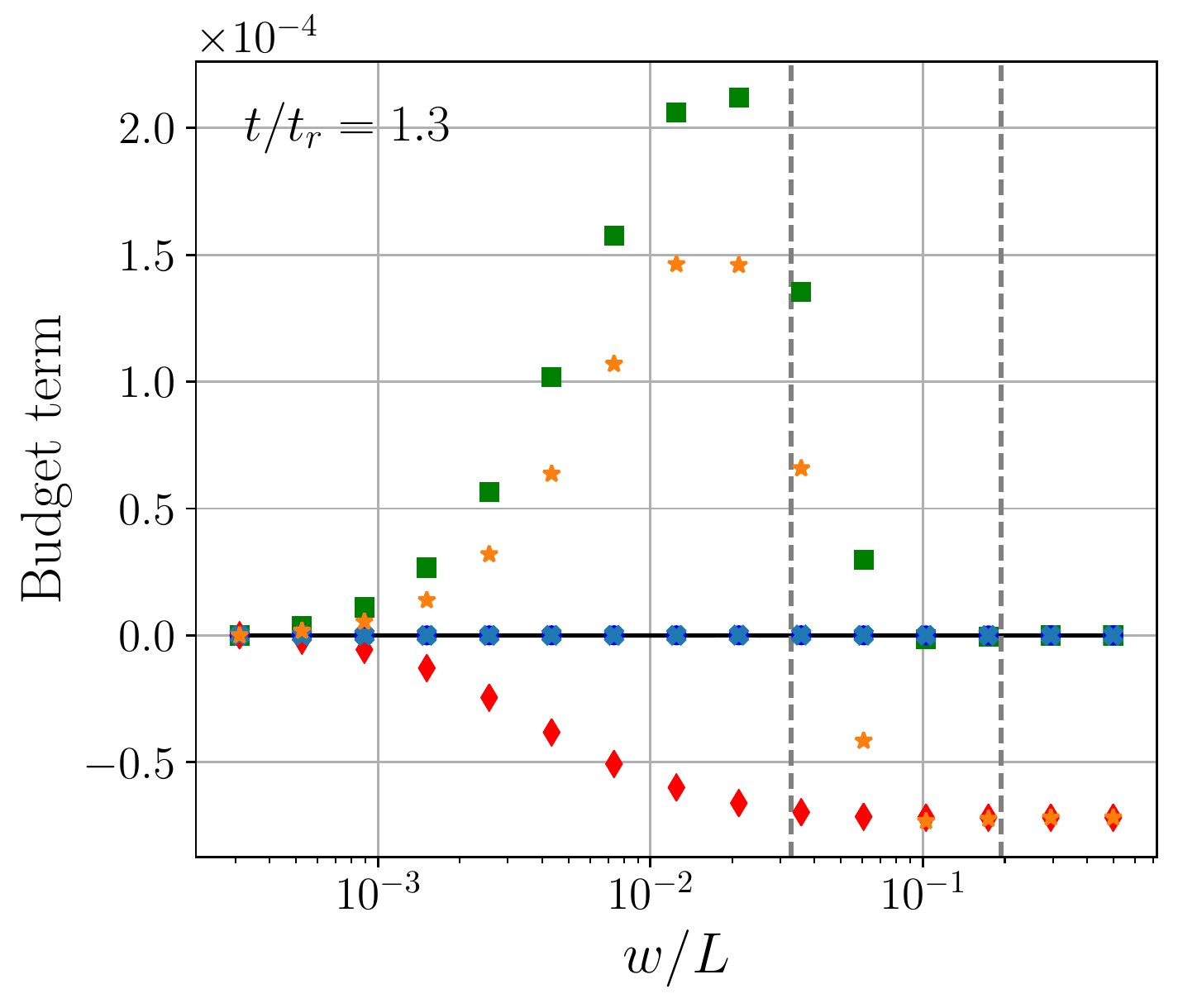}
  \includegraphics[trim=28 32 0 0, clip, width=0.28\textwidth]{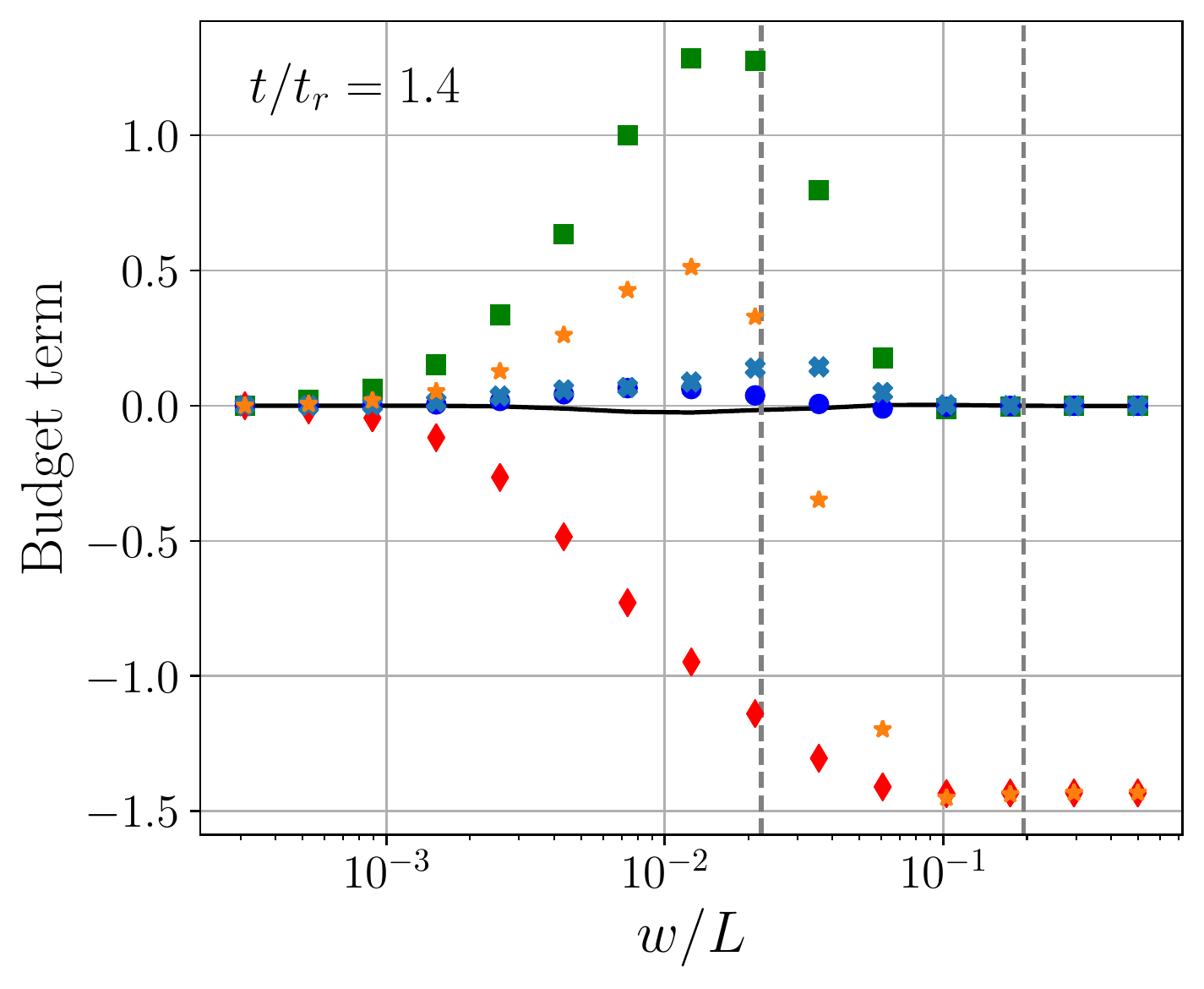}\\
  \hspace{0.19\textwidth}
  \includegraphics[trim=0 32 0 0, clip, width=0.29\textwidth]{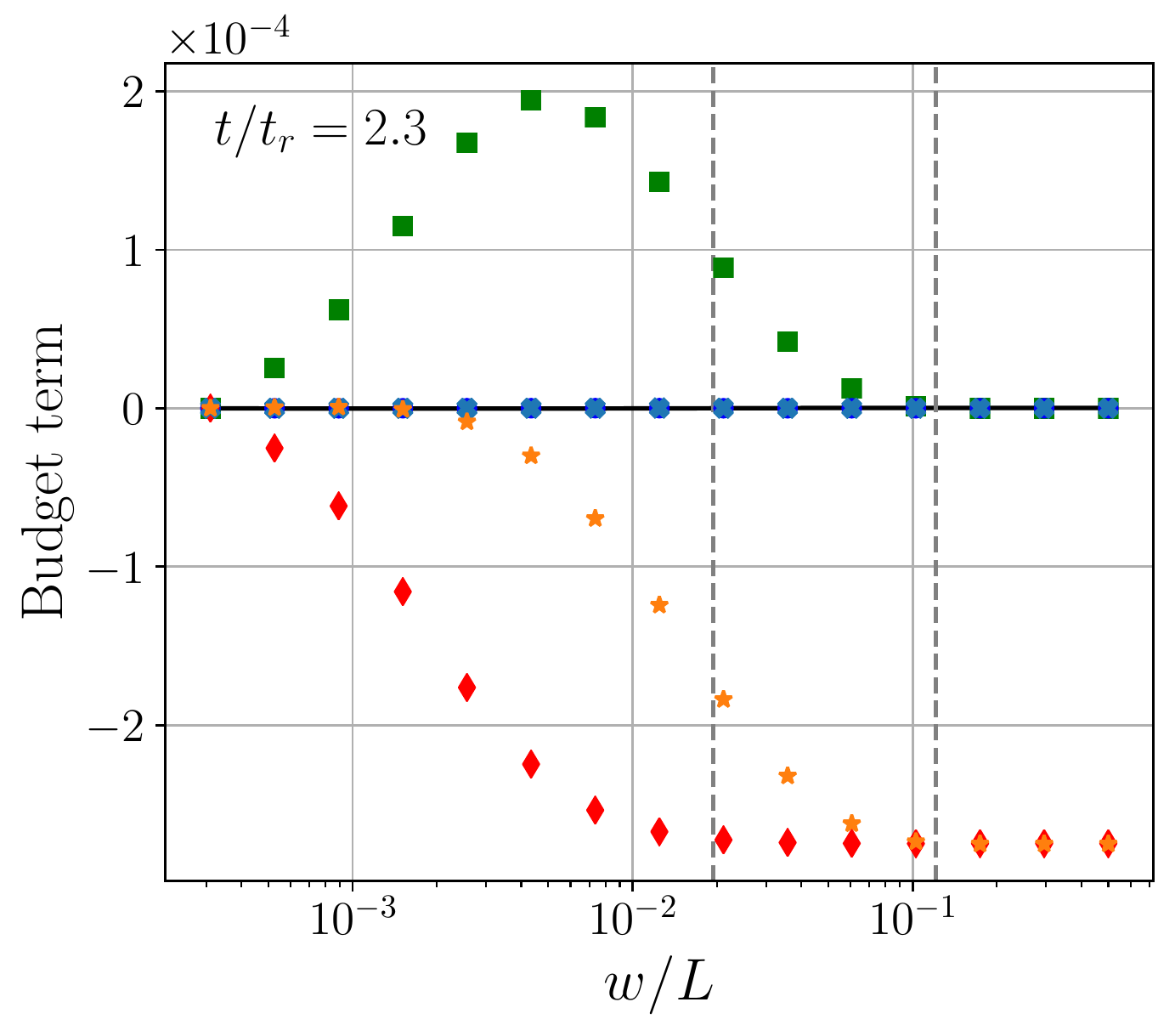}
  \includegraphics[trim=28 32 0 0, clip, width=0.28\textwidth]{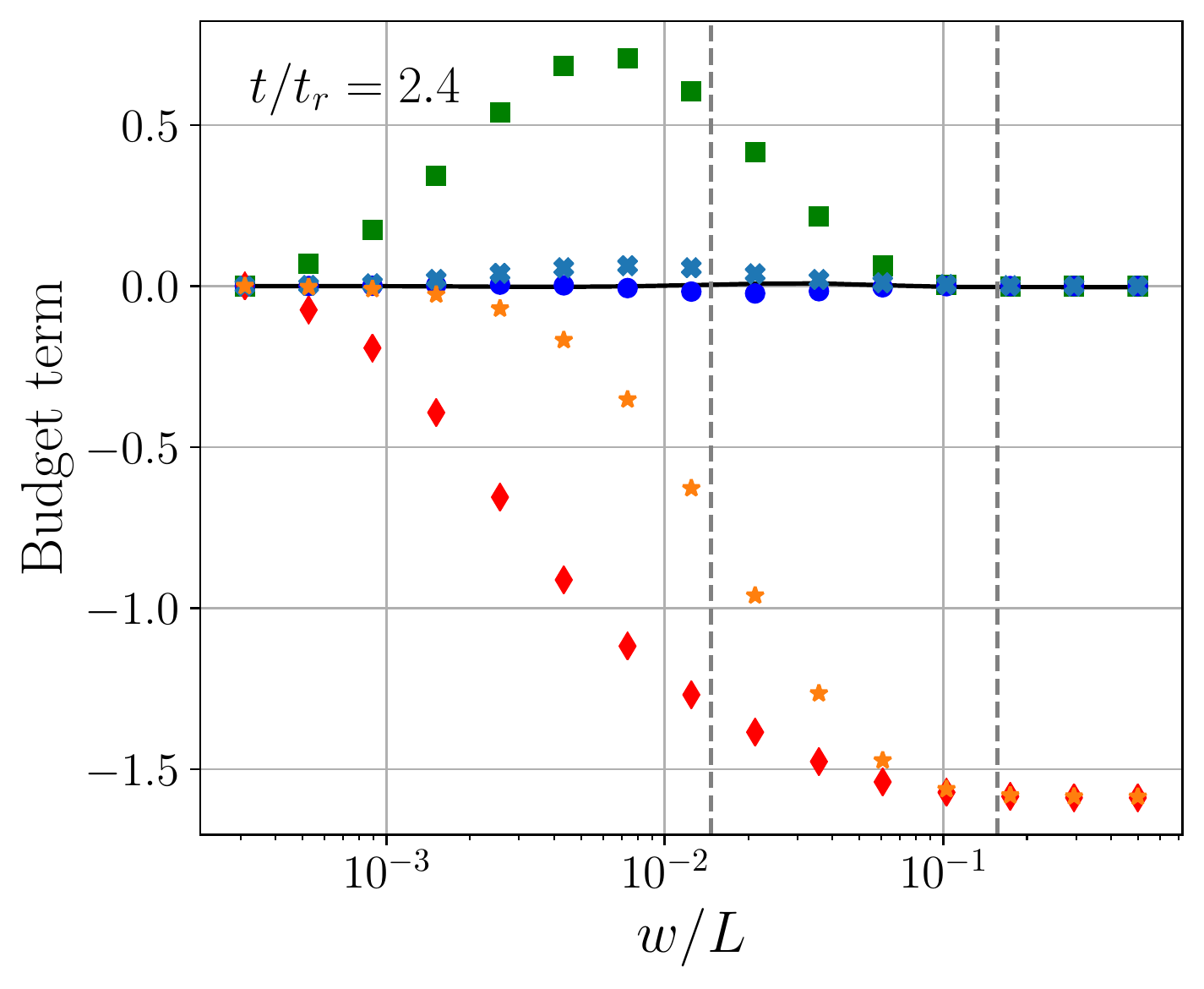}\\
  \hspace{0.18\textwidth}
  \includegraphics[trim=0 32 0 0, clip, width=0.3\textwidth]{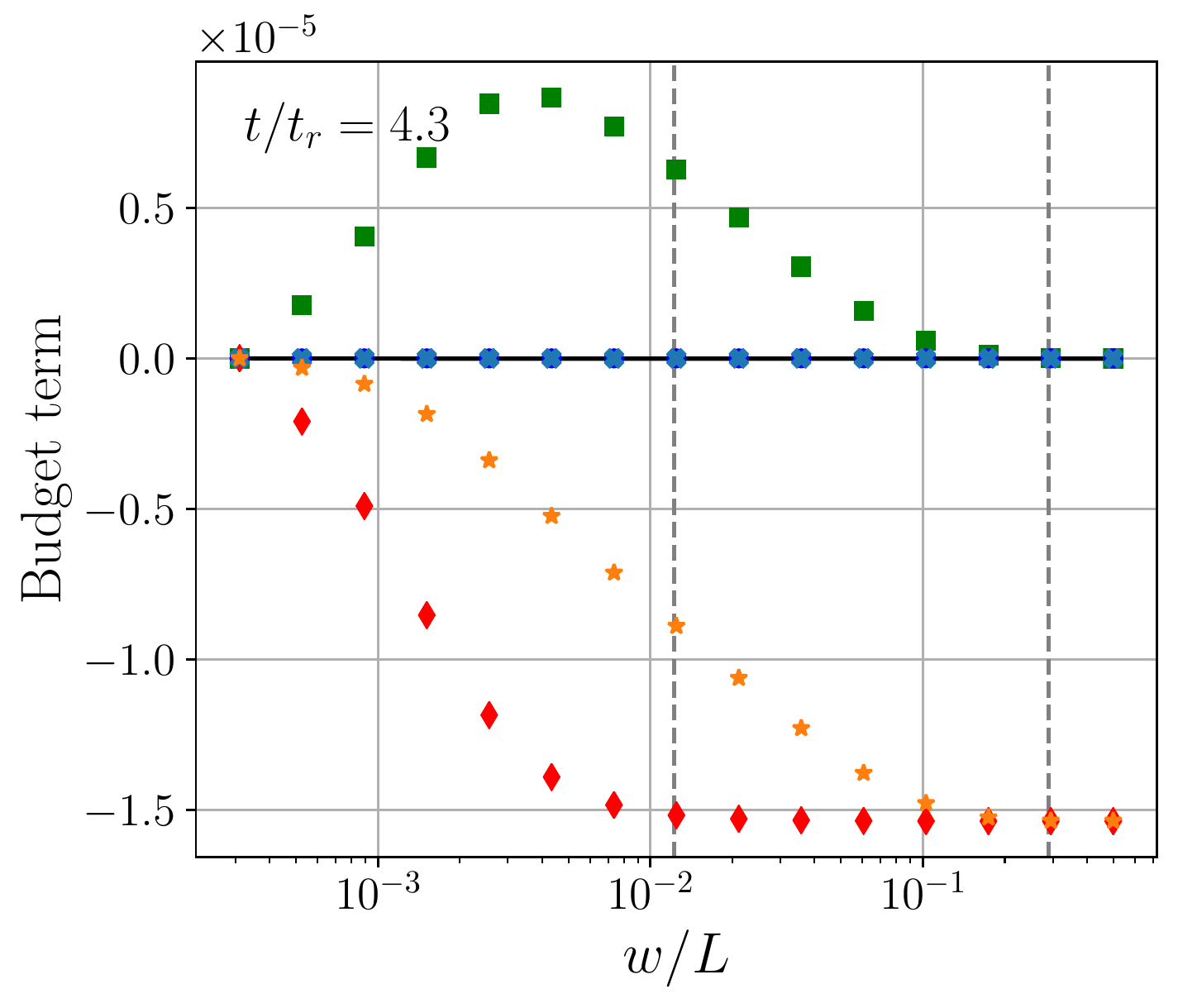}
  \includegraphics[trim=28 32 0 0, clip, width=0.28\textwidth]{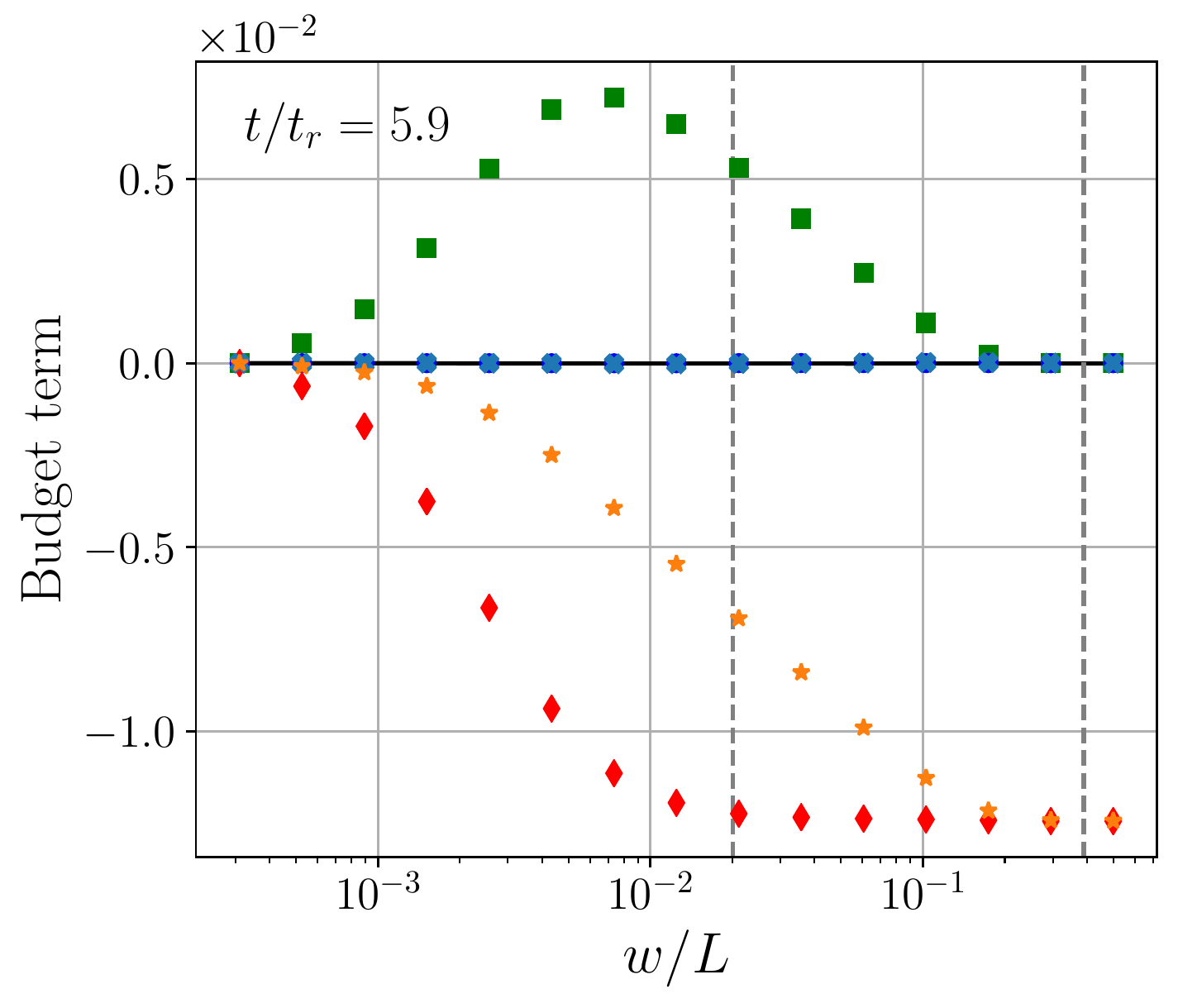}\\
  \hspace{0.18\textwidth}
  \includegraphics[trim=0 0 0 0, clip, width=0.3\textwidth]{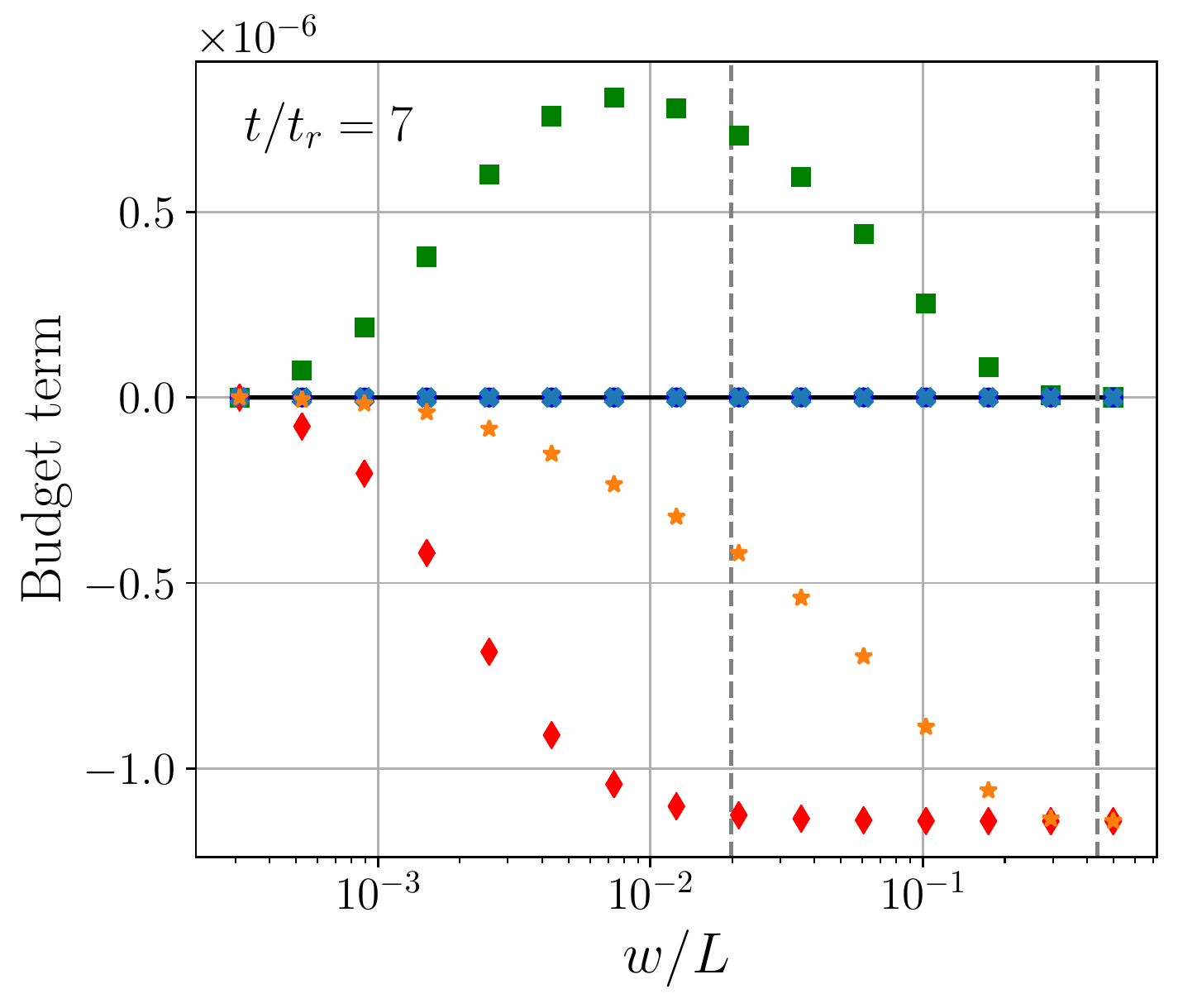}
  \includegraphics[trim=28 0 0 0, clip, width=0.28\textwidth]{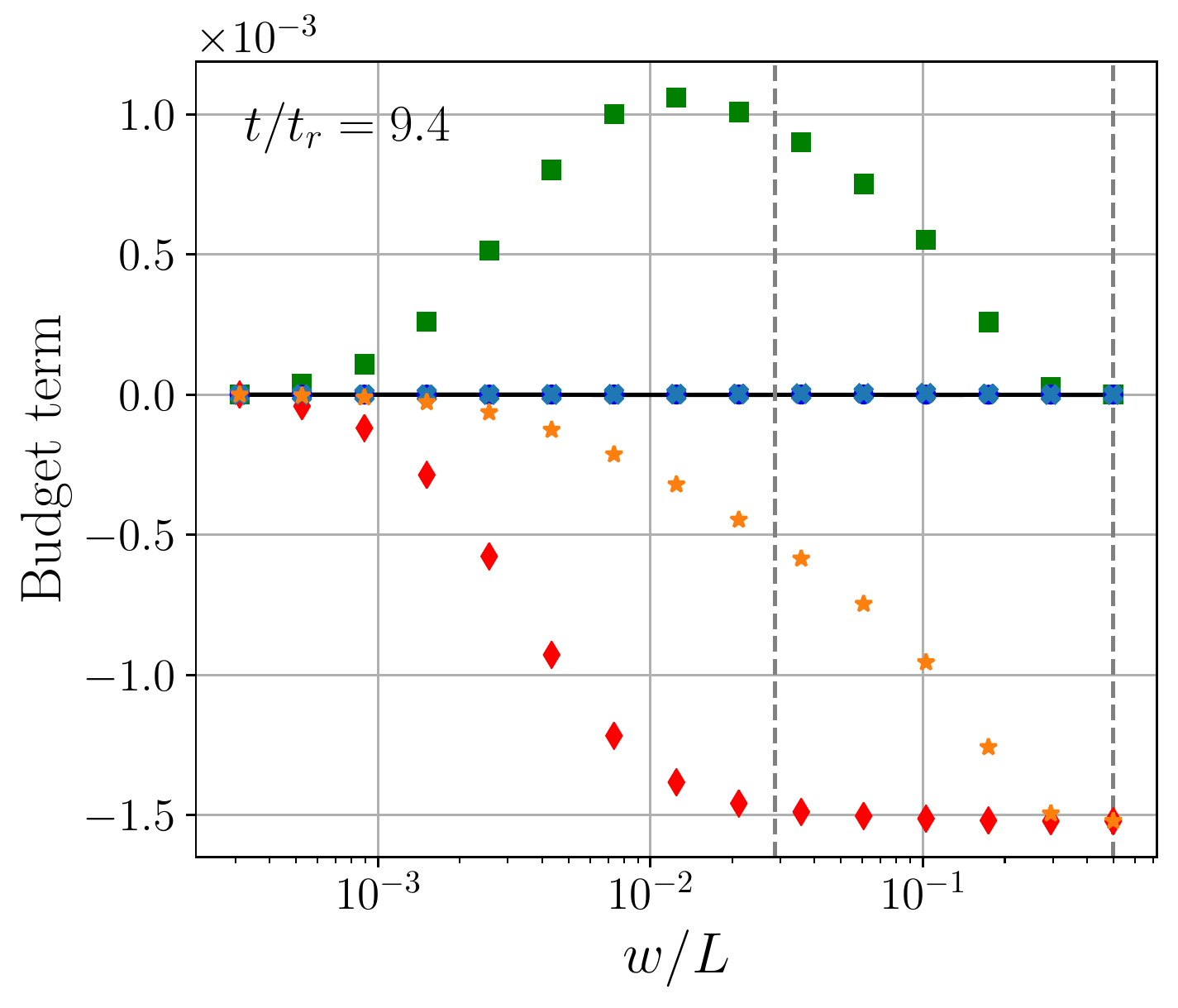}
  \caption{Volume averaged budget terms  for $b$ transport equation (\ref{eq:trans_b_volume_integrated}), as a function of filter width normalized by the box size, $w/L$. Note the difference in scales. Left and right columns correspond to $A=0.05,\ 0.75$ cases, respectively. Taylor micro-scale $\lambda_h$ and integral length scale $\mathcal{L}_v$, with $\lambda_h < \mathcal{L}_v$, are shown with vertical dashed lines. Black line indicates the residual. }
  \label{fig:b_budget}
\end{figure}

\subsection{Volume integrated budgets}

We now investigate the governing equation for $b$. Volume-integrating equation (\ref{eq:trans_b}) results in
\begin{eqnarray}
\left\langle \ddt{\ol \rho b } \right\rangle
= - \left\langle 2 (b+1) a_k \ol \rho_{,k} \right\rangle
+ \left\langle 2 \ol \rho a_k b_{,k} \right\rangle
+ \left\langle \ol \rho^2 \left( \frac{ \varphi(\rho, v, u_k)}{\ol \rho} \right)_{,k} \right\rangle
+ \left\langle 2 \ol{\rho}^2 \varphi(v, u_{k,k}) \right\rangle,
\label{eq:trans_b_volume_integrated}
\end{eqnarray}
where the terms on the right hand side correspond to net, volume integrated production, redistribution, turbulent transport, and destruction, respectively, and where the volume integrated advection term becomes zero due to spatial homogeneity of the flow.
These terms are plotted in Figure \ref{fig:b_budget} as they appear in in equation (\ref{eq:trans_b_volume_integrated}), and the residual is shown as the black line, indicating that the budget is closed to an excellent level of accuracy.
From the realizability condition for $b$ in equation (\ref{eq:realizability_b}), and since $\ol \rho \ge \rho_1>0$ as discussed at the end of section \ref{sec:innerprod_realizability}, $\ol \rho b$ is always positive.
Consequently, budget terms that cause $\ol \rho b$ to increase and decrease appear as positive and negative terms in Figure \ref{fig:b_budget}, respectively, and negative $\partial \ol \rho b / \partial t$ indicates a decrease of $\ol \rho b$ with time.
As with $b$, and in general all the SR quantities (Figure \ref{fig:vol_avg_stat_vars}), the terms governing the rate of change of $b$ in time transition smoothly between zero at the NS limit and their RANS values in the RANS limit.
In the latter, we recover the RANS budget for HVDT, in which $b$ is set by the configurational value given by the initial blob distribution, and then decays monotonically in time (see e.g. \cite{livescu_ristorcelli_2007}).
This way, the only term in the budget of the RANS description that is active is the destruction term, consistent with Figure \ref{fig:b_budget}.
The volume integrated destruction term varies monotonically between 0 in the NS limit and its RANS value in the RANS limit. 
However, during the time of peak conversion of PE to TKE, or at the end of the explosive growth regime, the volume integrated rate of change of $\ol \rho b$ does not change monotonically between the NS and RANS limits.
This is because the net, volume integrated production of $\ol \rho b$ is non-zero and can be larger than the rate of destruction at intermediate filter widths.
During the explosive growth regime, stirring occurs first at large scales, which is followed in time by the formation of structures at progressively smaller scales \cite{aslangil_etal_2020}. 
Consequently, the generalized variance in equation (\ref{eq:b_inner_expectedvalue}) is at first larger at larger scales, and at small scales it increases with time.
At later times after the end of the explosive growth regime, as turbulence becomes more developed, this strong variance is destroyed by mixing of the two fluids.
As a result, the destruction of $\ol \rho b$ at first is equal, and than greater than, the production as the filter width increases from the mesh size in the NS limit. 
This leads to a monotonically increasing (in magnitude) net decay of $\left\langle \ol \rho b \right\rangle$ as the filter width increases.
For the high Atwood number case, during the first two regimes leading up to the time of peak kinetic energy, the volume integrated transport of $\ol \rho b$ is positive, but small.

\begin{figure}[htb]
  \centering
  \includegraphics[width=0.15\textwidth]{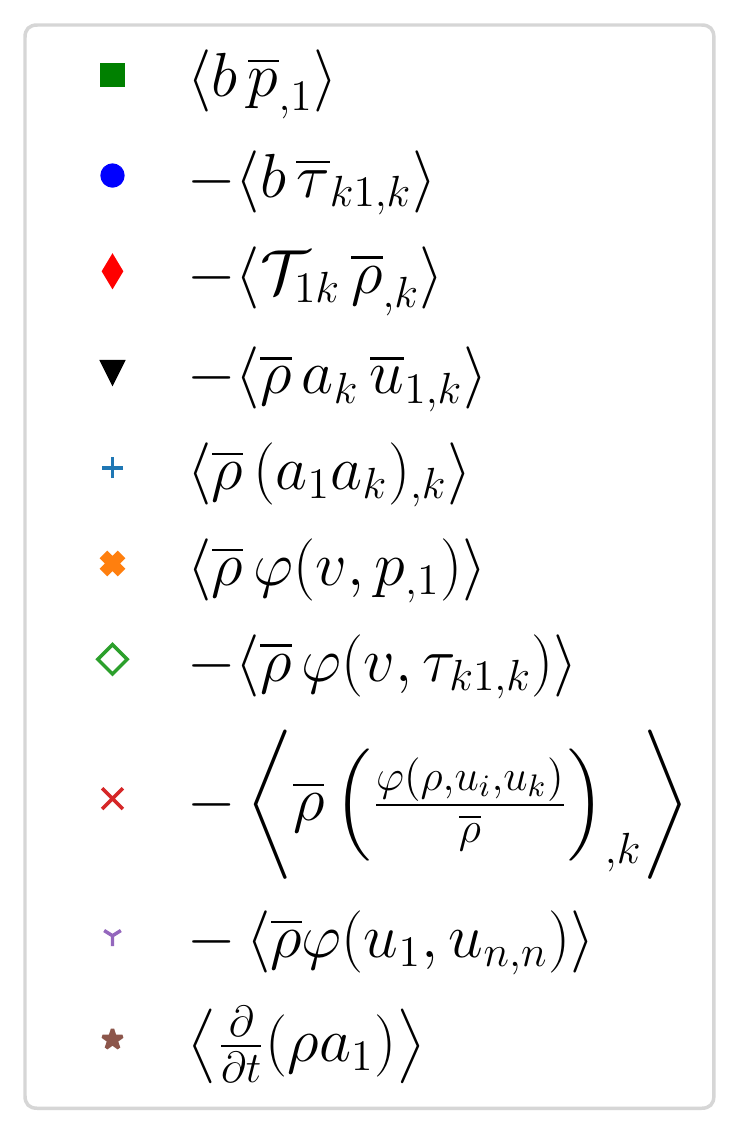}
  \includegraphics[trim=0 32 0 0, clip, width=0.3\textwidth]{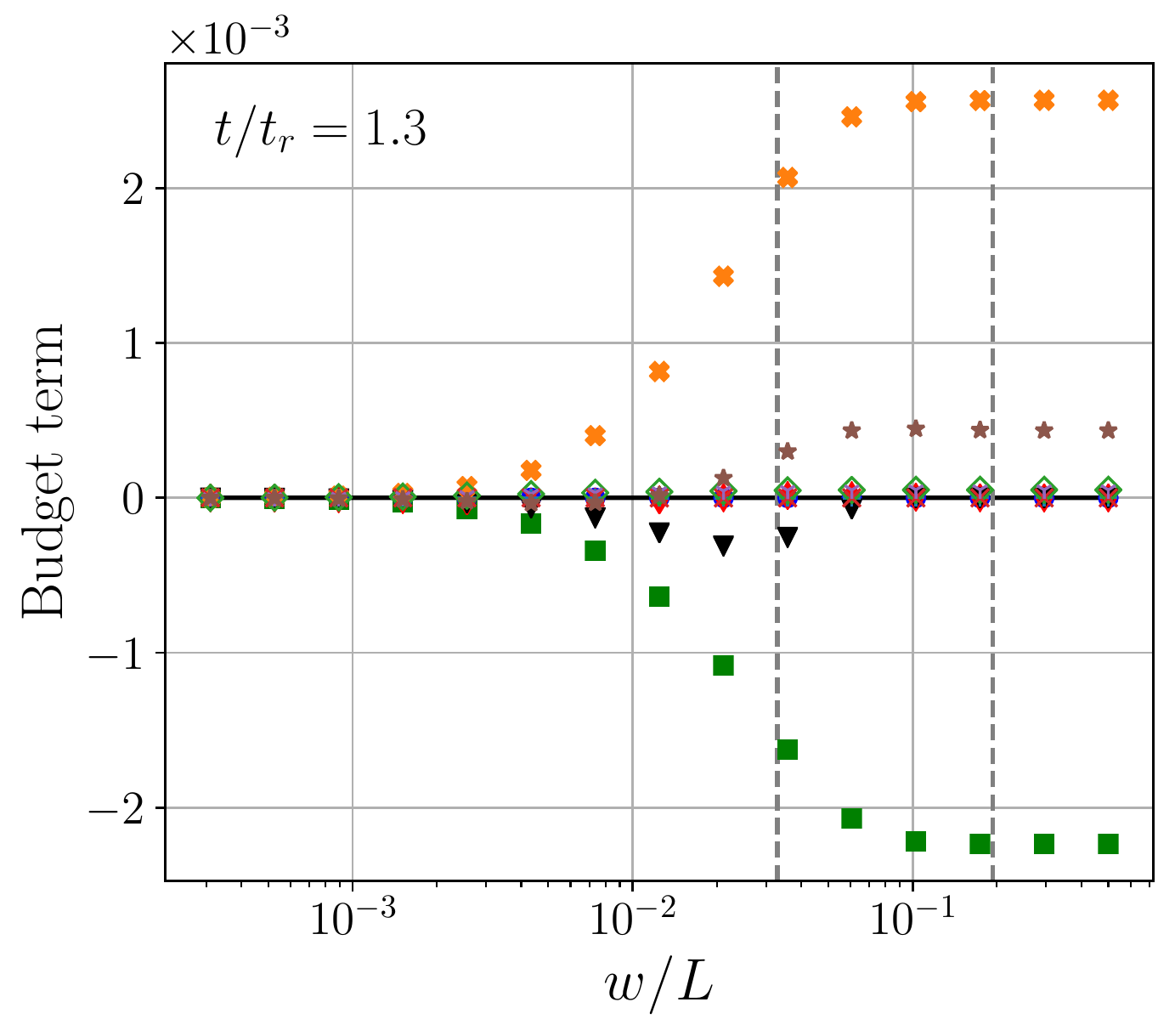}
  \includegraphics[trim=29 32 0 0, clip, width=0.28\textwidth]{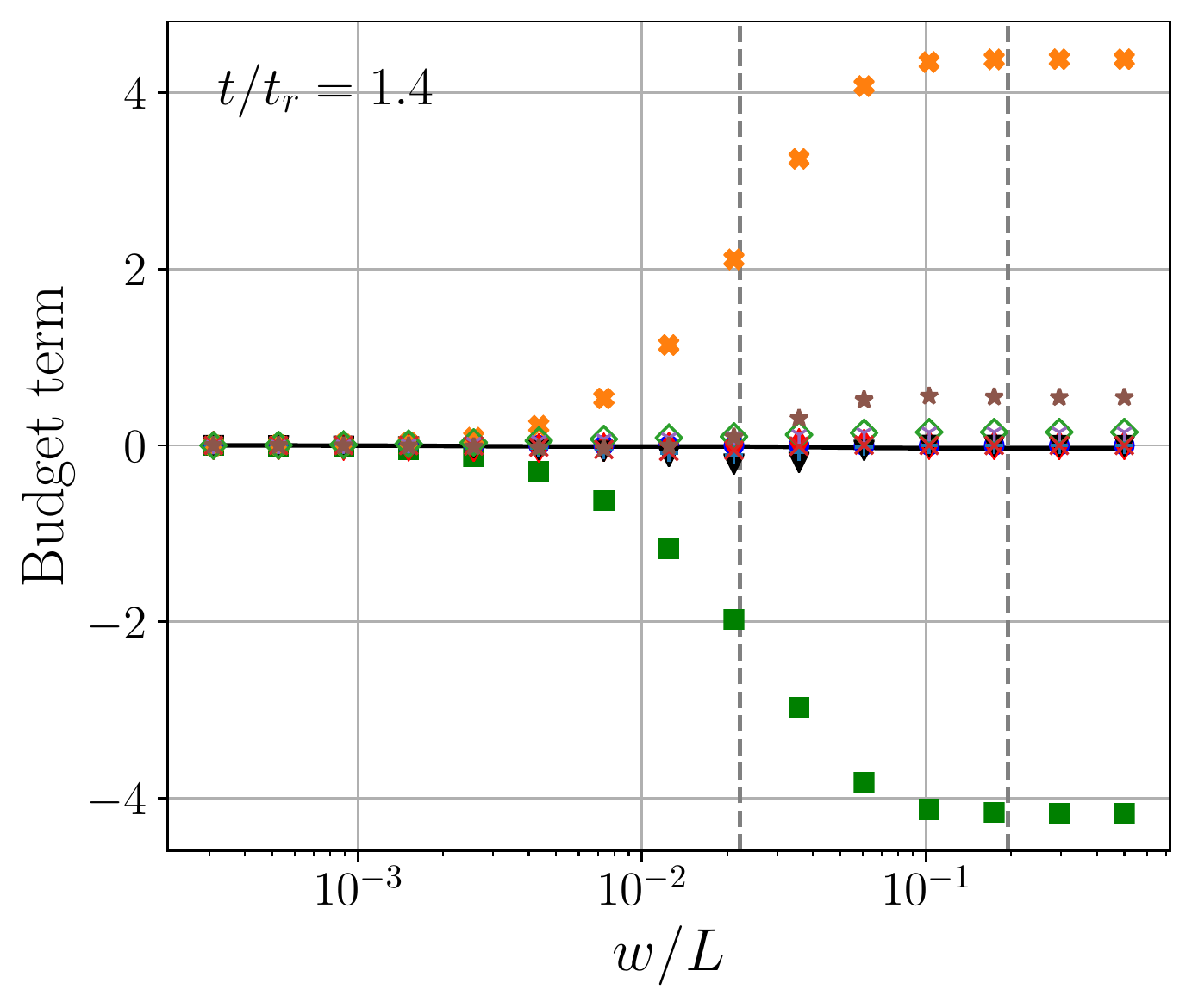}\\
  \hspace{0.15\textwidth}
  \includegraphics[trim=0 32 0 0, clip, width=0.31\textwidth]{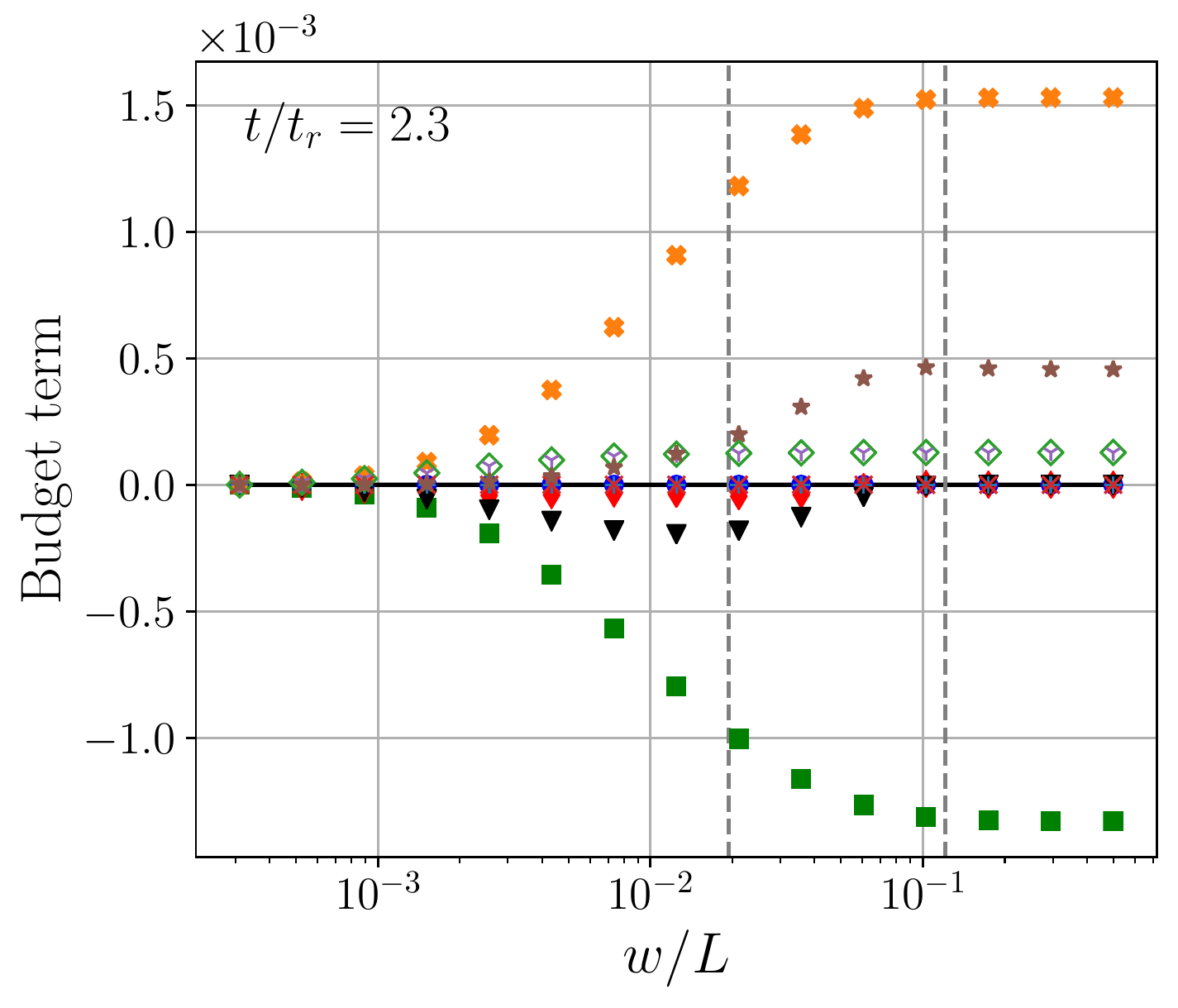}
  \includegraphics[trim=30 32 0 0, clip, width=0.29\textwidth]{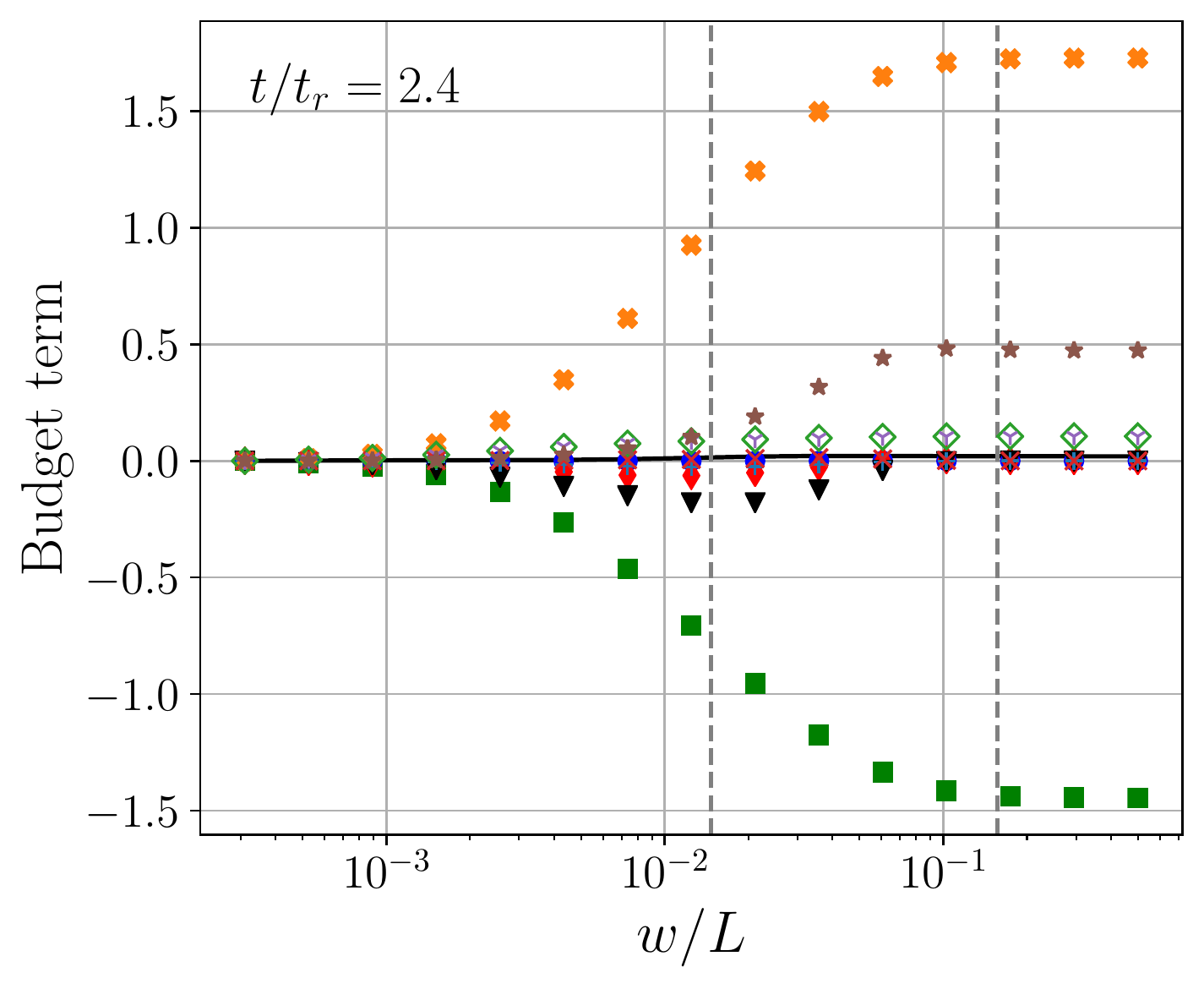}\\
  \hspace{0.13\textwidth}
  \includegraphics[trim=0 32 0 0, clip, width=0.32\textwidth]{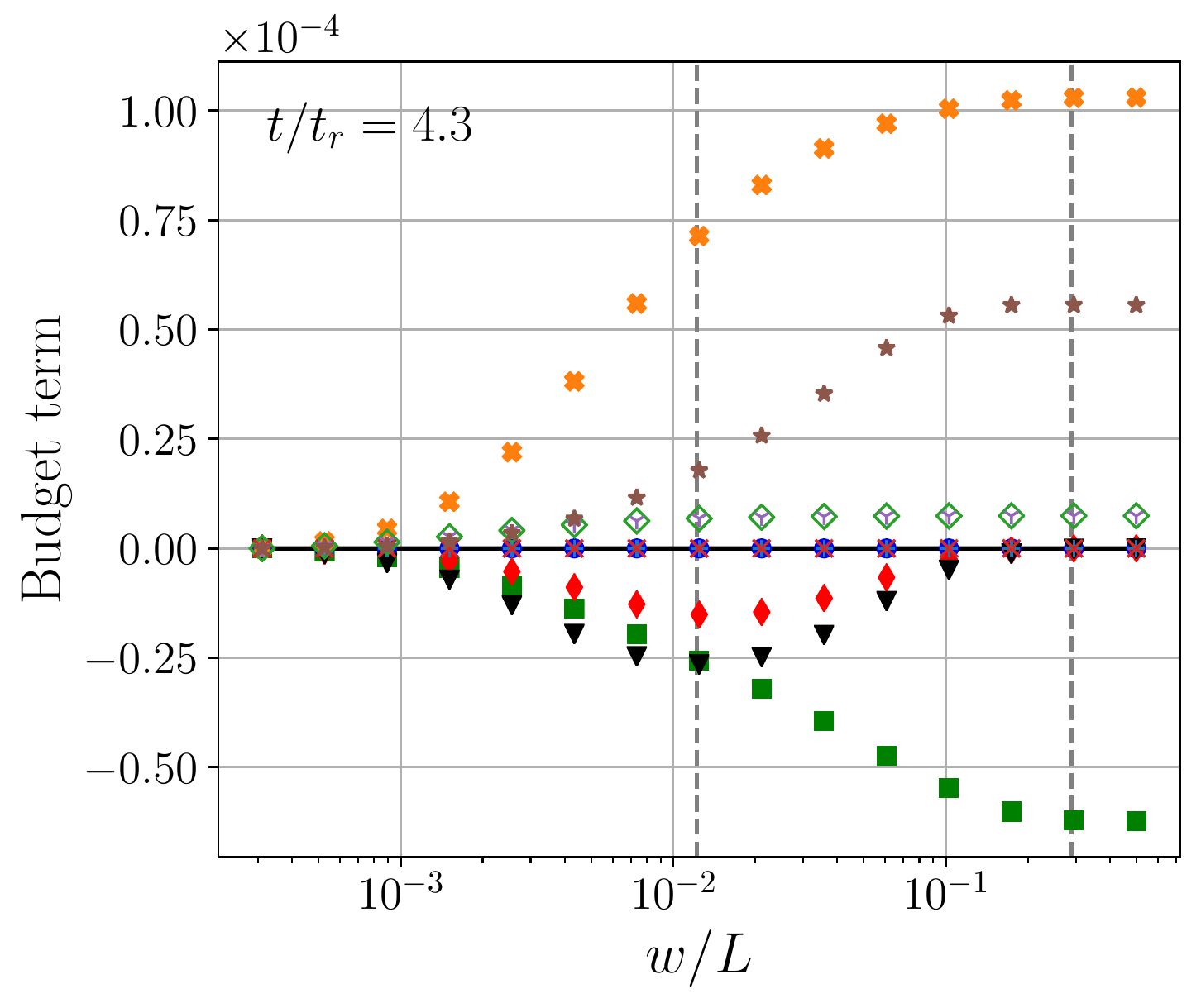}
  \includegraphics[trim=29 32 0 0, clip, width=0.29\textwidth]{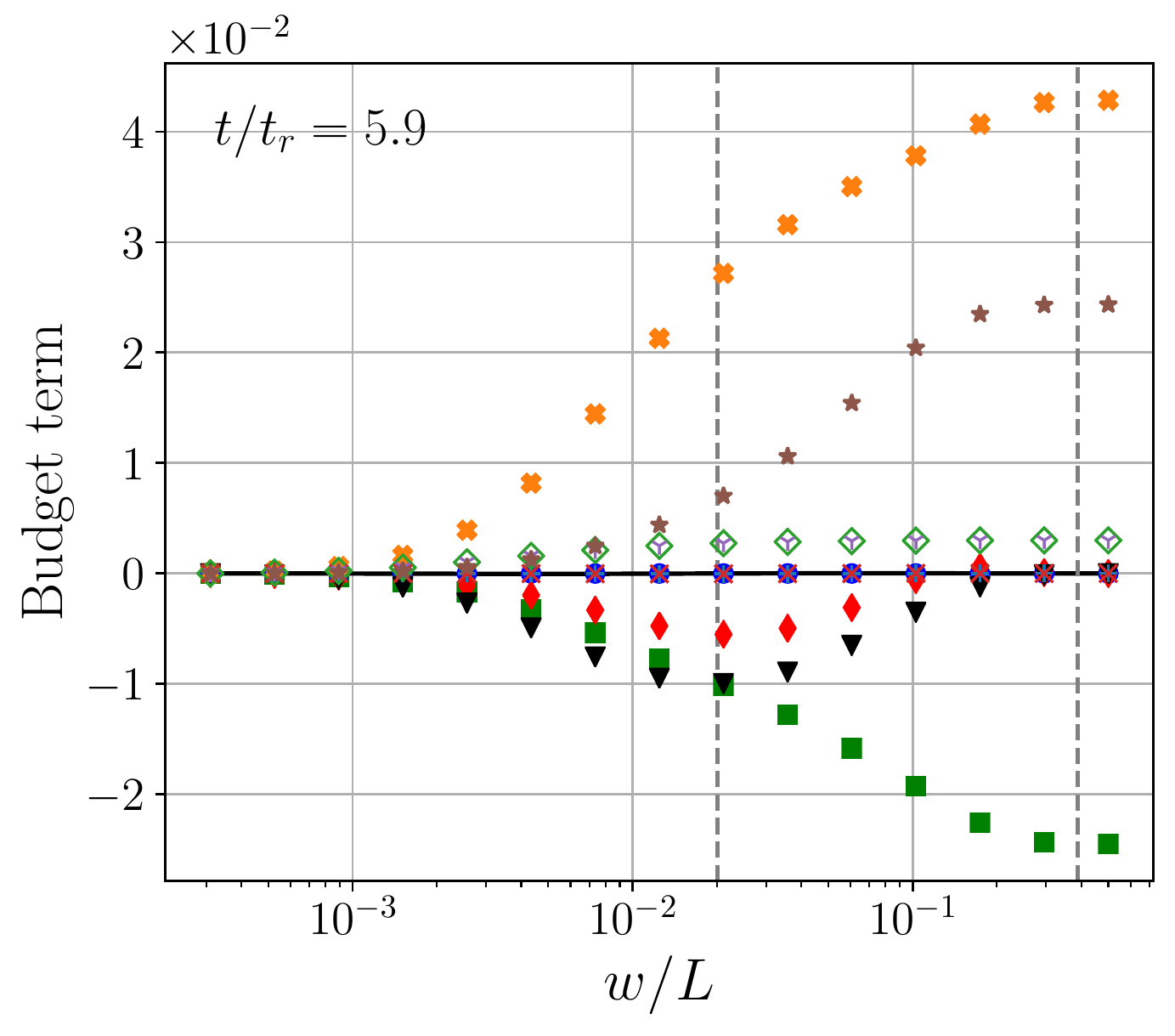}\\
  \hspace{0.15\textwidth}
  \includegraphics[trim=0 0 0 0, clip, width=0.31\textwidth]{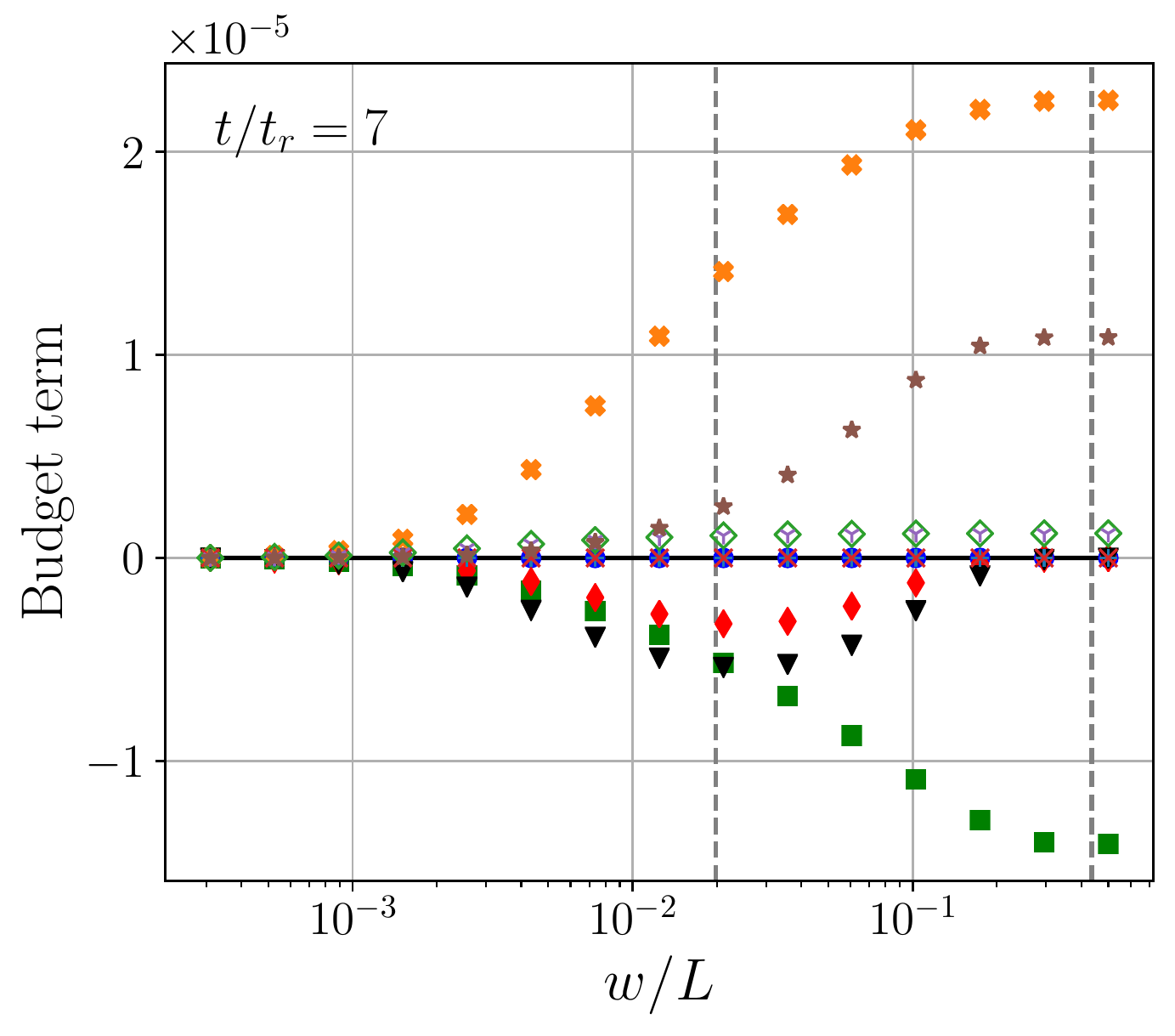}
  \includegraphics[trim=29 0 0 0, clip, width=0.3\textwidth]{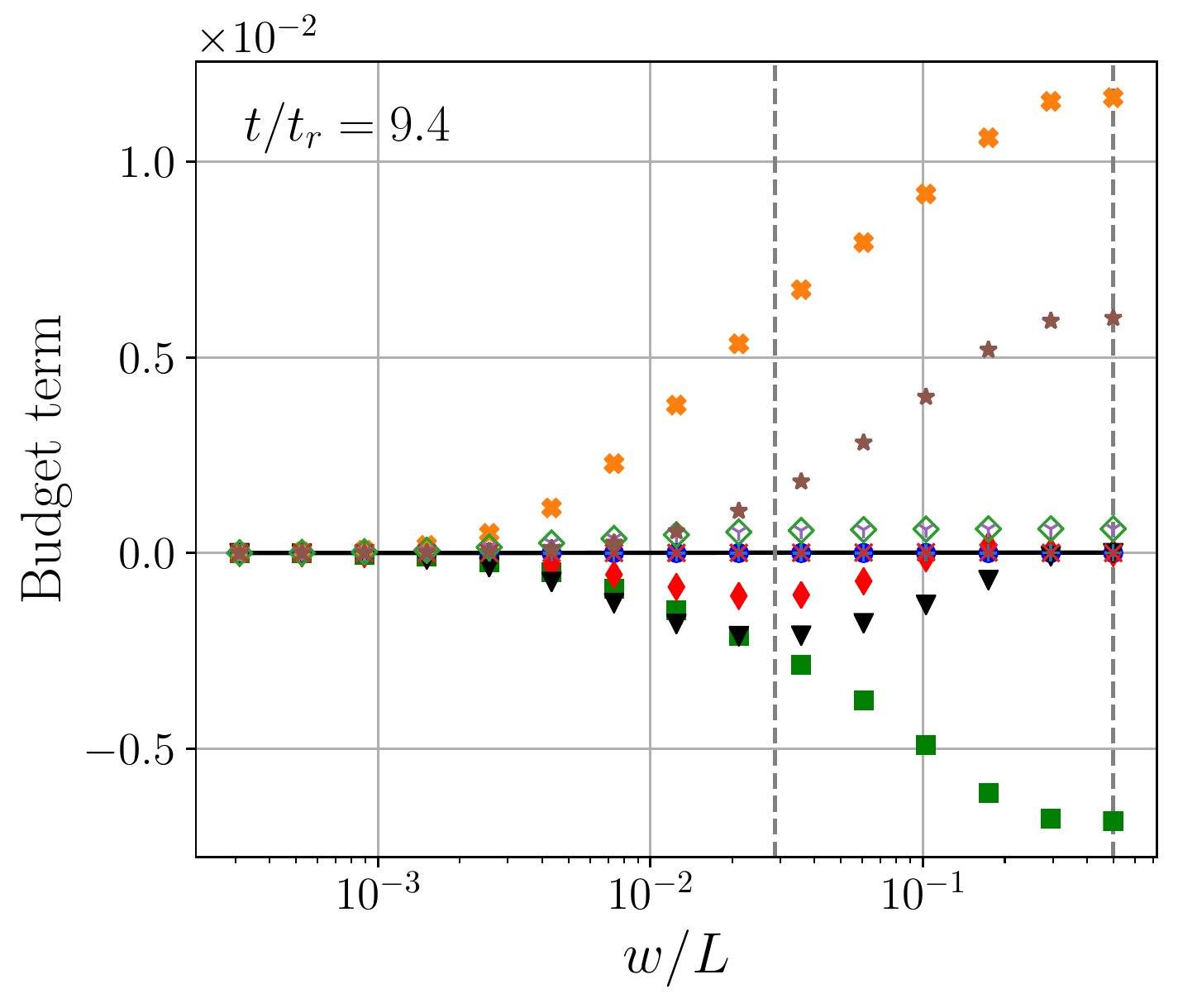}
  \caption{Volume averaged budget terms in (\ref{eq:trans_ai_volume_avg}) for $a_1$, as a function of filter width normalized by the box size, $w/L$. Note the difference in scales. Left and right columns correspond to $A=0.05,\ 0.75$ cases, respectively. Taylor micro-scale $\lambda_h$ and integral length scale $\mathcal{L}_v$, with $\lambda_h < \mathcal{L}_v$, are shown as vertical dashed lines. Black line indicates the residual.}
  \label{fig:a1_budget}
\end{figure}

The volume integrated budget for the dominating SR turbulent mass flux $a_1$, following equation (\ref{eq:trans_ai}), is
\begin{eqnarray}
\left \langle \ddt{\ol \rho a_1 } \right \rangle
&=& \left \langle b ( \ol p_{,1} -\ol \tau_{k1,k})  \right \rangle
- \left \langle \mathcal{T}_{1k} \ol \rho_{,k}  \right \rangle
- \left \langle \ol \rho a_k (\tilde u_1 - a_1)_{,k}  \right \rangle
+ \left \langle \ol \rho (a_1 a_k)_{,k}  \right \rangle\nonumber \\
&+& \left \langle \ol \rho \left( \frac{\varphi(\rho, u_1, u_k)}{\ol \rho} \right)_{,k}  \right \rangle
+ \left \langle  \ol \rho \, \varphi (v, p_{,1})  \right \rangle
- \left \langle \ol \rho \, \varphi(v,\tau_{k1,k})  \right \rangle
\label{eq:trans_ai_volume_avg}
\end{eqnarray}
where the terms on the right hand side represent production by pressure gradient, viscous stresses, turbulent stresses, redistribution, self-advection, turbulent transport and two destruction terms, respectively.
These are plotted in Figure \ref{fig:a1_budget}, where the residual is represented by the solid black line, again indicating the excellent closure of the budget. 
With the coordinate frame used for this flow, gravity $g$ and the RANS turbulent mass flux velocity point in the $-x_1$ direction, so that $a_{\mathrm{e}1}<0$.
As a result, production and destruction are represented as negative and positive values in Figure \ref{fig:a1_budget}, respectively, and positive $\partial (\ol \rho a_1) \partial t > 0$ corresponds to $\ol \rho a_1$ decaying in magnitude.
Similar to the $\ol \rho b$ budget, all budget terms are zero at the NS limit, and, at the RANS limit, the budget corresponds to the budget governing the RANS variable $a_{\mathrm{e}1}$, where time time rate of change is dominated by four terms: the $\ol \rho \varphi(v, p_{,1})$ destruction term, the production due to the pressure gradient $b \ol p_{,1}$ (itself dominated by the volume-mean pressure gradient), and, to a lesser though non-negligible extent, the destruction term due to viscous stresses $\ol \rho \varphi(v, \tau_{k1,1})$ and the commonly ignored dilatation term $\ol \rho \varphi(u_1, u_{n,n})$.
The terms that are active, in the volume integrated sense, at the RANS limit all decay monotonically to zero as the filter width is decreased to the grid size in the NS limit.
The redistribution term $\ol \rho a_k \ol u_{1,k}$ and the production term $\mathcal{T}_{1k} \ol \rho_{,k}$ are zero at the NS and RANS limits, but non-zero at intermediate scales as they peak at filter widths comparable to the horizontal Taylor micro-scale $\lambda_h$.
These peak values are small but non-zero at the end of the explosive growth regime, but they play an increasingly important role as time advances and turbulence develops.

\begin{figure}[htb]
  \centering
  \includegraphics[width=0.15\textwidth]{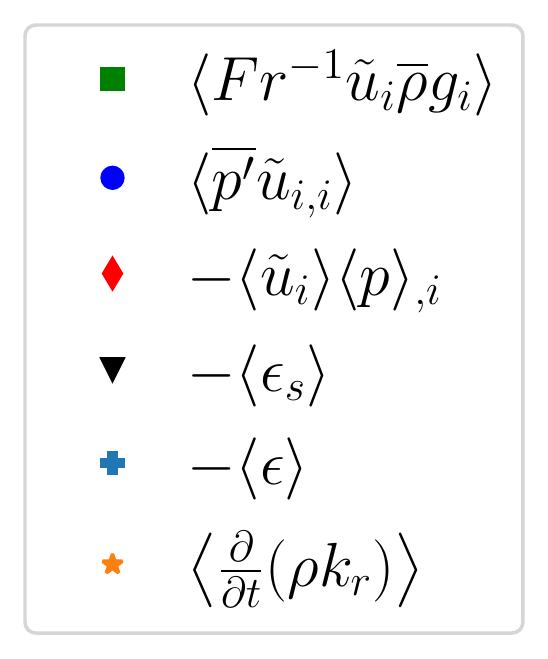}
  \includegraphics[trim=0 32 0 0, clip, width=0.3\textwidth]{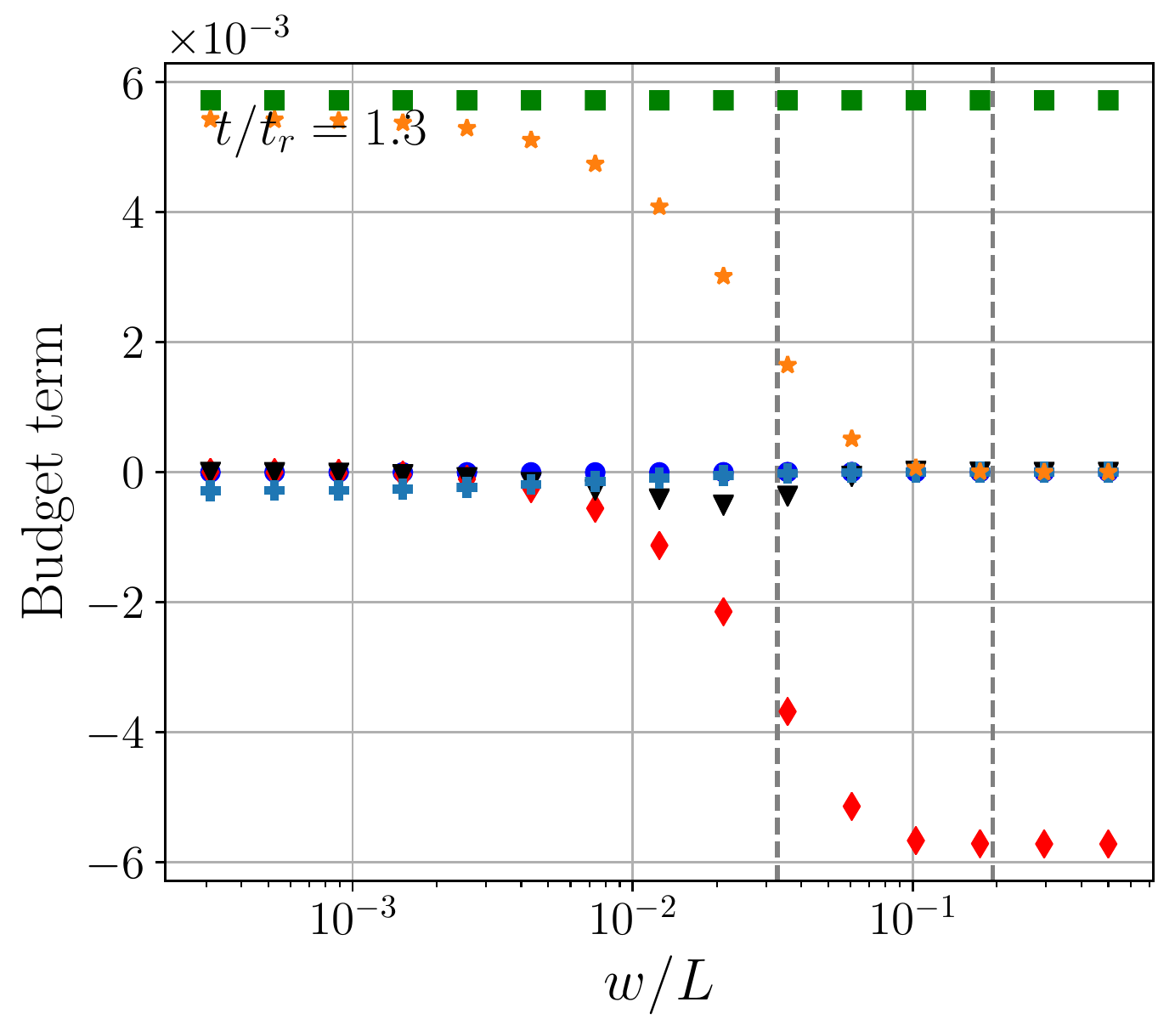}
  \includegraphics[trim=29 32 0 0, clip, width=0.28\textwidth]{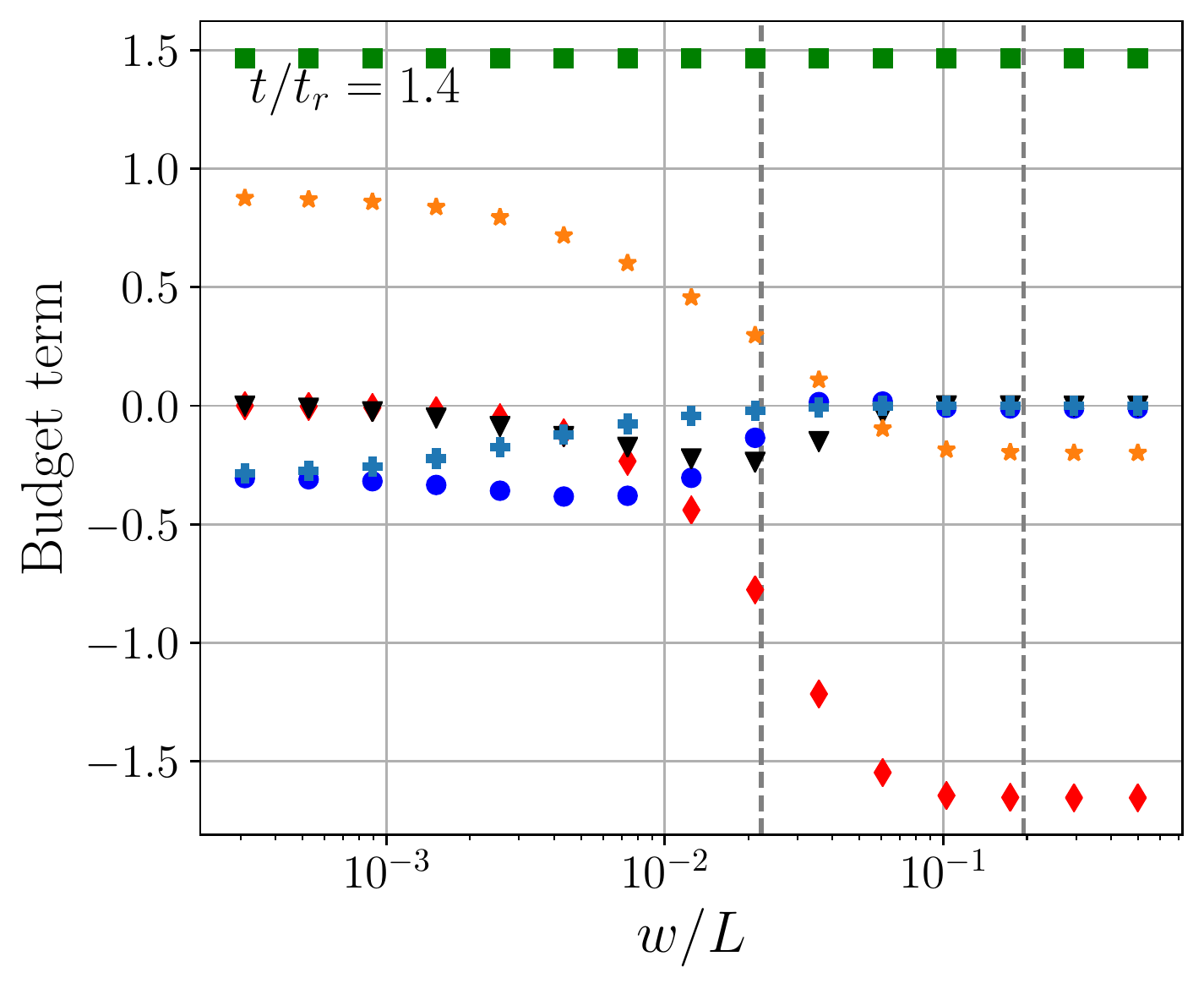}\\
  \hspace{0.15\textwidth}
  \includegraphics[trim=0 32 0 0, clip, width=0.31\textwidth]{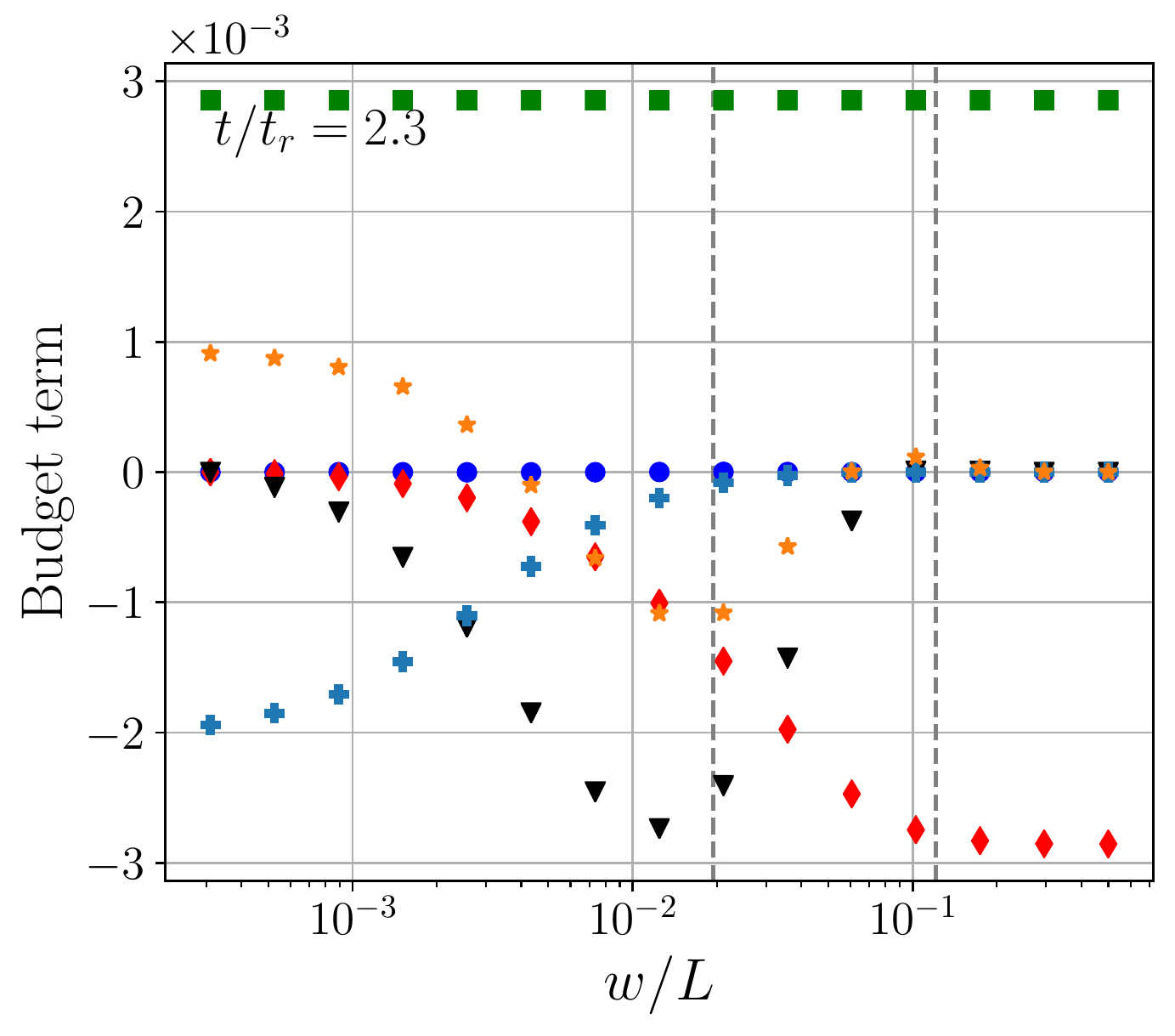}
  \includegraphics[trim=30 32 0 0, clip, width=0.29\textwidth]{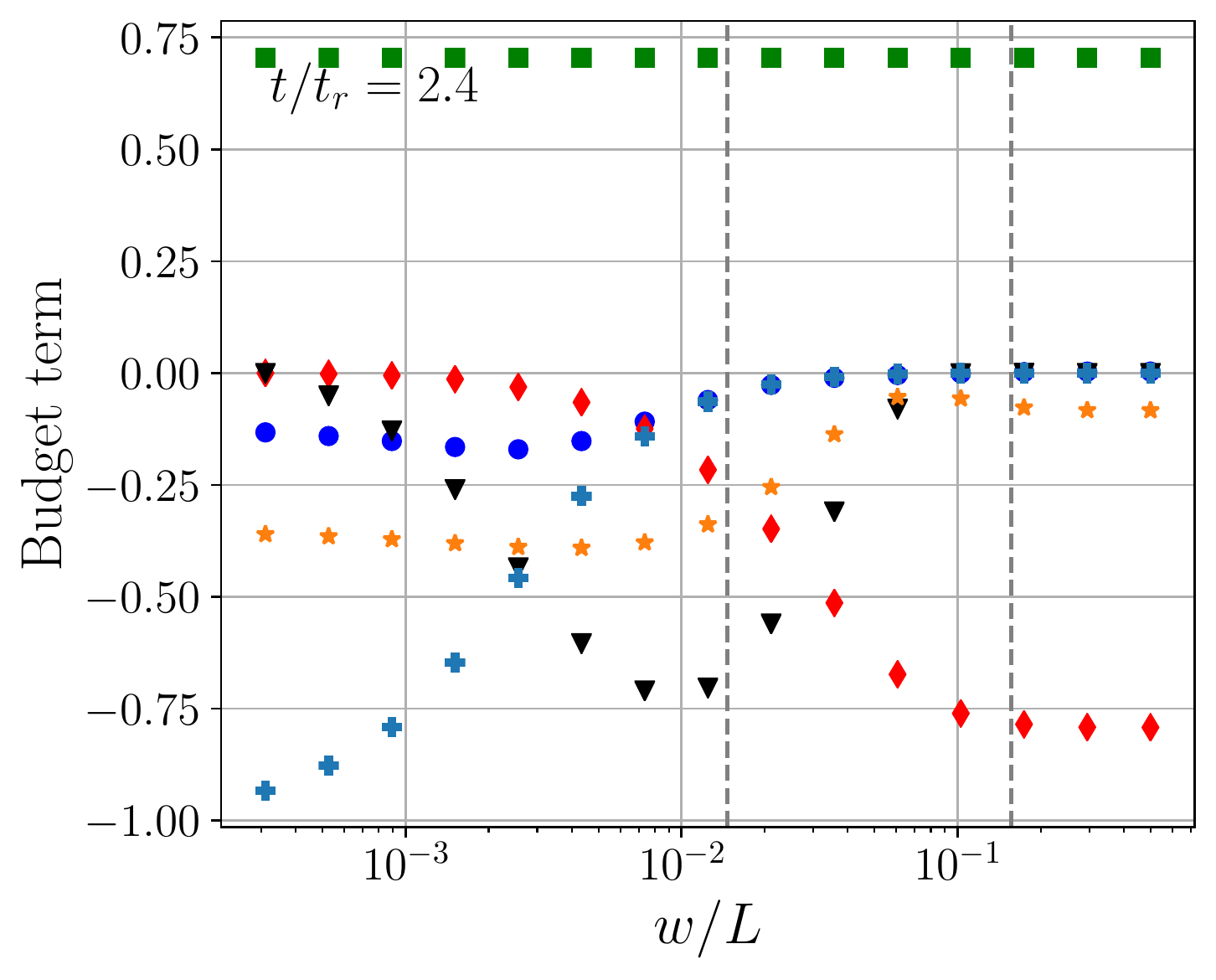}\\
  \hspace{0.13\textwidth}
  \includegraphics[trim=0 32 0 0, clip, width=0.32\textwidth]{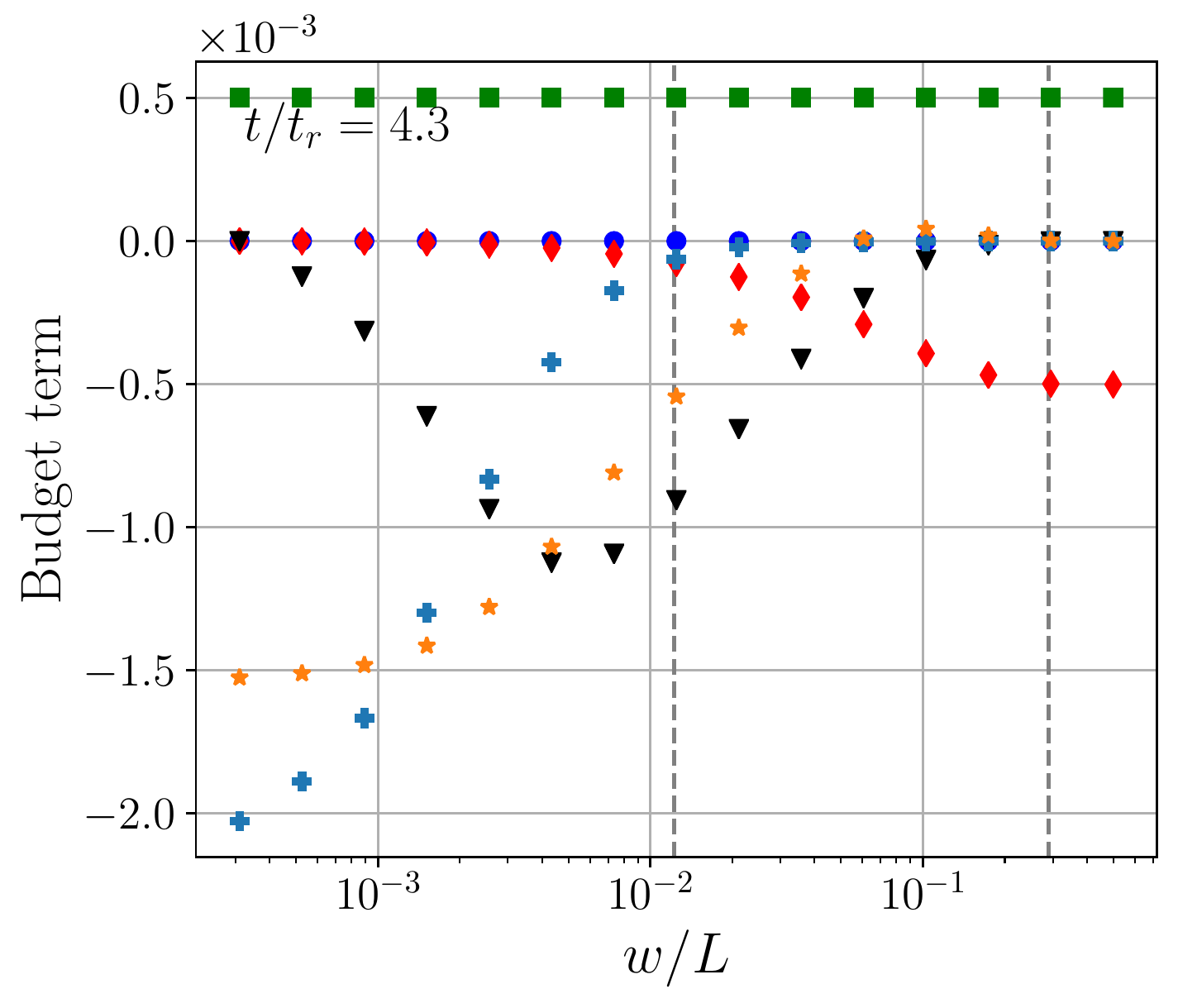}
  \includegraphics[trim=29 32 0 0, clip, width=0.29\textwidth]{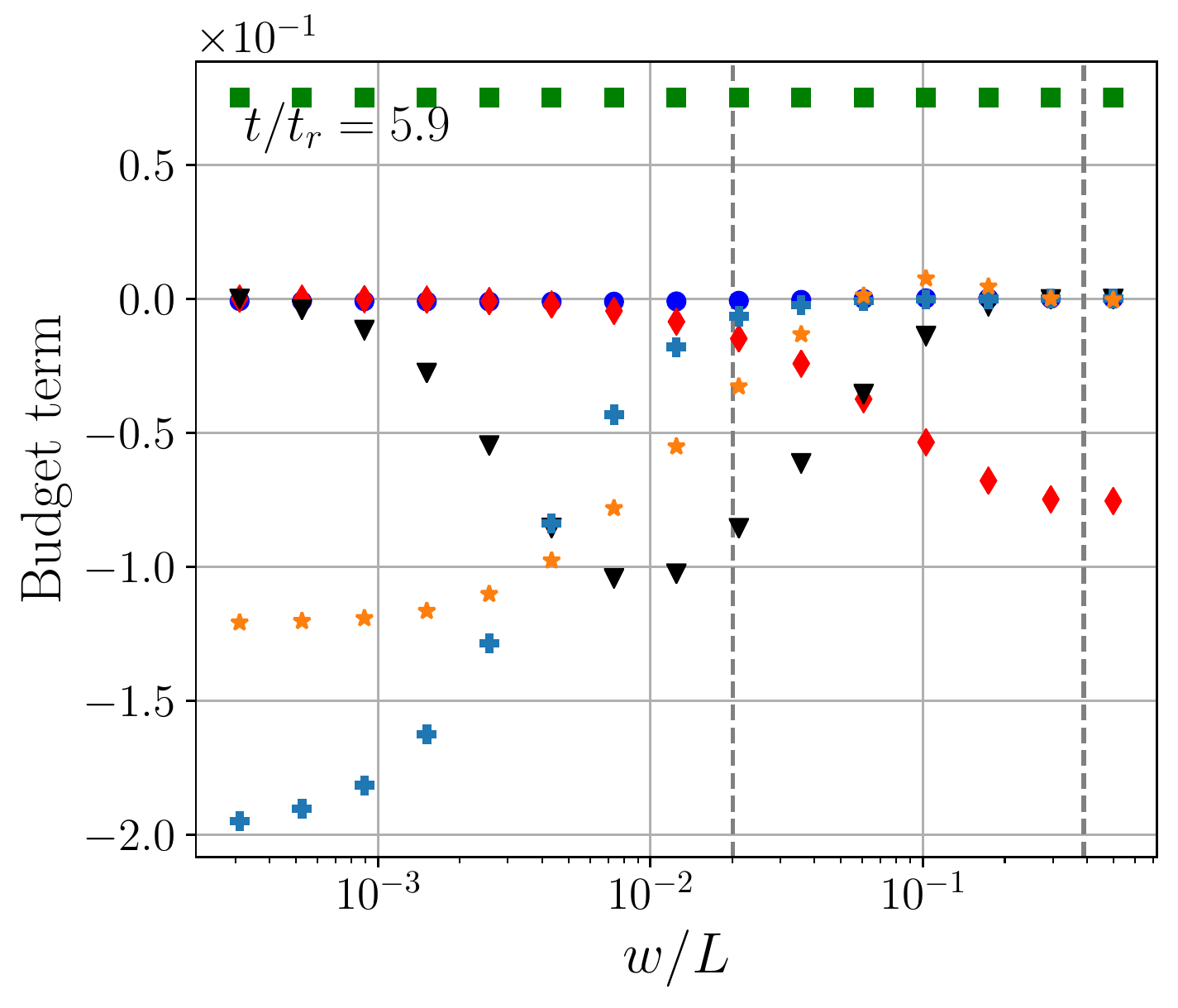}\\
  \hspace{0.15\textwidth}
  \includegraphics[trim=0 0 0 0, clip, width=0.31\textwidth]{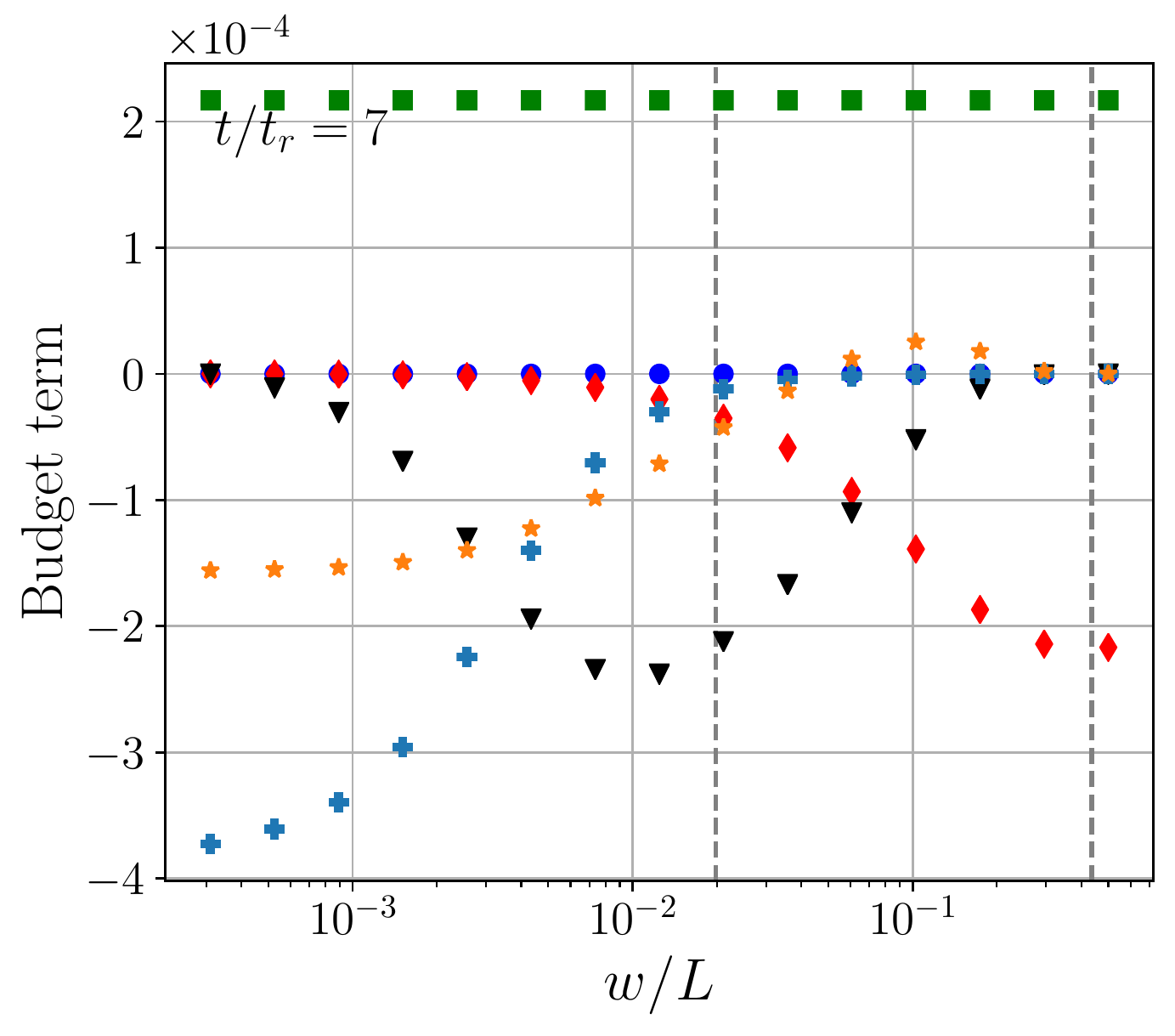}
  \includegraphics[trim=29 0 0 0, clip, width=0.3\textwidth]{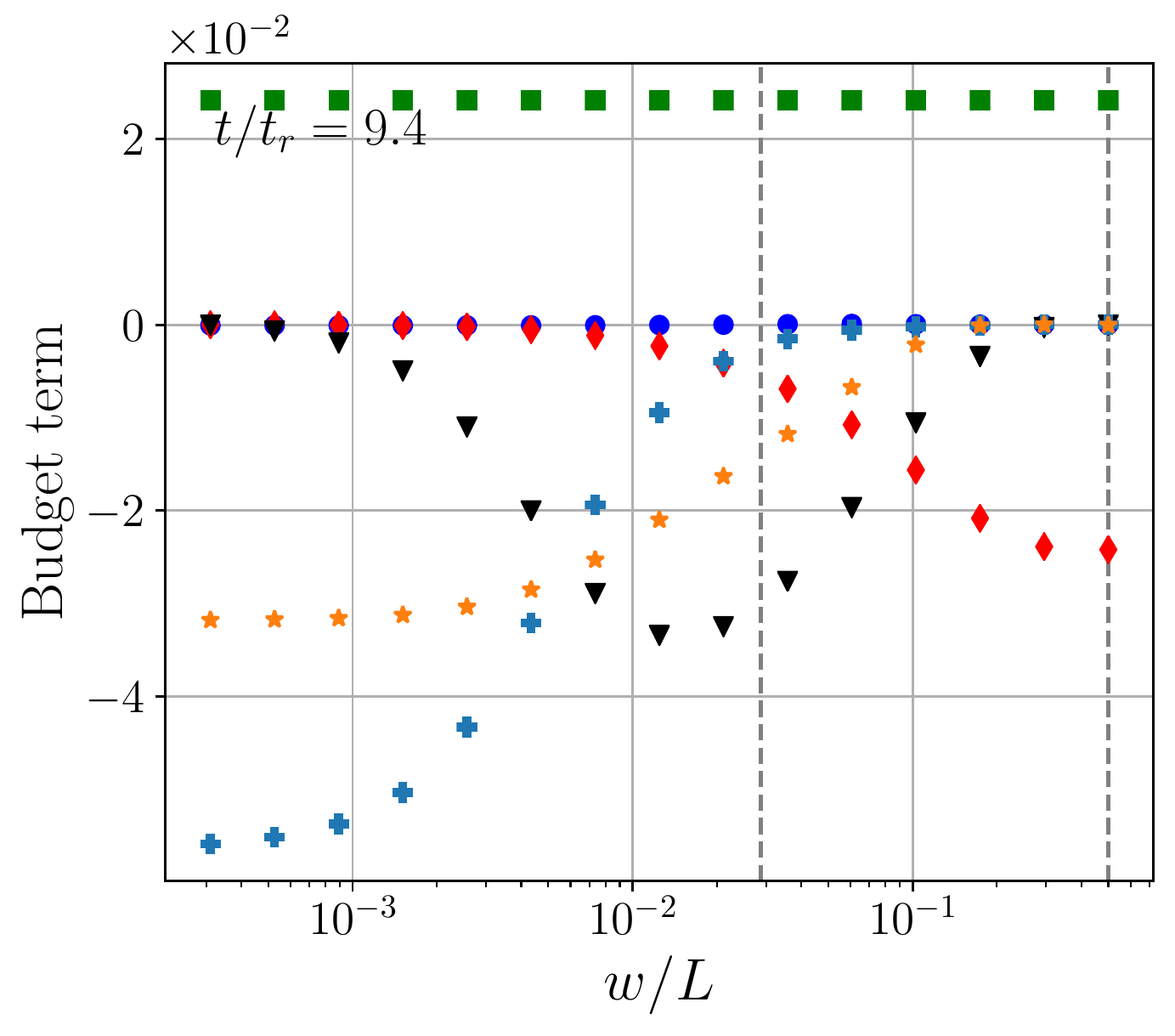}
  \caption{Volume averaged budget terms for the resolved kinetic energy in (\ref{eq:trans_tke_res_volume_avg}), as a function of filter width normalized by the box size, $w/L$. Note the difference in scales. Left and right columns correspond to $A=0.05,\ 0.75$ cases, respectively. Taylor micro-scale $\lambda_h$ and integral length scale $\mathcal{L}_v$, with $\lambda_h < \mathcal{L}_v$, are shown as vertical dashed lines.}
  \label{fig:tke_budget}
\end{figure}

We now investigate the resolved kinetic energy budget (\ref{eq:scaleresolvedke}) averaged over the volume,
\begin{equation}
  \left \langle \ddt{\ol \rho k_r} \right \rangle
= 
  \left \langle \frac{1}{Fr^2} \tilde u_i \overline \rho g_i \right \rangle
+ \left \langle \ol{p'} \frac{\partial \widetilde u_i}{\partial x_i} \right \rangle
- \left \langle \widetilde u_i \frac{\partial \langle p \rangle}{\partial x_i} \right \rangle
- \left \langle \epsilon_{s} \right \rangle
- \left \langle \epsilon \right \rangle,
\label{eq:trans_tke_res_volume_avg}
\end{equation}
where the terms on the right hand side represent production by conversion of potential energy to kinetic energy, work by the pressure fluctuation on the dilatation of the flow, work by the mean pressure gradient on Favre velocity, conversion to small scale kinetic energy by the residual stresses, and dissipation by molecular viscosity, respectively. 
The terms in equation (\ref{eq:trans_tke_res_volume_avg}) are plotted in Figure \ref{fig:tke_budget}.
The volume average of the advection and transport terms in (\ref{eq:scaleresolvedke}) is zero since the domain is periodic.
The pressure projection method in the time advancement scheme used for the DNS gives the average pressure over a given time step, not the instantaneous pressure needed in (\ref{eq:trans_tke_res_volume_avg}) to close the budget.
For this reason, we calculate the work by fluctuating pressure on dilatation as a residual using the other terms in (\ref{eq:trans_tke_res_volume_avg}), which we calculate from the DNS.
As a result, we do not plot a residual to the balance in (\ref{eq:trans_tke_res_volume_avg}) in Figure \ref{fig:tke_budget}.

Conversion of potential energy to kinetic energy is constant, independent of length scale or filter width, at all times for both Atwood numbers. This is consistent with the discussions in [\onlinecite{livescu_ristorcelli_2007,aluie_2013}], where it is shown that this conversion takes place at the scale of the domain. Work by the mean pressure gradient on the Favre velocity $\langle \tilde u_i \rangle \langle p \rangle_i$, is small at scales similar to or smaller than the horizontal Taylor micro-scale, where it has a net effect of transferring energy from the resolved scales to the small scales.
The transfer of kinetic energy $\langle \epsilon_s \rangle$ is from resolved scales to small scales, in the volume integrated sense, it is zero at the NS and RANS limits, and it peaks at scales similar to the horizontal Taylor micro-scale. We will see later that $\langle \epsilon_s \rangle$ locally can transfer energy upwards or downwards between the resolved and subscale kinetic energies.
Viscous dissipation of kinetic energy $\langle \epsilon \rangle$ is important at scales smaller than the horizontal Taylor micro-scale.
Work from fluctuating pressure on the dilatation of the flow $\langle \ol{p'} \tilde u_{i,i}\rangle$ is non-zero, though small, only during the explosive and gradual growth regimes of the $A=0.75$ flow. 
After the end of the explosive growth regime, large scale production, work by the mean pressure gradient, kinetic energy transfer between scales and viscous dissipation coexist at scales smaller than $\lambda_h$, the first three are present at scales between $\lambda_h$ and the the vertical integral length scale $\mathcal{L}_v$, and only the first two are present at scales larger than $\mathcal{L}_v$ and in the RANS limit.

To recapitulate, we observe two salient features regarding the terms in the volume integrated budget equations for $\ol \rho b$ and $\ol \rho a_1$: 1) they are zero in the NS limit, are dominated by the RANS budgets in the RANS limit, but 2) at intermediate scales, these budgets have important contributions from other terms that are not active in the RANS limit. 

\begin{figure}[htb]
  \centering
  \includegraphics[trim=0 0 0 0, clip, width=0.315\textwidth]{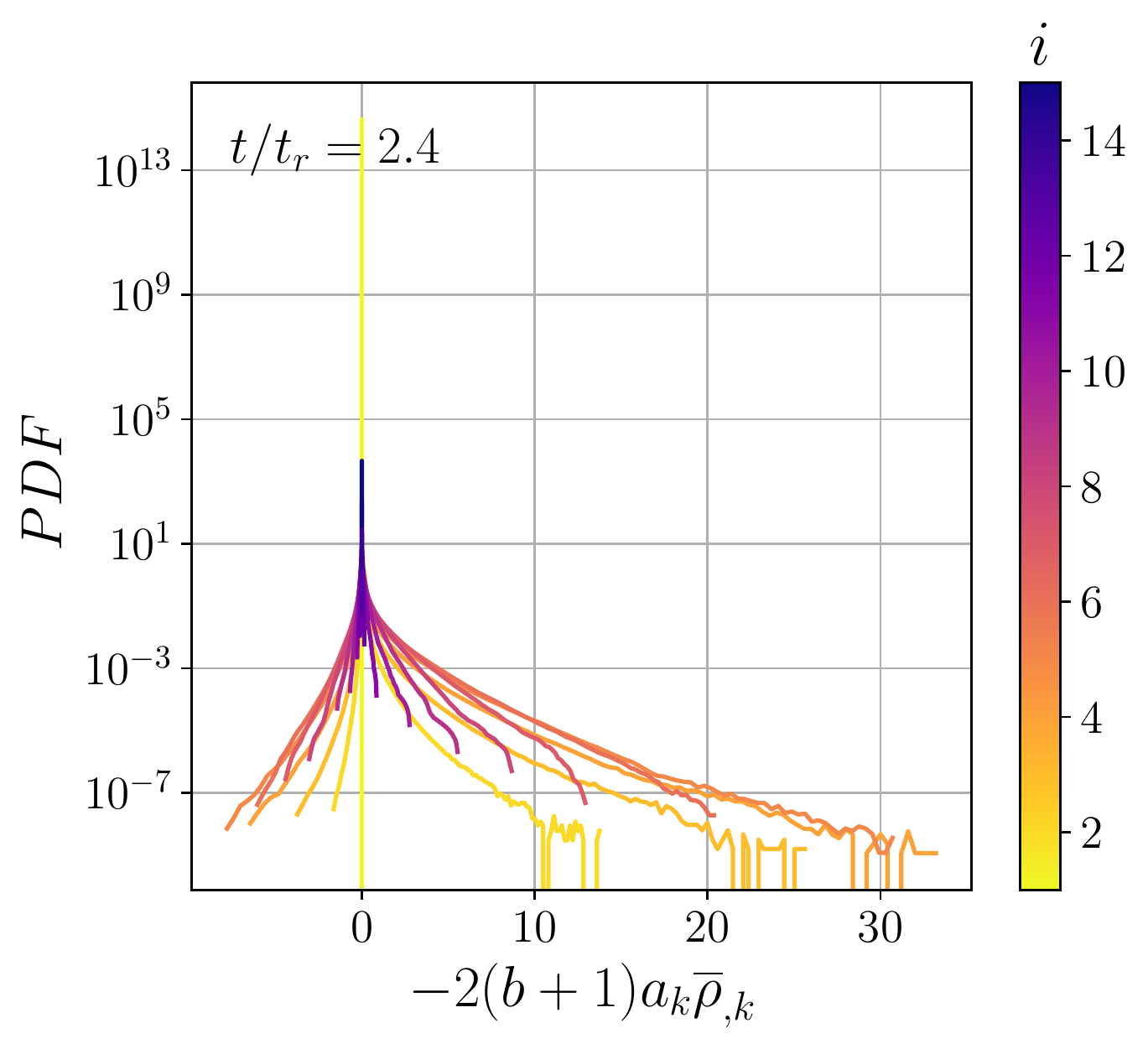}
  \includegraphics[trim=22 0 0 0, clip, width=0.3\textwidth]{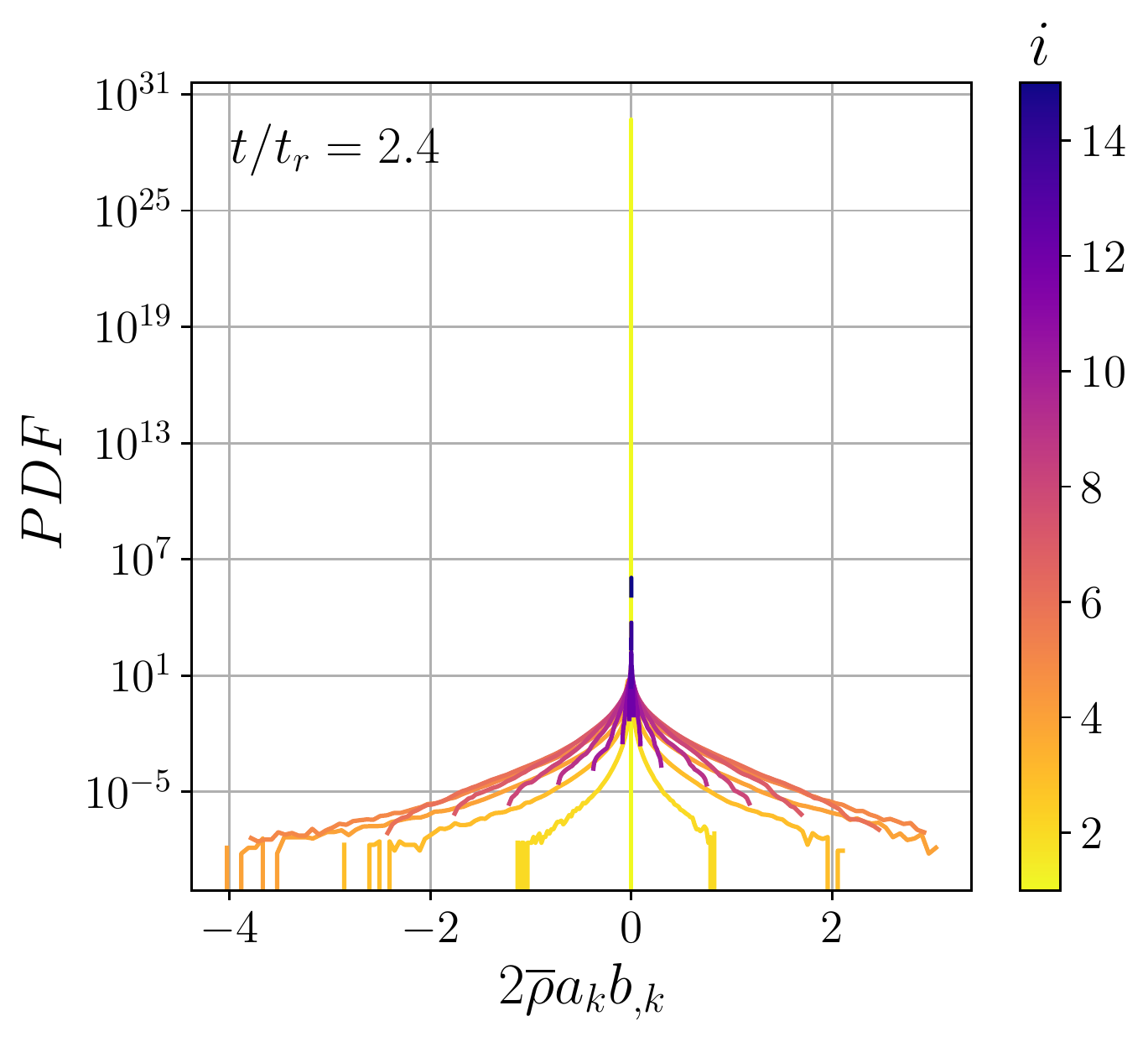} \\
  \includegraphics[trim=0 0 0 0, clip, width=0.315\textwidth]{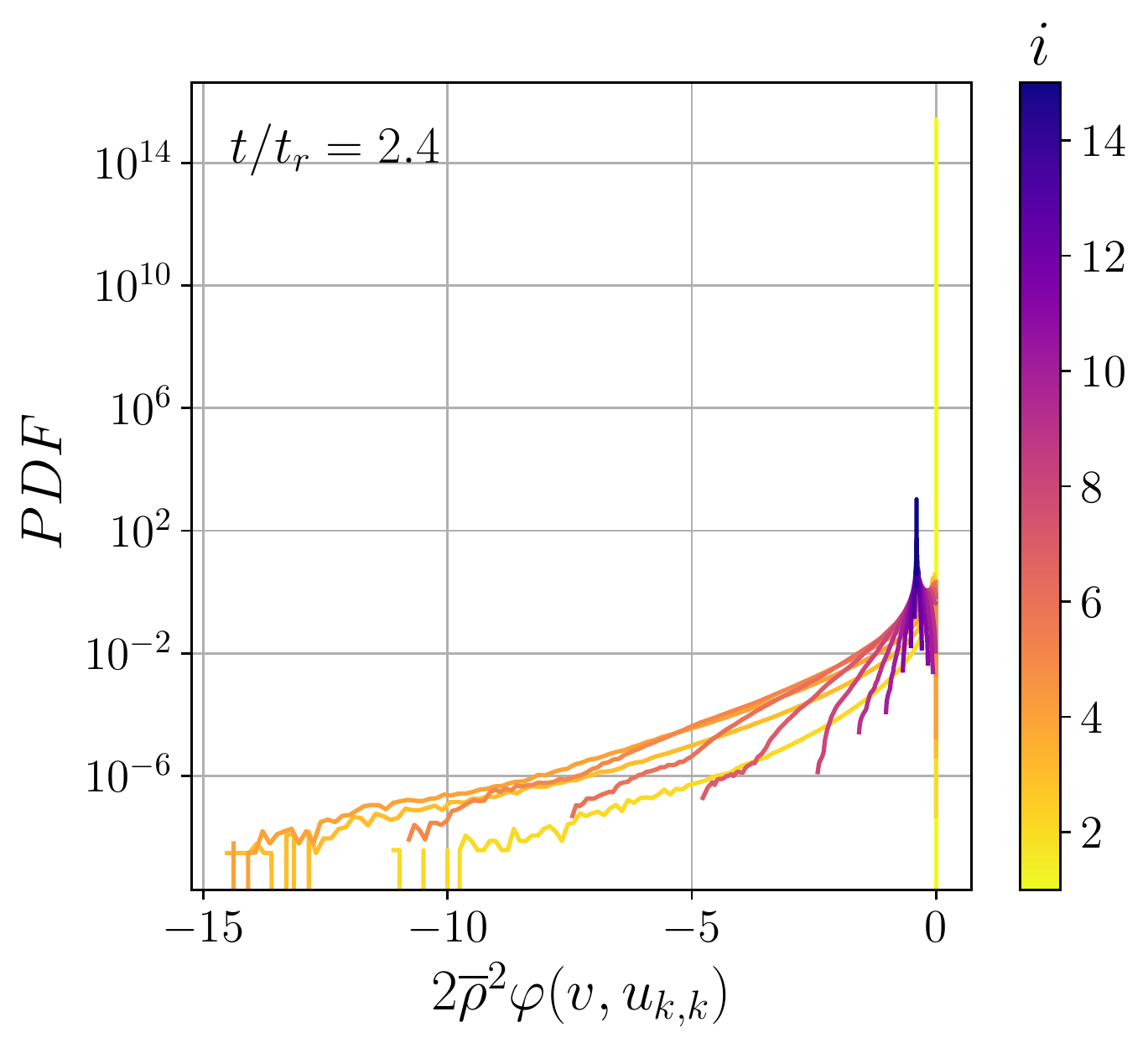}
  \includegraphics[trim=22 0 0 0, clip, width=0.3\textwidth]{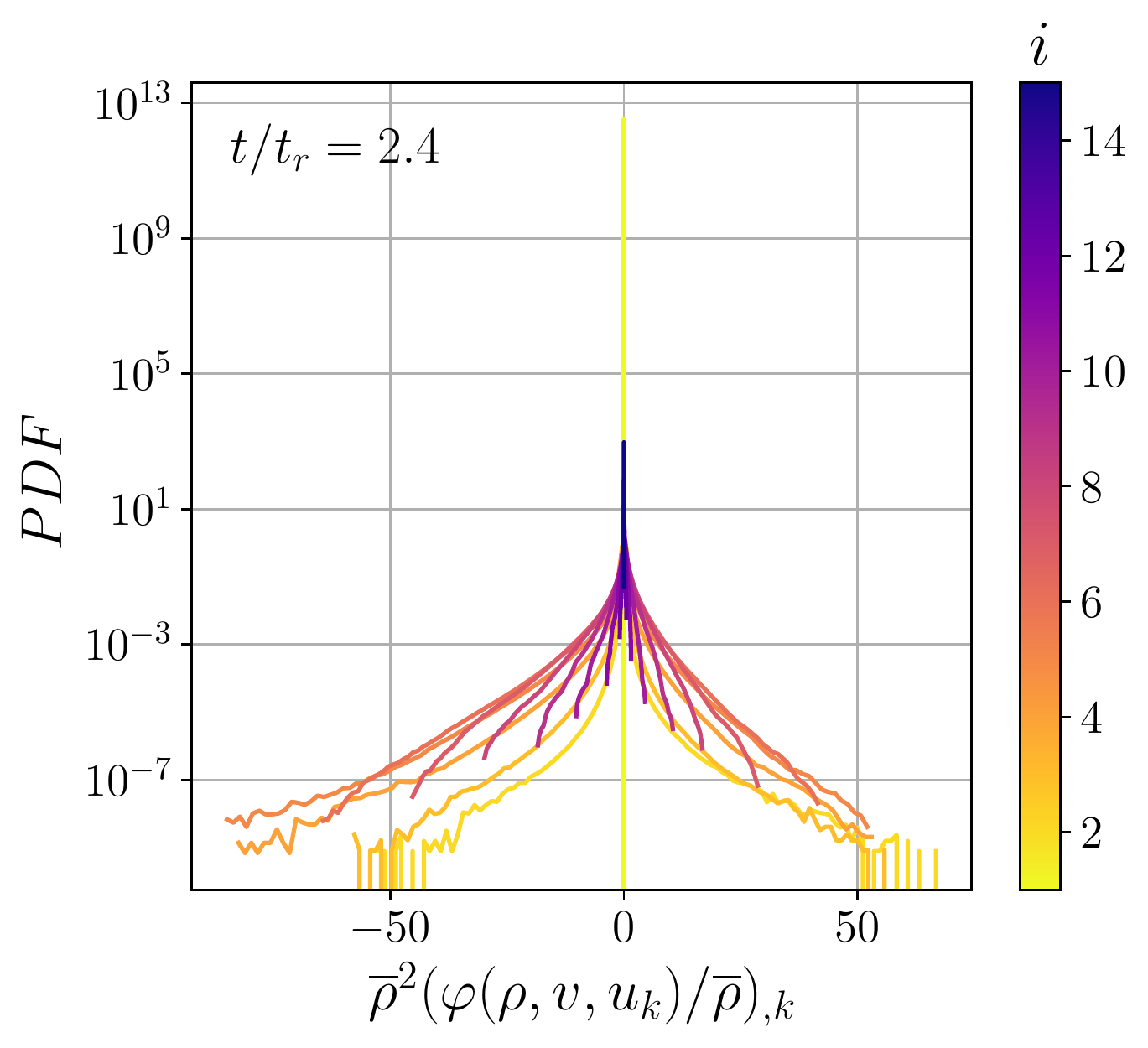}
  \caption{Probability density function of the production $-2 (b+1) a_k \overline \rho_{,k}$, redistribution $2 \overline \rho a_k b_{,k}$, destruction $2 \overline{\rho}^2 \varphi(v, u_{k,k})$ and transport $\overline{\rho}^2 (\varphi(\rho, v, u_k)/\overline{\rho})_{,k}$ terms, as they appear in equation (\ref{eq:trans_b_volume_integrated}), at $t/t_r = 2.4$ when $\mathcal{R}_{ii}$ peaks, for $A=0.75$. The colorbar indicates the filter width $i$ used, according to (\ref{eq:w_of_i}), Table \ref{tab:filter_width_list}, where $i=1$ corresponds to the smallest filter width $w/L = (\Delta x/\pi)/L=3.1\times10^{-4}$,  and $i=15$ corresponds to the largest filter width $w/L = 1/2$.}
  \label{fig:pdf_bterms}
\end{figure}

To illustrate the spatial variability of budget terms, we look at the terms governing the evolution of $b$ by plotting their PDF in Figure \ref{fig:pdf_bterms}.
For homogeneous variable density flows, one can use (\ref{eq:div_and_rho}) and integration by parts to show that the destruction of $b$ is negative.
%
Similar to the PDFs of SR variables shown in Figure \ref{fig:pdf_b_a1_T11_T12}, there is a smooth transition between the NS limit, where the budget terms are zero, and the RANS limit, where the budget equation for $b$ tends to the budget equation for $b_\mathrm{e}$ \cite{besnard_etal_1992}.
Also in Figure \ref{fig:pdf_b_a1_T11_T12}, there is a large spread of values, indicating the existence of rare events that can have values that are two or more orders of magnitude larger than the RANS quantity corresponding to $i=15$, $w/L=1/2$.
The transport term has a spread that is particularly large for filter widths $i=2-5$, $w/L=5.3\times10^{-4}-2.6\times10^{-3}$, and decreases for larger filter widths.
This term also has a reasonably symmetric PDF, indicating that the zero volume integrated net production is the result of the summation of large positive and negative terms.
This indicates that in the flow, at these intermediate scales, there is a lot of spatial variability in the transport of $\ol \rho b$.
The production term, which is zero in the RANS limit but nonzero at intermediate scales, also has large variability in the PDF, with positive and negative values, indicating that it creates and destroys variability in $ ( \p \rho / (\rho \ol \rho), \p \rho )_x$, see equation (\ref{eq:b_inner_expectedvalue}), due to variability in the alignment between the turbulence mass flux velocity $a_k$ and the density gradient $\ol \rho_{,k}$.
In the RANS limit, the destruction term plays a dominant role, and it has large variability at intermediate scales as well.
Redistribution has a zero volume integral across scales that results from positive and negative values due to the variability in the alignment between the mass flux, $a_k$, and gradient of b, $b_{,\ k}$.

\begin{figure}[htb]
  \centering
  \includegraphics[trim=0 32 0 0, clip, width=0.3\textwidth]{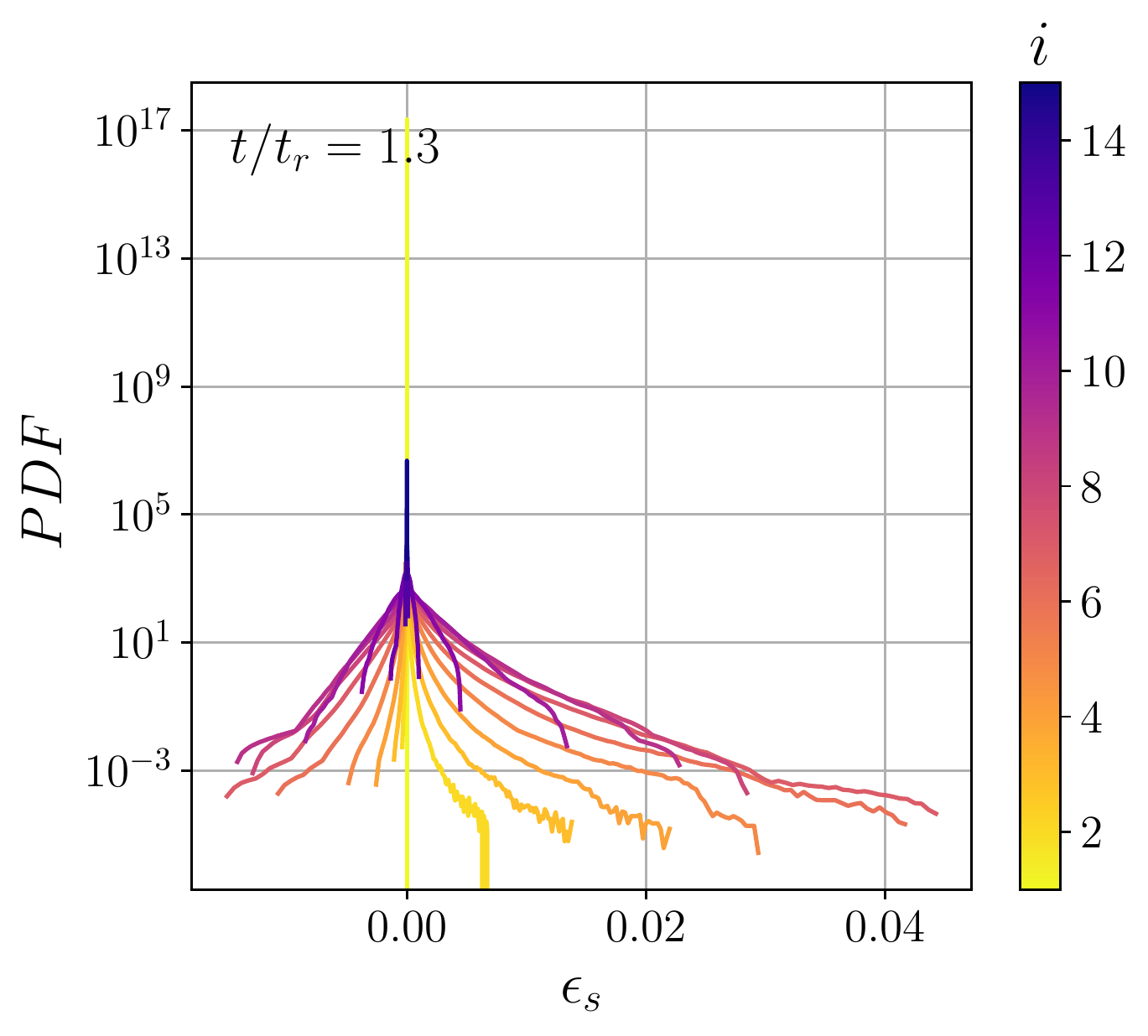}
  \includegraphics[trim=29 32 0 0, clip, width=0.28\textwidth]{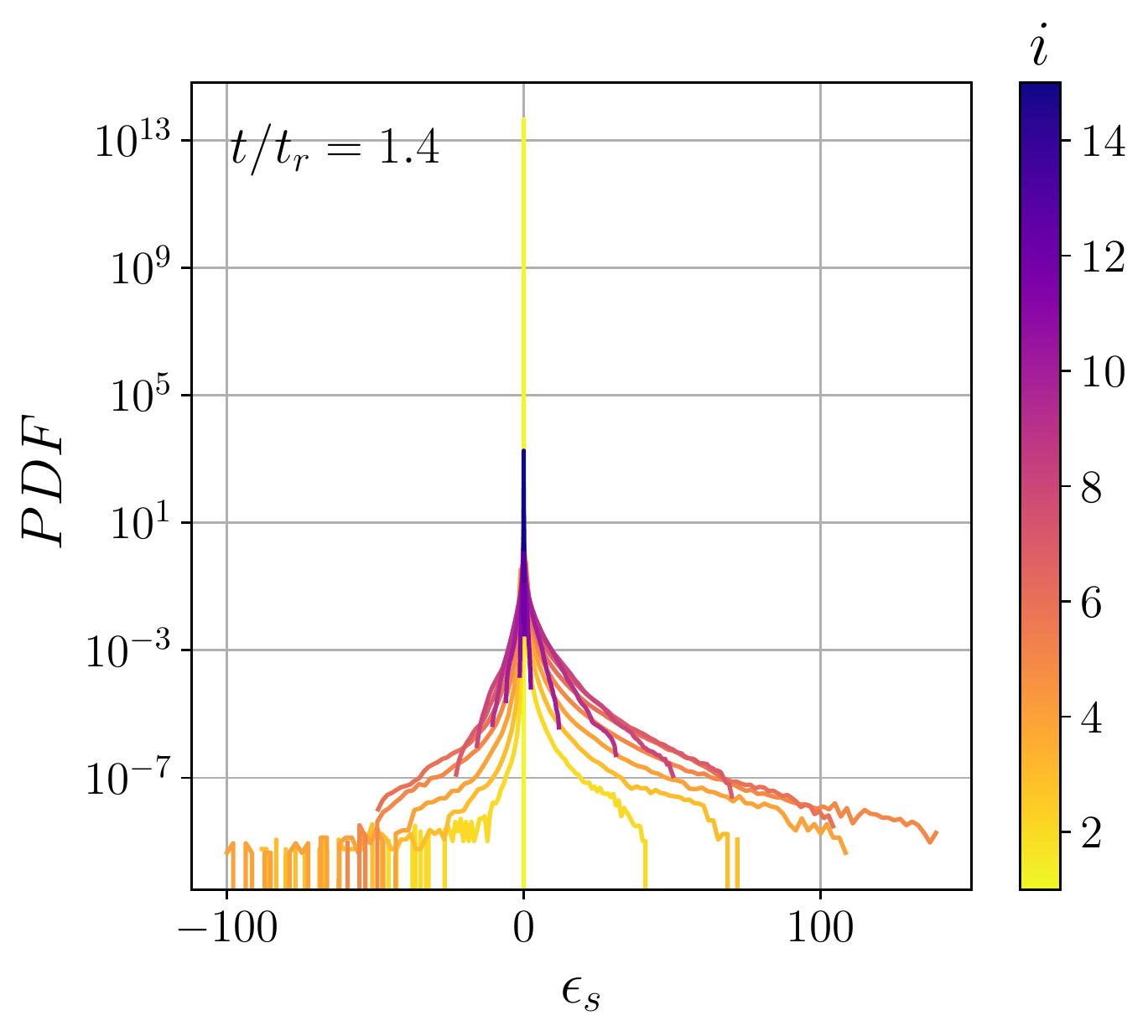}\\
  \includegraphics[trim=0 32 0 0, clip, width=0.31\textwidth]{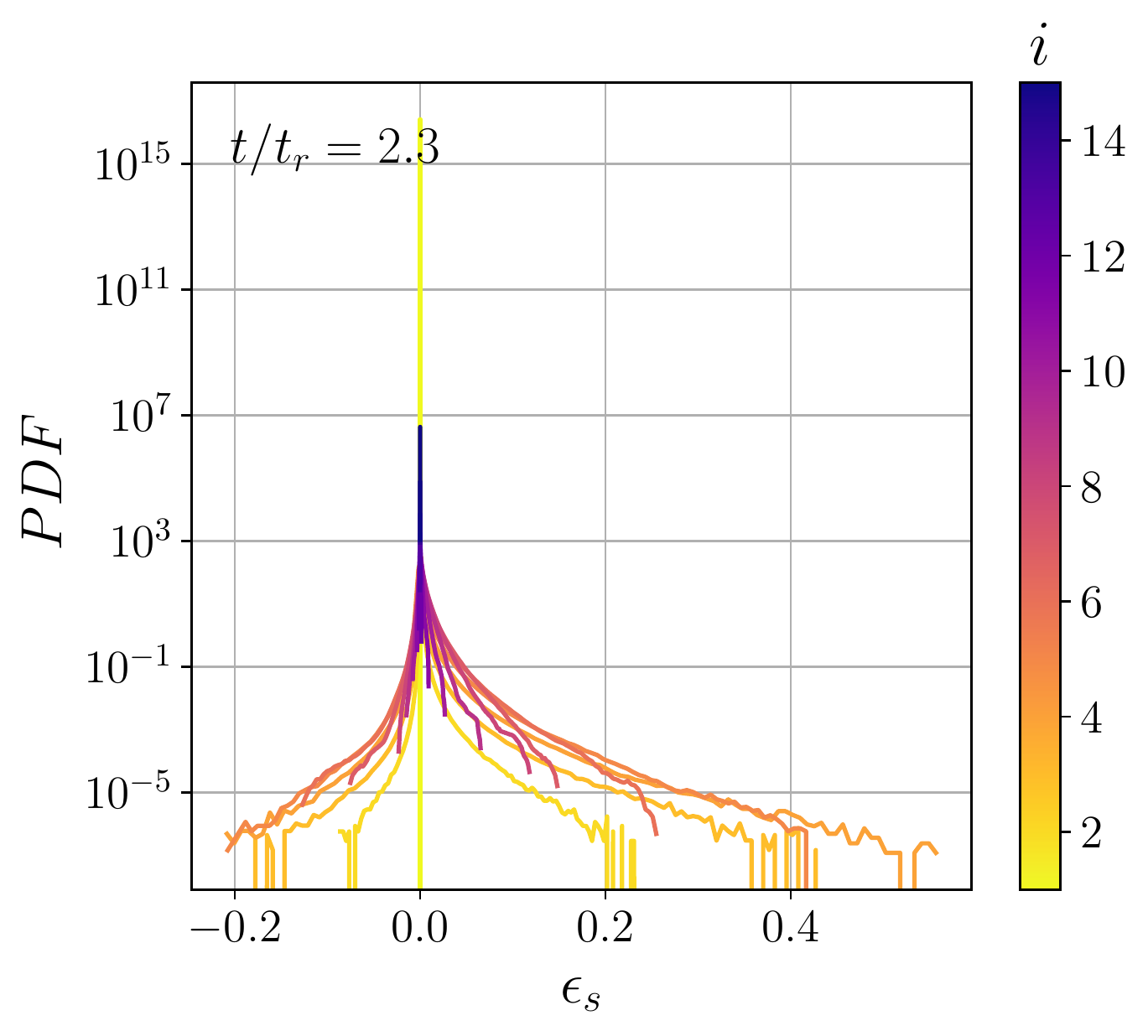}
  \includegraphics[trim=30 32 0 0, clip, width=0.29\textwidth]{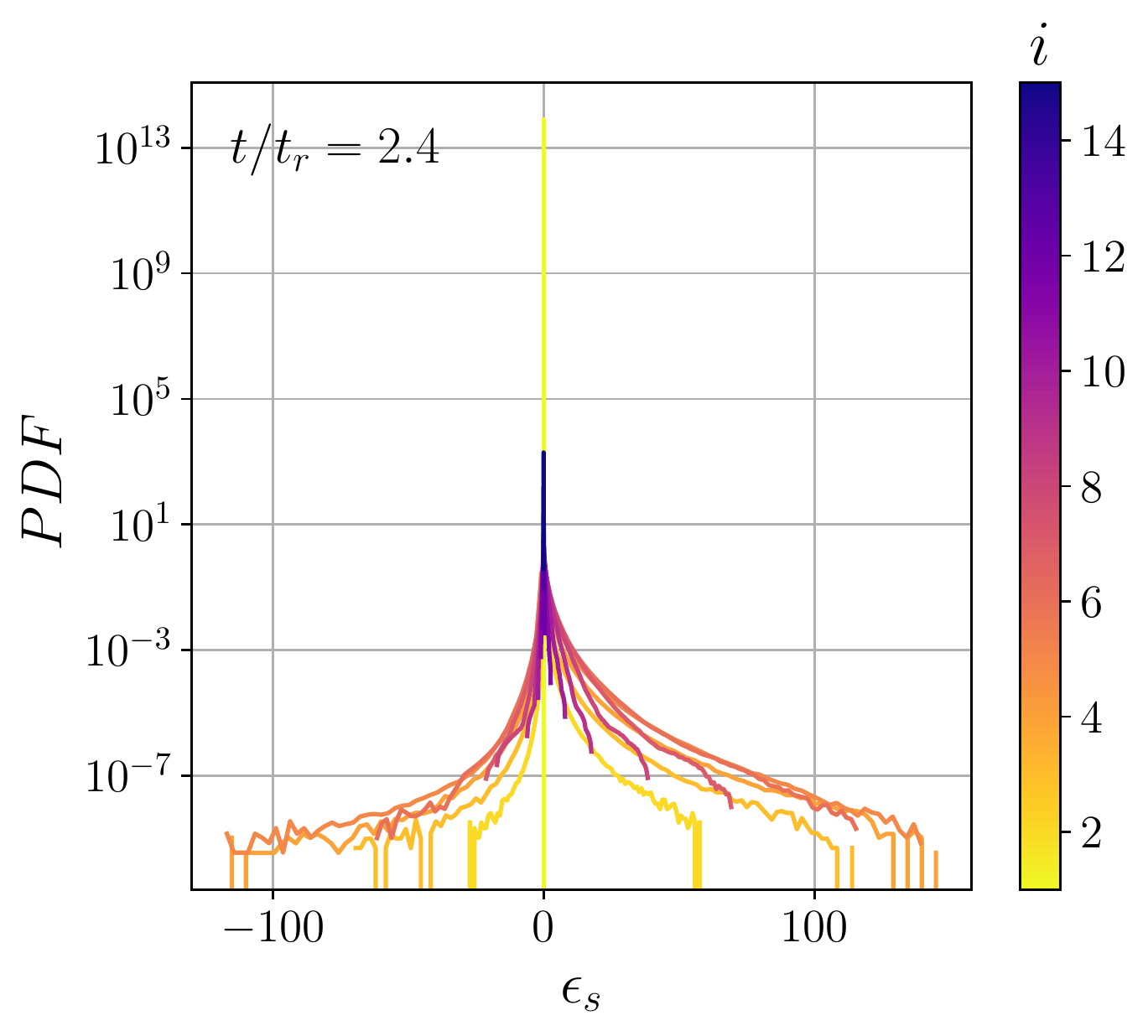}\\
  \includegraphics[trim=0 32 0 0, clip, width=0.32\textwidth]{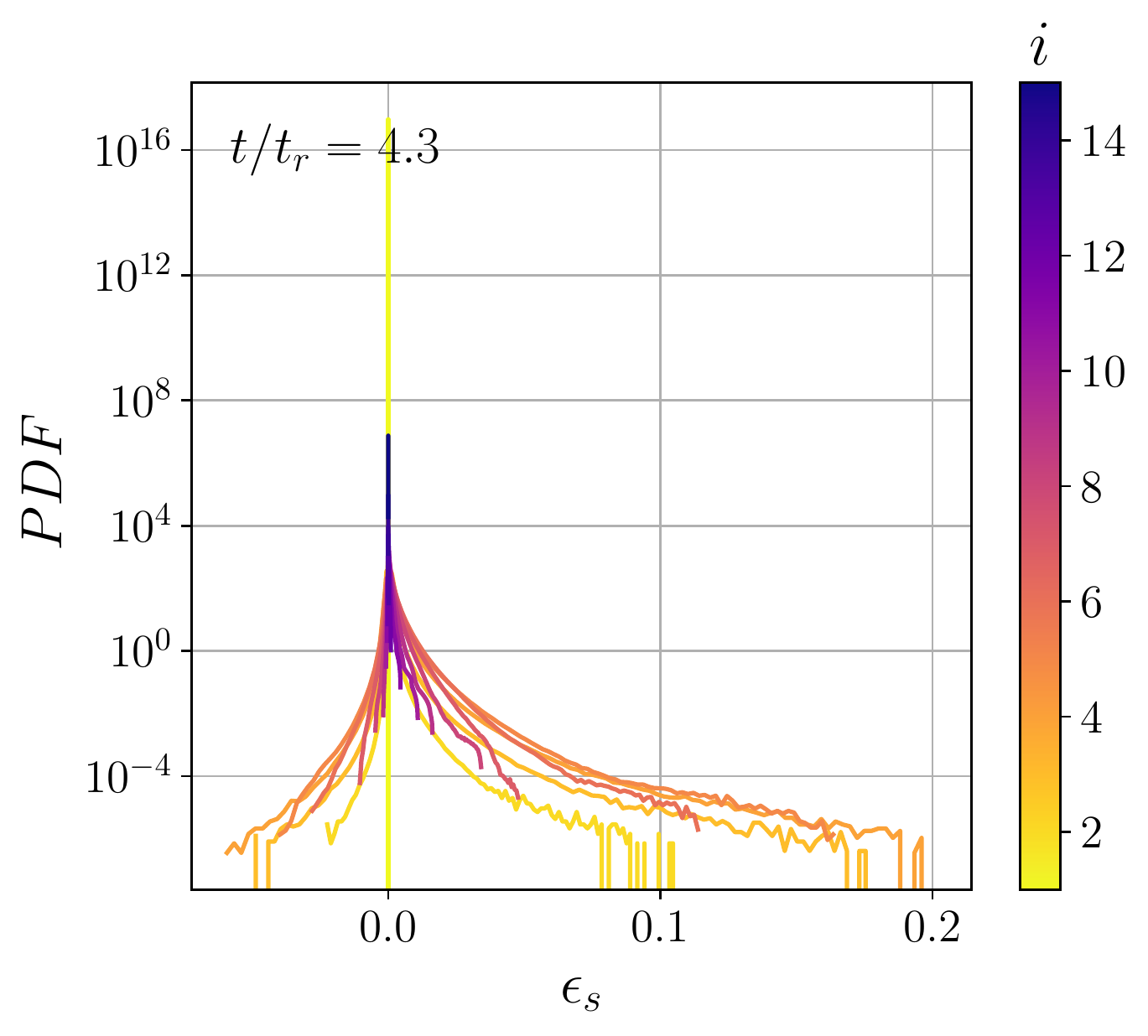}
  \includegraphics[trim=29 32 0 0, clip, width=0.29\textwidth]{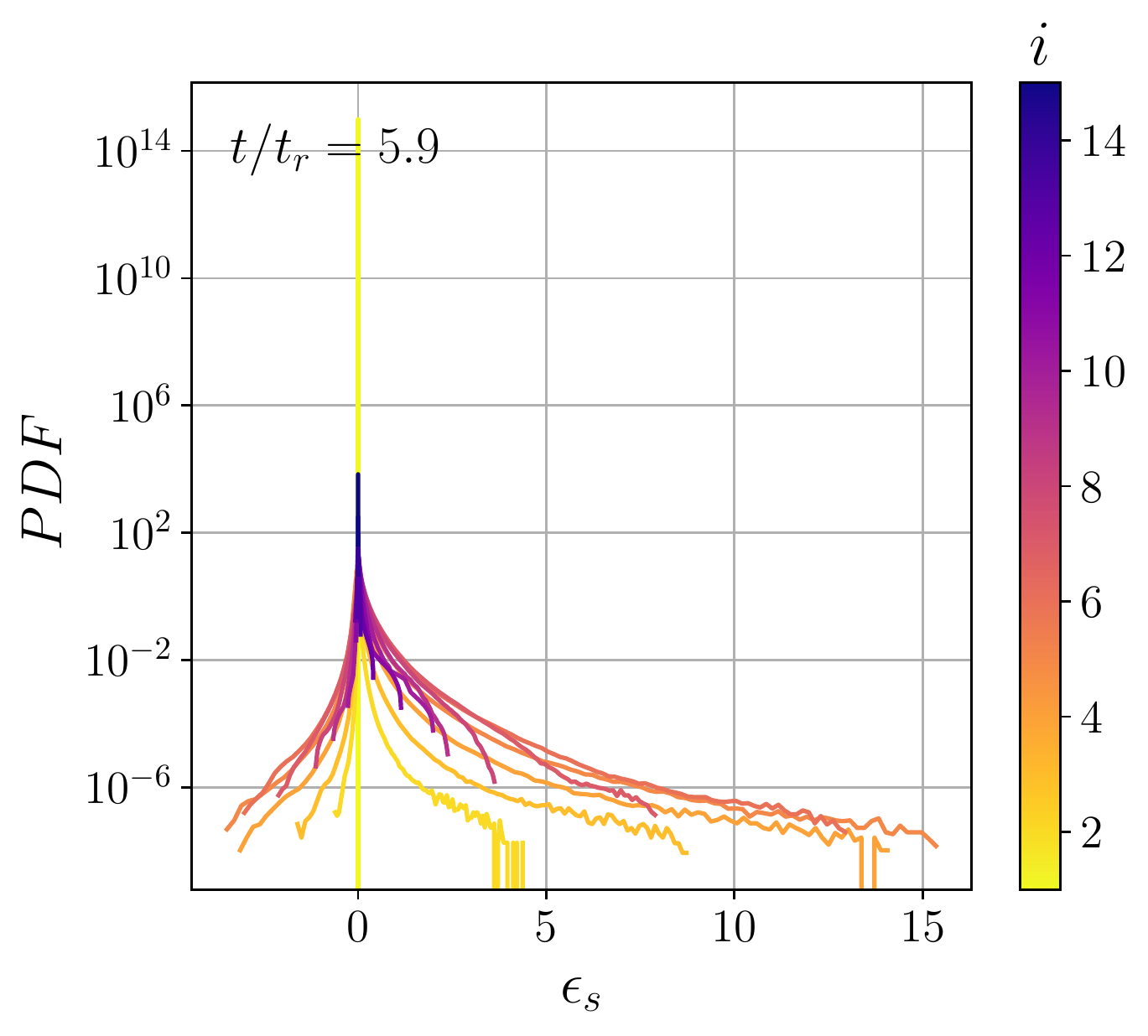}\\
  \includegraphics[trim=0 0 0 0, clip, width=0.31\textwidth]{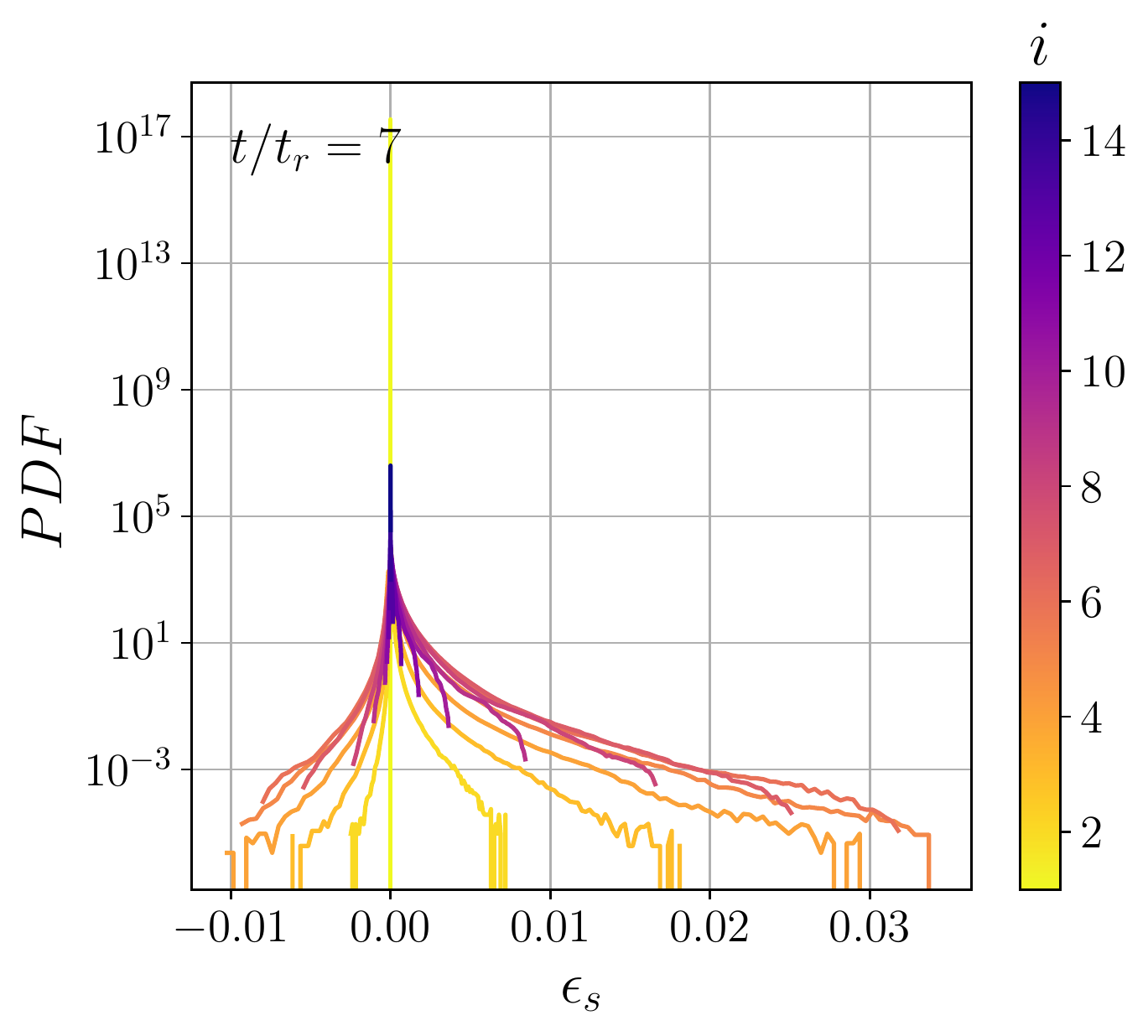}
  \includegraphics[trim=29 0 0 0, clip, width=0.3\textwidth]{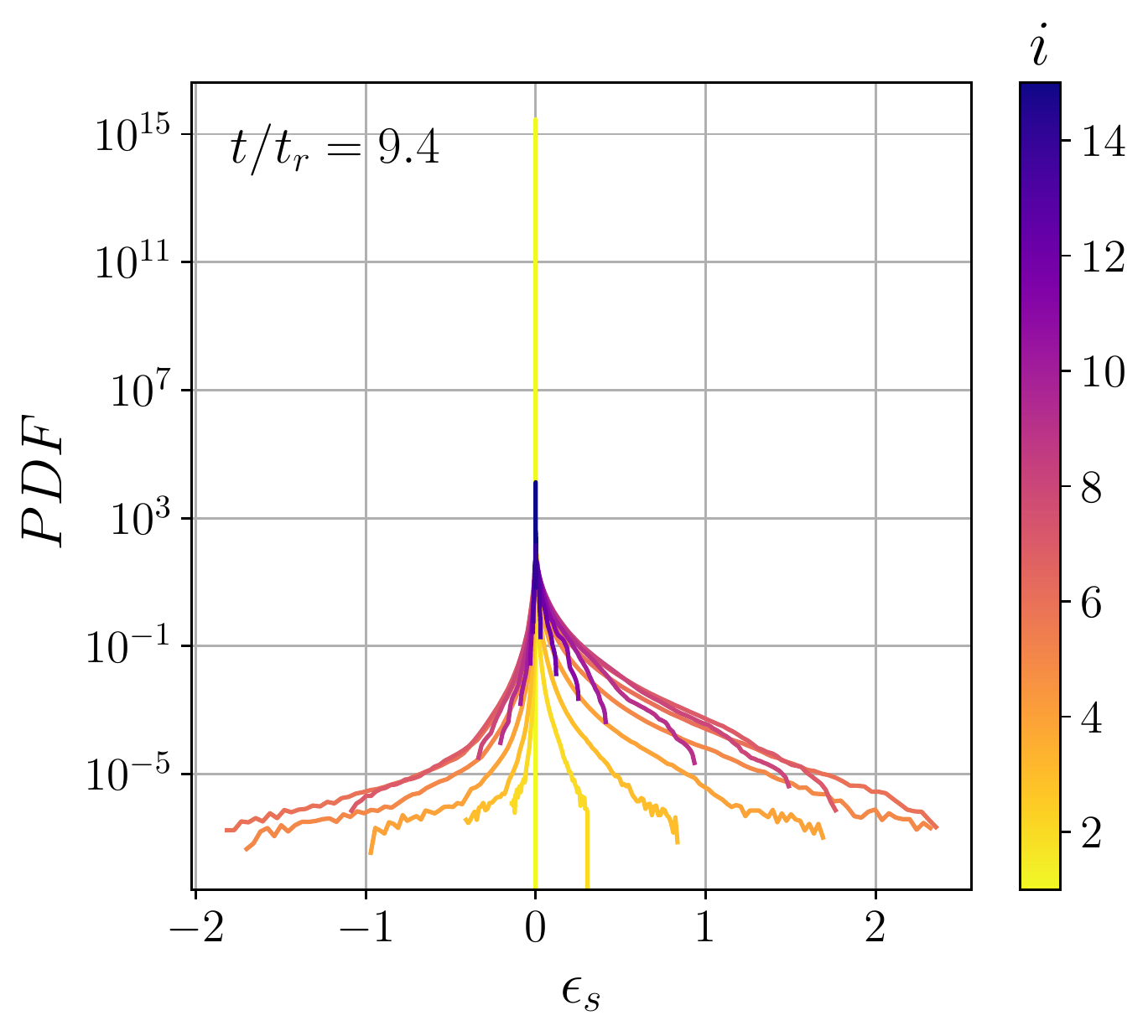}
  \caption{Probability density function of the transfer from resolved to small scale kinetic energy $\epsilon_{s} = -\overline{\rho} \, \mathcal{T}_{ij} \widetilde{S}_{ij}$ in (\ref{eq:scaleresolvedke}) and (\ref{eq:trans_tke_res_volume_avg}), as a function of filter width normalized by the box size, $w/L$ at different times. Left and right columns correspond to $A=0.05,0.75$, respectively. The colorbar indicates the filter width $i$ used, according to (\ref{eq:w_of_i}), Table \ref{tab:filter_width_list}, where $i=1$ corresponds to the smallest filter width $w/L = (\Delta x/\pi)/L$  and $i=15$ corresponds to the largest filter width $w/L = 1/2$.}
  \label{fig:kinetic_energy_transfer_pdf}
\end{figure}

The transfer of kinetic energy between the resolved scales and the sub-filter scales $\epsilon_s$ is important for understanding the physical nature of turbulent flows, as well as for the development of subgrid scale models for large eddy simulations and for scale resolving simulations. In Figure \ref{fig:kinetic_energy_transfer_pdf}, we plot probability density functions $\epsilon_s$ for the different filter widths used, for the four regimes we consider here. Recall from the discussion of Figure \ref{fig:tke_budget} that the net volume integrated transfer $\langle \epsilon_s \rangle$ is zero in the NS and RANS limits, and $\langle \epsilon_s \rangle > 0$ at intermediate scales, which is consistent with Figure \ref{fig:kinetic_energy_transfer_pdf}.
In the NS and RANS limits, $\epsilon_s = 0$ everywhere in the flow, while at intermediate scales it can be positive where kinetic energy is transferred to small scales, and negative where there is backscatter, i.e. transfer of kinetic energy from small scales to the resolved scales \cite{livescu_li_2017}. In LES approaches, backscatter acts as a source term in the kinetic energy equation and poses significant difficulties in maintaining stable computations. Many of the simple subgrid scale models do not account for backscatter, and properly describing this phenomenon is an active area of research \cite{Obrien_etal_2014}.
We have observed (not shown here) that, for both Atwood numbers, the fraction of the domain volume where backscatter occurs is roughly between 30\%-40\% at early times, during the turbulence growth regimes, and that this fraction is smaller, between 20\%-30\%, at later times during the decay regimes.
The range of values of $\epsilon_s$ increases with time until the kinetic energy peaks at the end of the saturated growth regime, after which it decreases. 
At the end of the explosive growth regime, the largest values of $\epsilon_s$ occur at filter widths $i = 6-7$, $w/L=4.3\times10^{-3}-7.4\times10^{-3}$ for $A=0.05$ and $i = 4-6$, $w/L=1.5\times10^{-3}-4.3\times10^{-3}$ for $A=0.75$. These filter widths decrease somewhat at the end of the saturated growth and the fast decay regimes, and then increase during the gradual decay regime. 
Rates of transfer of kinetic energy between resolved and sub-filter scales are orders of magnitude larger for the $A=0.75$ case than for the $A=0.05$ case.

\begin{figure}[htb]
  \centering
  \includegraphics[trim=0 32 0 0, clip, width=0.35\textwidth]{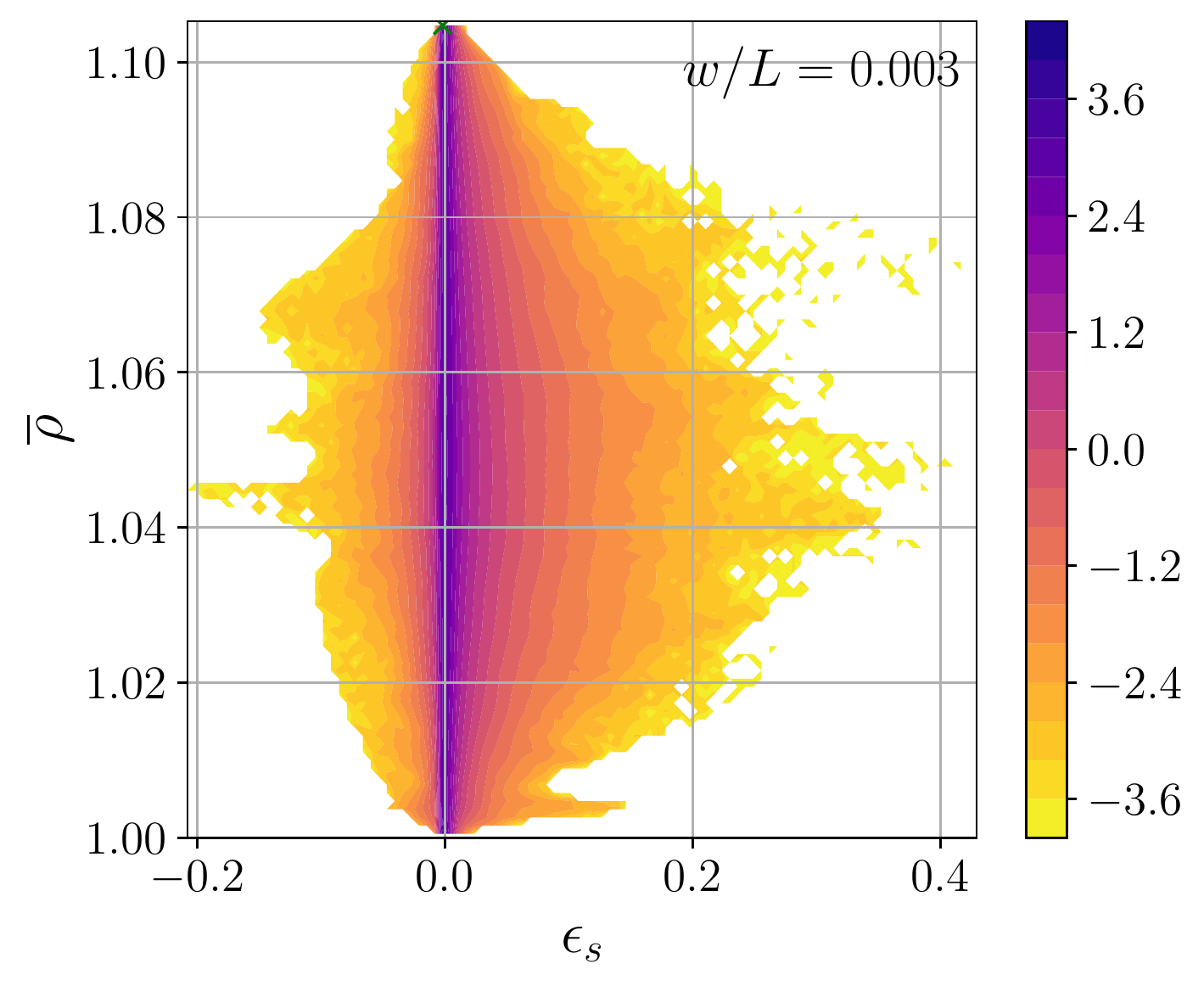}
  \includegraphics[trim=25 32 0 0, clip, width=0.31\textwidth]{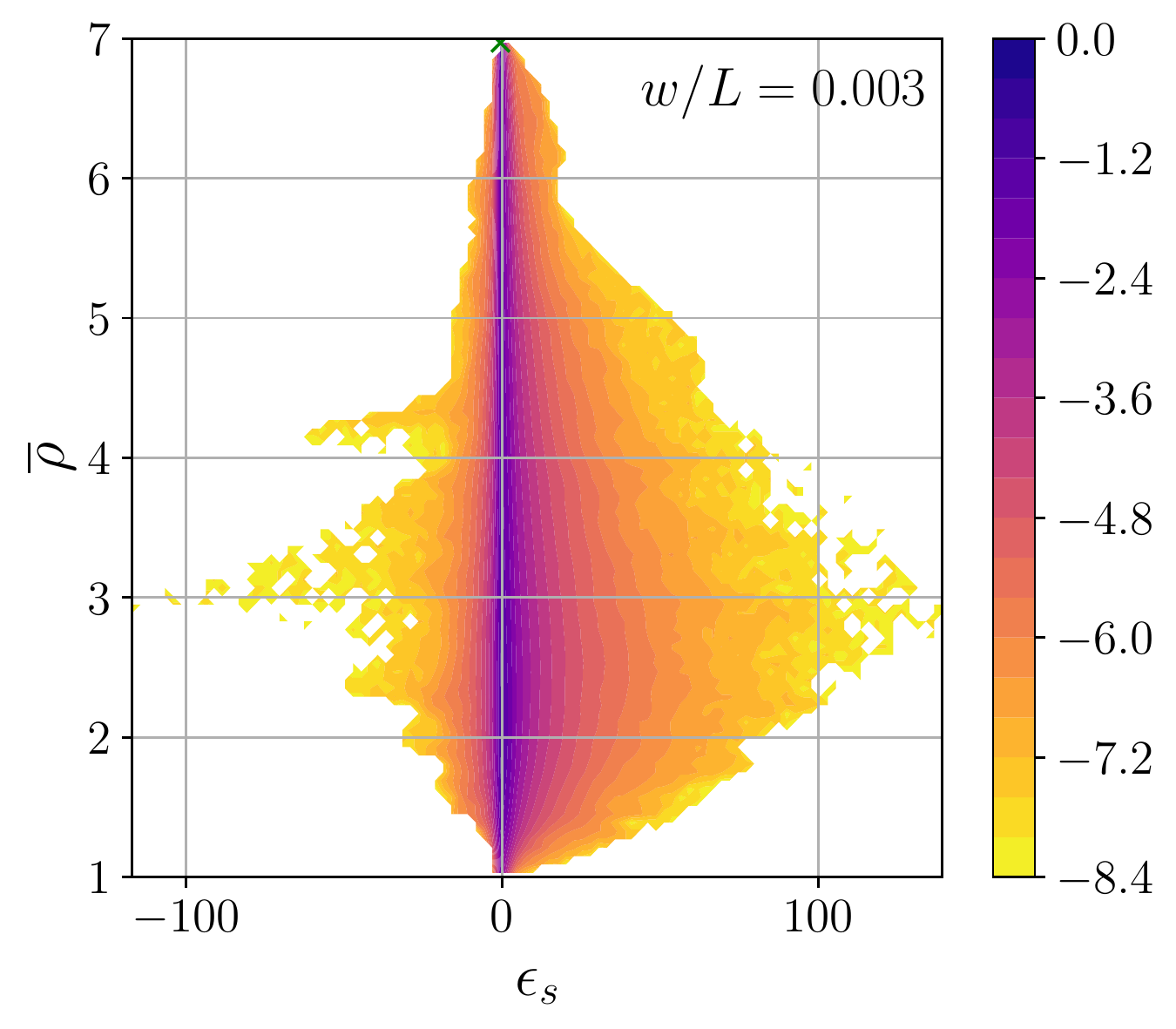}\\
  \includegraphics[trim=0 32 0 0, clip, width=0.35\textwidth]{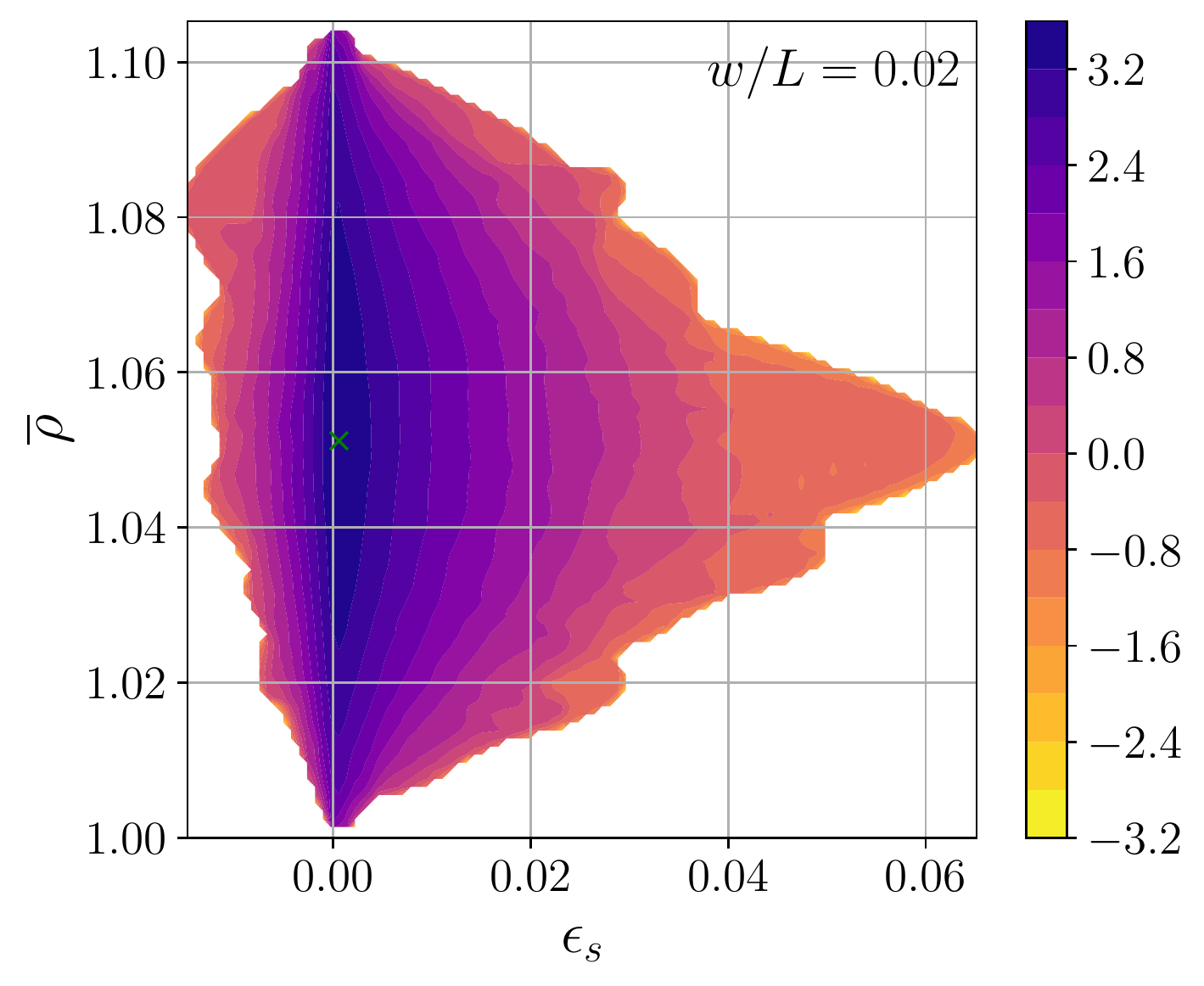}
  \includegraphics[trim=25 32 0 0, clip, width=0.31\textwidth]{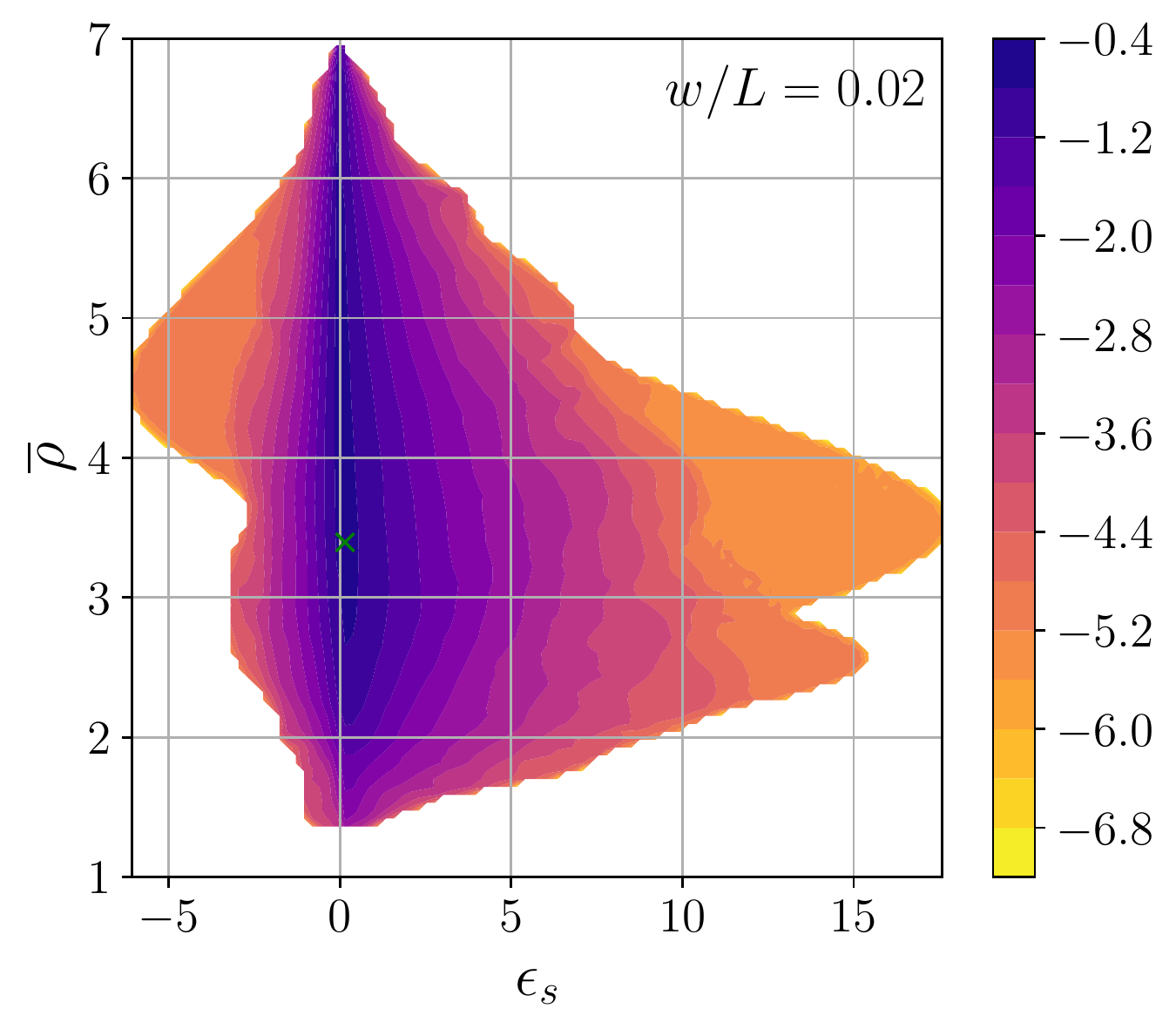} \\
  \includegraphics[trim=0 0 0 0, clip, width=0.34\textwidth]{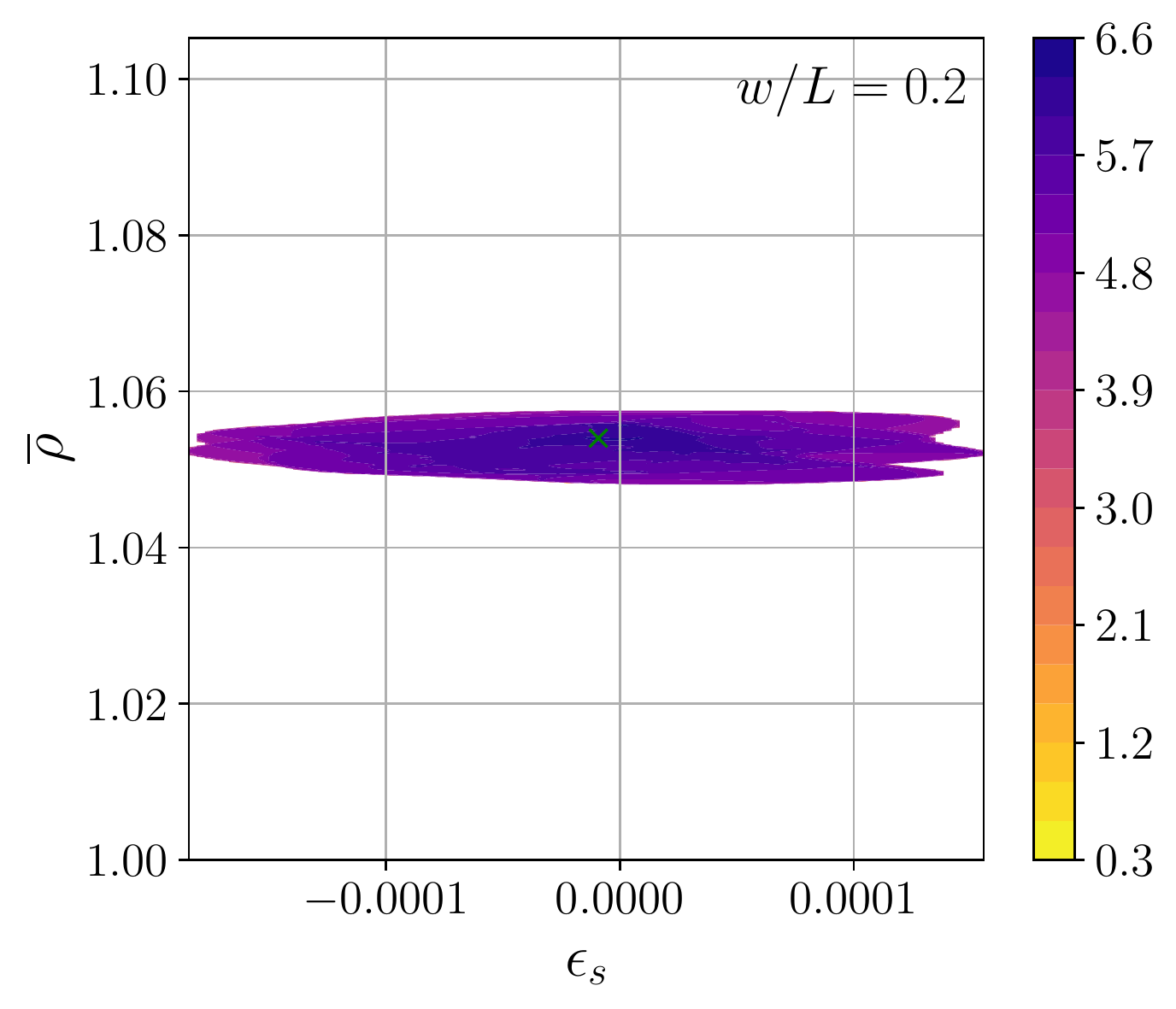}
  \includegraphics[trim=25 0 0 0, clip, width=0.31\textwidth]{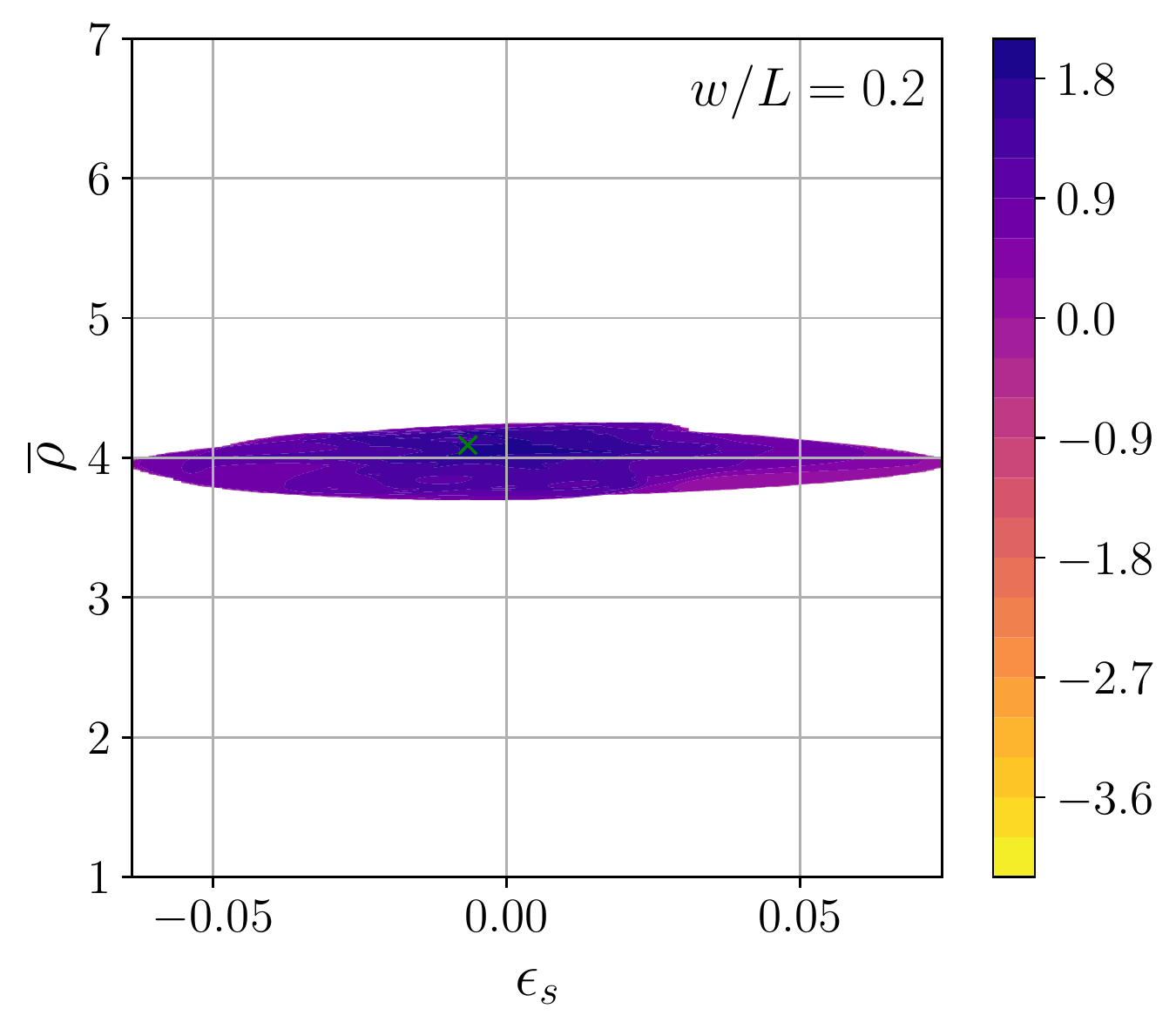}
  \caption{Joint probability density function for $\epsilon_s, \ol \rho$, at $t/t_r = 2.3$ and $t/t_r = 2.4$, when $\mathcal{R}_{ii}$ peaks, for $A=0.05$ (left) and $A=0.75$ (right), at three different filter widths $w/L=0.003, 0.02, 0.2$, as indicated.}
  \label{fig:rho_epssgs_pdf}
\end{figure}

The rate of transfer of kinetic energy between scales is largely affected by variable density effects, as illustrated in Figure \ref{fig:rho_epssgs_pdf}, where we plot the joint probability density function for $(\epsilon_s, \ol rho)$ at the end of the saturated growth regime period for three filter widths $w/L=0.0026, 0.021, 0.17$ corresponding to $i=5,9,13$, respectively, in Figure \ref{fig:kinetic_energy_transfer_pdf}.
For the flow with $A=0.05$, the joint PDF is mostly symmetrical with respect to the mean density $\langle \rho \rangle = 1.05$, while for $A=0.75$, the joint PDF is skewed towards densities smaller than the mean density $\langle \rho \rangle = 4$.
This happens because, due to inertial effects, turbulence is more energetic in the lighter fluid than in the heavier fluid \cite{livescu_ristorcelli_2007,livescu_ristorcelli_2008,aslangil_etal_2020}.
As the filter width increases between these three filter widths, the range of variability in $\epsilon_s$ decreases by several orders of magnitudes.

\section{Discussion and Conclusions}
\label{sec:duscussion_conclusions}

We have formulated a set of generalized, scale resolving (SR) variables for variable density turbulence, $b, a_i, \mathcal{T}_{ij}$, in equations (\ref{eq:b_inner_expectedvalue}-\ref{eq:Tij_inner_expectedvalue}), which are presented and discussed in section \ref{sec:filtering_and_RANSaveraging}.
These variables are written as inner products of the fluctuations of a quantity $q$ of the form 
\begin{eqnarray}
  \pp q_i(x) \equiv q_i(\xi) - \widetilde{u}_i(x), \\
  \p q_i(x)  \equiv q_i(\xi) - \hat{q}_i(x),
\end{eqnarray} 
for velocity, $\pp u(\xi,x)$ in (\ref{eq:gen_fluctuation}) and $\p u(\xi,x)$ in (\ref{eq:u_gen_fluctuation_one_prime}), and density, $\p \rho(\xi,x)$ in (\ref{eq:b_gen_fluctuation_one_prime}). The generalized fluctuating quantities represent fluctuations of a field variable $q_i(\xi)$ at points $\xi$, with respect to its filtered value $\tilde{q}_i(x)$ or $\ol{q}_i(x)$ at a point $x$.
The realizability conditions for the SR variables are a generalization of the realizability conditions for their RANS counterparts, and in the limit of large length scales, the latter are a special case of the former.
Evolution equations for $b, a_i, \mathcal{T}_{ij}$ are presented  in (\ref{eq:trans_b}, \ref{eq:trans_ai}, \ref{eq:trans_Tij}) in section \ref{sec:generalized_gov_eqns}.
We showed how the volume integrated SR variables $b, a_1, k_r$ and their budget terms vary smoothly between zero in the NS limit and their RANS counterparts in the RANS limit, as a function of length scale, or filter width, for homogeneous variable density turbulence.
These properties hold for filter kernels that have a positive stencil in space, which also ensures the preservation of scalar bounds, thus ensuring the consistency between the SR statistical description and the RANS statistical description of the flow.
For these reasons, we use the Gaussian filter, however, the analysis can be performed with other positive definite filters.

To illustrate these ideas, we perform diagnostics of the SR-equivalent of RANS variables that are used to investigate \cite{livescu_ristorcelli_2007,livescu_ristorcelli_2008,baltzer_livescu_2020,aslangil_etal_2020,aslangil_etal_2020b} and model \cite{besnard_etal_1992,stalsberg-zarling_gore_2011, schwarzkopf_etal_2011,schwarzkopf_etal_2015} variable density turbulence, namely the density-specific volume covariance $b$, the turbulent mass-flux velocity $a_i$, the Reynolds stress $\mathcal{T}_{ij}$, and the resolved kinetic energy $k_r$, defined in equations (\ref{eq:b_inner_expectedvalue}), (\ref{eq:a1_inner_expectedvalue}), (\ref{eq:Tij_inner_expectedvalue}), (\ref{eq:resolved_ke}), respectively, using theory and diagnostics from DNS of homogeneous variable density turbulence at times that are representative of different dynamical regimes in this flow.

In particular, in the RANS limit where the resolved scales similar to or larger than the dominating integral length scale, {\it i} the SR variables converge to the RANS variables; {\it ii} the governing equations of the SR variables converge to the governing equations of the RANS variables; {\it iii} inner products of random generalized fluctuations, equivalent to the expected value of their product, become the expected value of fluctuating quantities, or the statistical moments, in the RANS framework.
The terms dominating the balance equations for the SR variables include dynamical processes that are not active in the RANS balance equations.
For example, in the RANS limit, the only active term in the balance equation for $b_\mathrm{e}$ is the destruction term, and thus $b_\mathrm{e}$ is described by a purely decaying process.
The SR balance equation for $b$, on the other hand, is dominated by production, redistribution, transport and destruction terms, and $b$ can grow or decay, depending on the scale being considered and the stage of the flow.
At early stages of the flow, volume integrated production of $\ol \rho b$ is non-zero and can be larger than the rate of destruction at intermediate filter widths.
The flow is initialized with length scales $l$ between $1/5 \le l/L \le 1/3$, which are slightly larger than the integral length scale for the flow.
For this reason, at the onset of the flow, $b=0$ for filter widths $w/L < 1/5$.
Stirring occurs first at large scales, which is followed in time by the formation of structures and generation of density-specific volume covariance ($b$) at progressively smaller scales \cite{aslangil_etal_2020}. 
Consequently, the generalized variance in equation (\ref{eq:b_inner_expectedvalue}) is at first larger at larger scales, and at small scales it increases with time.
At later times after the end of the explosive growth regime, as turbulence becomes more developed, this strong variance is destroyed by mixing of the two fluids.
As a result, the destruction of $\ol \rho b$ at first is equal to, and then greater than the production as the filter width increases from the mesh size in the NS limit. 
This leads to a monotonically increasing (in magnitude) net decay of $\left\langle \ol \rho b \right \rangle$ as the filter width increases.

In summary, the dynamics at intermediate length scales are richer than the dynamics in the RANS description of the flow.
This has important implications for modeling, as it means that the dynamics at resolutions between LES and RANS resolutions are not just a modulated version of the dynamics represented by the RANS equations, an assumption used in some hybrid RANS/LES strategies.

This work supports the notion of a generalized, length-scale adaptive model in terms of the SR variables, that converges to DNS at high resolutions, and to classical RANS statistics at coarse resolutions
(e.g. [\onlinecite{perot_gadebusch_2007}] and [\onlinecite{perot_gadebusch_2009}])
. 
We believe that our work is a step towards formalizing the concept of such self-adaptive models, putting this concept on firmer footing. 




\begin{acknowledgments}
J.A.S. was supported by the Advanced Simulation and Computing (ASC) program through the Physics and Engineering Models - Mix and Burn project, and D.A. and D.L. were supported by the Office of Experimental Sciences program at Los Alamos National Laboratory (LANL). High-performance computing resources were provided by ASC and LANL Institutional Computing Program. This work was performed under the auspices of the US DOE/NNSA at LANL under Contract No. DE-AC52-06NA25396.
\end{acknowledgments}


\bibliography{master_bibliography}

\begin{thebibliography}{40}%
\makeatletter
\providecommand \@ifxundefined [1]{%
 \@ifx{#1\undefined}
}%
\providecommand \@ifnum [1]{%
 \ifnum #1\expandafter \@firstoftwo
 \else \expandafter \@secondoftwo
 \fi
}%
\providecommand \@ifx [1]{%
 \ifx #1\expandafter \@firstoftwo
 \else \expandafter \@secondoftwo
 \fi
}%
\providecommand \natexlab [1]{#1}%
\providecommand \enquote  [1]{``#1''}%
\providecommand \bibnamefont  [1]{#1}%
\providecommand \bibfnamefont [1]{#1}%
\providecommand \citenamefont [1]{#1}%
\providecommand \href@noop [0]{\@secondoftwo}%
\providecommand \href [0]{\begingroup \@sanitize@url \@href}%
\providecommand \@href[1]{\@@startlink{#1}\@@href}%
\providecommand \@@href[1]{\endgroup#1\@@endlink}%
\providecommand \@sanitize@url [0]{\catcode `\\12\catcode `\$12\catcode
  `\&12\catcode `\#12\catcode `\^12\catcode `\_12\catcode `\%12\relax}%
\providecommand \@@startlink[1]{}%
\providecommand \@@endlink[0]{}%
\providecommand \url  [0]{\begingroup\@sanitize@url \@url }%
\providecommand \@url [1]{\endgroup\@href {#1}{\urlprefix }}%
\providecommand \urlprefix  [0]{URL }%
\providecommand \Eprint [0]{\href }%
\providecommand \doibase [0]{http://dx.doi.org/}%
\providecommand \selectlanguage [0]{\@gobble}%
\providecommand \bibinfo  [0]{\@secondoftwo}%
\providecommand \bibfield  [0]{\@secondoftwo}%
\providecommand \translation [1]{[#1]}%
\providecommand \BibitemOpen [0]{}%
\providecommand \bibitemStop [0]{}%
\providecommand \bibitemNoStop [0]{.\EOS\space}%
\providecommand \EOS [0]{\spacefactor3000\relax}%
\providecommand \BibitemShut  [1]{\csname bibitem#1\endcsname}%
\let\auto@bib@innerbib\@empty
\bibitem [{\citenamefont {Pope}(2001)}]{pope_2001}%
  \BibitemOpen
  \bibfield  {author} {\bibinfo {author} {\bibfnamefont {S.~B.}\ \bibnamefont
  {Pope}},\ }\href@noop {} {\enquote {\bibinfo {title} {Turbulent flows},}\ }
  (\bibinfo {year} {2001})\BibitemShut {NoStop}%
\bibitem [{\citenamefont {Menter}, \citenamefont {Kuntz},\ and\ \citenamefont
  {Bender}(2003)}]{menter_etal_2012}%
  \BibitemOpen
  \bibfield  {author} {\bibinfo {author} {\bibfnamefont {F.}~\bibnamefont
  {Menter}}, \bibinfo {author} {\bibfnamefont {M.}~\bibnamefont {Kuntz}}, \
  and\ \bibinfo {author} {\bibfnamefont {R.}~\bibnamefont {Bender}},\ }\enquote
  {\bibinfo {title} {A scale-adaptive simulation model for turbulent flow
  predictions},}\ in\ \href {\doibase 10.2514/6.2003-767} {\emph {\bibinfo
  {booktitle} {41st Aerospace Sciences Meeting and Exhibit}}}\ (\bibinfo
  {publisher} {AIAA},\ \bibinfo {year} {2003})\ \Eprint
  {http://arxiv.org/abs/https://arc.aiaa.org/doi/pdf/10.2514/6.2003-767}
  {https://arc.aiaa.org/doi/pdf/10.2514/6.2003-767} \BibitemShut {NoStop}%
\bibitem [{\citenamefont {Chaouat}(2017)}]{chaouat_2017}%
  \BibitemOpen
  \bibfield  {author} {\bibinfo {author} {\bibfnamefont {B.}~\bibnamefont
  {Chaouat}},\ }\bibfield  {title} {\enquote {\bibinfo {title} {{The State of
  the Art of Hybrid RANS/LES Modeling for the Simulation of Turbulent
  Flows}},}\ }\href {\doibase 10.1007/s10494-017-9828-8} {\bibfield  {journal}
  {\bibinfo  {journal} {{Flow, Turbulence and Combustion}}\ }\textbf {\bibinfo
  {volume} {99}},\ \bibinfo {pages} {pages 279 -- 327} (\bibinfo {year}
  {2017})}\BibitemShut {NoStop}%
\bibitem [{\citenamefont {Heinz}(2020)}]{heinz_2020}%
  \BibitemOpen
  \bibfield  {author} {\bibinfo {author} {\bibfnamefont {S.}~\bibnamefont
  {Heinz}},\ }\bibfield  {title} {\enquote {\bibinfo {title} {A review of
  hybrid rans-les methods for turbulent flows: Concepts and applications},}\
  }\href {\doibase https://doi.org/10.1016/j.paerosci.2019.100597} {\bibfield
  {journal} {\bibinfo  {journal} {Progress in Aerospace Sciences}\ }\textbf
  {\bibinfo {volume} {114}},\ \bibinfo {pages} {100597} (\bibinfo {year}
  {2020})}\BibitemShut {NoStop}%
\bibitem [{\citenamefont {Meneveau}\ and\ \citenamefont
  {Katz}(2000)}]{meneveau_katz_2000}%
  \BibitemOpen
  \bibfield  {author} {\bibinfo {author} {\bibfnamefont {C.}~\bibnamefont
  {Meneveau}}\ and\ \bibinfo {author} {\bibfnamefont {J.}~\bibnamefont
  {Katz}},\ }\bibfield  {title} {\enquote {\bibinfo {title} {Scale-invariance
  and turbulence models for large-eddy simulation},}\ }\href {\doibase
  10.1146/annurev.fluid.32.1.1} {\bibfield  {journal} {\bibinfo  {journal}
  {Annual Review of Fluid Mechanics}\ }\textbf {\bibinfo {volume} {32}},\
  \bibinfo {pages} {1--32} (\bibinfo {year} {2000})},\ \Eprint
  {http://arxiv.org/abs/https://doi.org/10.1146/annurev.fluid.32.1.1}
  {https://doi.org/10.1146/annurev.fluid.32.1.1} \BibitemShut {NoStop}%
\bibitem [{\citenamefont {Fasel}, \citenamefont {von Terzi},\ and\
  \citenamefont {Sandberg}(2006)}]{fasel_etal_2006}%
  \BibitemOpen
  \bibfield  {author} {\bibinfo {author} {\bibfnamefont {H.~F.}\ \bibnamefont
  {Fasel}}, \bibinfo {author} {\bibfnamefont {D.~A.}\ \bibnamefont {von
  Terzi}}, \ and\ \bibinfo {author} {\bibfnamefont {R.~D.}\ \bibnamefont
  {Sandberg}},\ }\bibfield  {title} {\enquote {\bibinfo {title} {A methodology
  for simulating compressible turbulent flows},}\ }\href
  {http://dx.doi.org/10.1115/1.2150231} {\bibfield  {journal} {\bibinfo
  {journal} {Journal of Applied Mechanics}\ }\textbf {\bibinfo {volume} {73}},\
  \bibinfo {pages} {405--412} (\bibinfo {year} {2006})}\BibitemShut {NoStop}%
\bibitem [{\citenamefont {Grinstein}\ \emph {et~al.}(2020)\citenamefont
  {Grinstein}, \citenamefont {Saenz}, \citenamefont {Rauenzahn}, \citenamefont
  {Germano},\ and\ \citenamefont {Israel}}]{grinstein_etal_2020}%
  \BibitemOpen
  \bibfield  {author} {\bibinfo {author} {\bibfnamefont {F.}~\bibnamefont
  {Grinstein}}, \bibinfo {author} {\bibfnamefont {J.}~\bibnamefont {Saenz}},
  \bibinfo {author} {\bibfnamefont {R.}~\bibnamefont {Rauenzahn}}, \bibinfo
  {author} {\bibfnamefont {M.}~\bibnamefont {Germano}}, \ and\ \bibinfo
  {author} {\bibfnamefont {D.}~\bibnamefont {Israel}},\ }\bibfield  {title}
  {\enquote {\bibinfo {title} {Dynamic bridging modeling for coarse grained
  simulations of shock driven turbulent mixing},}\ }\href {\doibase
  https://doi.org/10.1016/j.compfluid.2020.104430} {\bibfield  {journal}
  {\bibinfo  {journal} {Computers \& Fluids}\ }\textbf {\bibinfo {volume}
  {199}},\ \bibinfo {pages} {104430} (\bibinfo {year} {2020})}\BibitemShut
  {NoStop}%
\bibitem [{\citenamefont {Perot}\ and\ \citenamefont
  {Gadebusch}(2007)}]{perot_gadebusch_2007}%
  \BibitemOpen
  \bibfield  {author} {\bibinfo {author} {\bibfnamefont {J.~B.}\ \bibnamefont
  {Perot}}\ and\ \bibinfo {author} {\bibfnamefont {J.}~\bibnamefont
  {Gadebusch}},\ }\bibfield  {title} {\enquote {\bibinfo {title} {A
  self-adapting turbulence model for flow simulation at any mesh resolution},}\
  }\href {\doibase 10.1063/1.2780195} {\bibfield  {journal} {\bibinfo
  {journal} {Physics of Fluids}\ }\textbf {\bibinfo {volume} {19}},\ \bibinfo
  {pages} {115105} (\bibinfo {year} {2007})},\ \Eprint
  {http://arxiv.org/abs/https://doi.org/10.1063/1.2780195}
  {https://doi.org/10.1063/1.2780195} \BibitemShut {NoStop}%
\bibitem [{\citenamefont {Perot}\ and\ \citenamefont
  {Gadebusch}(2009)}]{perot_gadebusch_2009}%
  \BibitemOpen
  \bibfield  {author} {\bibinfo {author} {\bibfnamefont {J.~B.}\ \bibnamefont
  {Perot}}\ and\ \bibinfo {author} {\bibfnamefont {J.}~\bibnamefont
  {Gadebusch}},\ }\bibfield  {title} {\enquote {\bibinfo {title} {A stress
  transport equation model for simulating turbulence at any mesh resolution},}\
  }\href {\doibase 10.1007/s00162-009-0113-x} {\bibfield  {journal} {\bibinfo
  {journal} {Theoretical and Computational Fluid Dynamics}\ }\textbf {\bibinfo
  {volume} {23}},\ \bibinfo {pages} {271--286} (\bibinfo {year}
  {2009})}\BibitemShut {NoStop}%
\bibitem [{\citenamefont {Germano}(1992)}]{germano_1992}%
  \BibitemOpen
  \bibfield  {author} {\bibinfo {author} {\bibfnamefont {M.}~\bibnamefont
  {Germano}},\ }\bibfield  {title} {\enquote {\bibinfo {title} {Turbulence: the
  filtering approach},}\ }\href@noop {} {\bibfield  {journal} {\bibinfo
  {journal} {J. Fluid Mech.}\ }\textbf {\bibinfo {volume} {238}},\ \bibinfo
  {pages} {325--336} (\bibinfo {year} {1992})}\BibitemShut {NoStop}%
\bibitem [{\citenamefont {Besnard}\ \emph {et~al.}(1992)\citenamefont
  {Besnard}, \citenamefont {Harlow}, \citenamefont {Rauenzahn},\ and\
  \citenamefont {Zemach}}]{besnard_etal_1992}%
  \BibitemOpen
  \bibfield  {author} {\bibinfo {author} {\bibfnamefont {D.}~\bibnamefont
  {Besnard}}, \bibinfo {author} {\bibfnamefont {F.~H.}\ \bibnamefont {Harlow}},
  \bibinfo {author} {\bibfnamefont {R.~M.}\ \bibnamefont {Rauenzahn}}, \ and\
  \bibinfo {author} {\bibfnamefont {C.}~\bibnamefont {Zemach}},\ }\href@noop {}
  {\enquote {\bibinfo {title} {Turbulence transport equations for
  variable-density turbulence and their relationship to two-field models},}\
  }\bibinfo {type} {Tech. Rep.}\ (\bibinfo  {institution} {Los Alamos National
  Lab., NM (United States)},\ \bibinfo {year} {1992})\BibitemShut {NoStop}%
\bibitem [{\citenamefont {Aslangil}, \citenamefont {Livescu},\ and\
  \citenamefont {Banerjee}(2020{\natexlab{a}})}]{aslangil_etal_2020}%
  \BibitemOpen
  \bibfield  {author} {\bibinfo {author} {\bibfnamefont {D.}~\bibnamefont
  {Aslangil}}, \bibinfo {author} {\bibfnamefont {D.}~\bibnamefont {Livescu}}, \
  and\ \bibinfo {author} {\bibfnamefont {A.}~\bibnamefont {Banerjee}},\
  }\bibfield  {title} {\enquote {\bibinfo {title} {Effects of {A}twood and
  {R}eynolds numbers on the evolution of buoyancy-driven homogeneous
  variable-density turbulence},}\ }\href {\doibase 10.1017/jfm.2020.268}
  {\bibfield  {journal} {\bibinfo  {journal} {Journal of Fluid Mechanics}\
  }\textbf {\bibinfo {volume} {895}},\ \bibinfo {pages} {A12} (\bibinfo {year}
  {2020}{\natexlab{a}})}\BibitemShut {NoStop}%
\bibitem [{\citenamefont {Stalsberg-Zarling}\ and\ \citenamefont
  {Gore}(2011)}]{stalsberg-zarling_gore_2011}%
  \BibitemOpen
  \bibfield  {author} {\bibinfo {author} {\bibfnamefont {K.}~\bibnamefont
  {Stalsberg-Zarling}}\ and\ \bibinfo {author} {\bibfnamefont {R.~A.}\
  \bibnamefont {Gore}},\ }\href@noop {} {\enquote {\bibinfo {title} {The bhr2
  turbulence model: Incompressible isotropic decay, rayleigh--taylor,
  kelvin--helmholtz and homogeneous variable density turbulence},}\ }\bibinfo
  {type} {Tech. Rep.}\ \bibinfo {number} {LA-UR 11-04773}\ (\bibinfo
  {institution} {Los Alamos National Laboratory},\ \bibinfo {year}
  {2011})\BibitemShut {NoStop}%
\bibitem [{\citenamefont {Schwarzkopf}\ \emph {et~al.}(2011)\citenamefont
  {Schwarzkopf}, \citenamefont {Livescu}, \citenamefont {Gore}, \citenamefont
  {Rauenzahn},\ and\ \citenamefont {Ristorcelli}}]{schwarzkopf_etal_2011}%
  \BibitemOpen
  \bibfield  {author} {\bibinfo {author} {\bibfnamefont {J.~D.}\ \bibnamefont
  {Schwarzkopf}}, \bibinfo {author} {\bibfnamefont {D.}~\bibnamefont
  {Livescu}}, \bibinfo {author} {\bibfnamefont {R.~A.}\ \bibnamefont {Gore}},
  \bibinfo {author} {\bibfnamefont {R.~M.}\ \bibnamefont {Rauenzahn}}, \ and\
  \bibinfo {author} {\bibfnamefont {J.~R.}\ \bibnamefont {Ristorcelli}},\
  }\bibfield  {title} {\enquote {\bibinfo {title} {Application of a
  second-moment closure model to mixing processes involving multicomponent
  miscible fluids},}\ }\href {\doibase 10.1080/14685248.2011.633084} {\bibfield
   {journal} {\bibinfo  {journal} {Journal of Turbulence}\ }\textbf {\bibinfo
  {volume} {12}},\ \bibinfo {pages} {N49} (\bibinfo {year} {2011})}\BibitemShut
  {NoStop}%
\bibitem [{\citenamefont {Schwarzkopf}\ \emph {et~al.}(2016)\citenamefont
  {Schwarzkopf}, \citenamefont {Livescu}, \citenamefont {Baltzer},
  \citenamefont {Gore},\ and\ \citenamefont
  {Ristorcelli}}]{schwarzkopf_etal_2015}%
  \BibitemOpen
  \bibfield  {author} {\bibinfo {author} {\bibfnamefont {J.~D.}\ \bibnamefont
  {Schwarzkopf}}, \bibinfo {author} {\bibfnamefont {D.}~\bibnamefont
  {Livescu}}, \bibinfo {author} {\bibfnamefont {J.~R.}\ \bibnamefont
  {Baltzer}}, \bibinfo {author} {\bibfnamefont {R.~A.}\ \bibnamefont {Gore}}, \
  and\ \bibinfo {author} {\bibfnamefont {J.~R.}\ \bibnamefont {Ristorcelli}},\
  }\bibfield  {title} {\enquote {\bibinfo {title} {A two-length scale
  turbulence model for single-phase multi-fluid mixing},}\ }\href {\doibase
  10.1007/s10494-015-9643-z} {\bibfield  {journal} {\bibinfo  {journal} {Flow,
  Turbulence and Combustion}\ }\textbf {\bibinfo {volume} {96}},\ \bibinfo
  {pages} {1--43} (\bibinfo {year} {2016})}\BibitemShut {NoStop}%
\bibitem [{\citenamefont {Livescu}\ \emph {et~al.}(2009)\citenamefont
  {Livescu}, \citenamefont {Ristorcelli}, \citenamefont {Gore}, \citenamefont
  {Dean}, \citenamefont {Cabot},\ and\ \citenamefont
  {Cook}}]{livescu_etal_2009}%
  \BibitemOpen
  \bibfield  {author} {\bibinfo {author} {\bibfnamefont {D.}~\bibnamefont
  {Livescu}}, \bibinfo {author} {\bibfnamefont {J.~R.}\ \bibnamefont
  {Ristorcelli}}, \bibinfo {author} {\bibfnamefont {R.~A.}\ \bibnamefont
  {Gore}}, \bibinfo {author} {\bibfnamefont {S.~H.}\ \bibnamefont {Dean}},
  \bibinfo {author} {\bibfnamefont {W.~H.}\ \bibnamefont {Cabot}}, \ and\
  \bibinfo {author} {\bibfnamefont {A.~W.}\ \bibnamefont {Cook}},\ }\bibfield
  {title} {\enquote {\bibinfo {title} {High-reynolds number rayleigh--taylor
  turbulence},}\ }\href {\doibase 10.1080/14685240902870448} {\bibfield
  {journal} {\bibinfo  {journal} {Journal of Turbulence}\ }\textbf {\bibinfo
  {volume} {10}},\ \bibinfo {pages} {N13} (\bibinfo {year} {2009})},\ \Eprint
  {http://arxiv.org/abs/http://dx.doi.org/10.1080/14685240902870448}
  {http://dx.doi.org/10.1080/14685240902870448} \BibitemShut {NoStop}%
\bibitem [{\citenamefont {Vreman}, \citenamefont {Geurts},\ and\ \citenamefont
  {Kuerten}(1994)}]{vreman_etal_1994}%
  \BibitemOpen
  \bibfield  {author} {\bibinfo {author} {\bibfnamefont {B.}~\bibnamefont
  {Vreman}}, \bibinfo {author} {\bibfnamefont {B.}~\bibnamefont {Geurts}}, \
  and\ \bibinfo {author} {\bibfnamefont {H.}~\bibnamefont {Kuerten}},\
  }\bibfield  {title} {\enquote {\bibinfo {title} {Realizability conditions for
  the turbulent stress tensor in large-eddy simulation},}\ }\href {\doibase
  10.1017/S0022112094003745} {\bibfield  {journal} {\bibinfo  {journal}
  {Journal of Fluid Mechanics}\ }\textbf {\bibinfo {volume} {278}},\ \bibinfo
  {pages} {351--362} (\bibinfo {year} {1994})}\BibitemShut {NoStop}%
\bibitem [{\citenamefont {Livescu}\ and\ \citenamefont
  {Ristorcelli}(2007)}]{livescu_ristorcelli_2007}%
  \BibitemOpen
  \bibfield  {author} {\bibinfo {author} {\bibfnamefont {D.}~\bibnamefont
  {Livescu}}\ and\ \bibinfo {author} {\bibfnamefont {J.~R.}\ \bibnamefont
  {Ristorcelli}},\ }\bibfield  {title} {\enquote {\bibinfo {title}
  {Buoyancy-driven variable-density turbulence},}\ }\href {\doibase
  10.1017/S0022112007008270} {\bibfield  {journal} {\bibinfo  {journal} {J.
  Fluid Mech.}\ }\textbf {\bibinfo {volume} {591}},\ \bibinfo {pages} {43--71}
  (\bibinfo {year} {2007})}\BibitemShut {NoStop}%
\bibitem [{\citenamefont {Buzzicotti}\ \emph {et~al.}(2018)\citenamefont
  {Buzzicotti}, \citenamefont {Linkmann}, \citenamefont {Aluie}, \citenamefont
  {Biferale}, \citenamefont {Brasseur},\ and\ \citenamefont
  {Meneveau}}]{buzzicotti_etal_2018}%
  \BibitemOpen
  \bibfield  {author} {\bibinfo {author} {\bibfnamefont {M.}~\bibnamefont
  {Buzzicotti}}, \bibinfo {author} {\bibfnamefont {M.}~\bibnamefont
  {Linkmann}}, \bibinfo {author} {\bibfnamefont {H.}~\bibnamefont {Aluie}},
  \bibinfo {author} {\bibfnamefont {L.}~\bibnamefont {Biferale}}, \bibinfo
  {author} {\bibfnamefont {J.}~\bibnamefont {Brasseur}}, \ and\ \bibinfo
  {author} {\bibfnamefont {C.}~\bibnamefont {Meneveau}},\ }\bibfield  {title}
  {\enquote {\bibinfo {title} {Effect of filter type on the statistics of
  energy transfer between resolved and subfilter scales from a-priori analysis
  of direct numerical simulations of isotropic turbulence},}\ }\href {\doibase
  10.1080/14685248.2017.1417597} {\bibfield  {journal} {\bibinfo  {journal}
  {Journal of Turbulence}\ }\textbf {\bibinfo {volume} {19}},\ \bibinfo {pages}
  {167--197} (\bibinfo {year} {2018})},\ \Eprint
  {http://arxiv.org/abs/https://doi.org/10.1080/14685248.2017.1417597}
  {https://doi.org/10.1080/14685248.2017.1417597} \BibitemShut {NoStop}%
\bibitem [{\citenamefont {Boyd}\ and\ \citenamefont
  {Vandenberghe}(2018)}]{boyd_vandenberghe_2018}%
  \BibitemOpen
  \bibfield  {author} {\bibinfo {author} {\bibfnamefont {S.}~\bibnamefont
  {Boyd}}\ and\ \bibinfo {author} {\bibfnamefont {L.}~\bibnamefont
  {Vandenberghe}},\ }\href {https://books.google.com/books?id=IApaDwAAQBAJ}
  {\emph {\bibinfo {title} {Introduction to Applied Linear Algebra: Vectors,
  Matrices, and Least Squares}}}\ (\bibinfo  {publisher} {Cambridge University
  Press},\ \bibinfo {year} {2018})\BibitemShut {NoStop}%
\bibitem [{\citenamefont {Batchelor}, \citenamefont {Canuto},\ and\
  \citenamefont {Chasnov}(1992)}]{batchelor_etal_1992}%
  \BibitemOpen
  \bibfield  {author} {\bibinfo {author} {\bibfnamefont {G.~K.}\ \bibnamefont
  {Batchelor}}, \bibinfo {author} {\bibfnamefont {V.~M.}\ \bibnamefont
  {Canuto}}, \ and\ \bibinfo {author} {\bibfnamefont {J.~R.}\ \bibnamefont
  {Chasnov}},\ }\bibfield  {title} {\enquote {\bibinfo {title} {Homogeneous
  buoyancy-generated turbulence},}\ }\href {\doibase 10.1017/S0022112092001149}
  {\bibfield  {journal} {\bibinfo  {journal} {J. Fluid Mech.}\ }\textbf
  {\bibinfo {volume} {235}},\ \bibinfo {pages} {349--378} (\bibinfo {year}
  {1992})}\BibitemShut {NoStop}%
\bibitem [{\citenamefont {Sandoval}, \citenamefont {Clark},\ and\ \citenamefont
  {Riley}(1997)}]{sandoval_etal_1997}%
  \BibitemOpen
  \bibfield  {author} {\bibinfo {author} {\bibfnamefont {D.~L.}\ \bibnamefont
  {Sandoval}}, \bibinfo {author} {\bibfnamefont {T.~T.}\ \bibnamefont {Clark}},
  \ and\ \bibinfo {author} {\bibfnamefont {J.~J.}\ \bibnamefont {Riley}},\
  }\bibfield  {title} {\enquote {\bibinfo {title} {Buoyancy-generated
  variable-density turbulence},}\ }in\ \href {\doibase
  10.1007/978-94-011-5474-1_22} {\emph {\bibinfo {booktitle} {{IUTAM} Symposium
  on Variable Density Low-Speed Turbulent Flows: Proceedings of the {IUTAM}
  Symposium held in Marseille, France, 8-10 July 1996}}},\ \bibinfo {editor}
  {edited by\ \bibinfo {editor} {\bibfnamefont {L.}~\bibnamefont {Fulachier}},
  \bibinfo {editor} {\bibfnamefont {J.~L.}\ \bibnamefont {Lumley}}, \ and\
  \bibinfo {editor} {\bibfnamefont {F.}~\bibnamefont {Anselmet}}}\ (\bibinfo
  {publisher} {Springer Netherlands},\ \bibinfo {address} {Dordrecht},\
  \bibinfo {year} {1997})\ pp.\ \bibinfo {pages} {173--180}\BibitemShut
  {NoStop}%
\bibitem [{\citenamefont {Livescu}\ and\ \citenamefont
  {Ristorcelli}(2008)}]{livescu_ristorcelli_2008}%
  \BibitemOpen
  \bibfield  {author} {\bibinfo {author} {\bibfnamefont {D.}~\bibnamefont
  {Livescu}}\ and\ \bibinfo {author} {\bibfnamefont {J.~R.}\ \bibnamefont
  {Ristorcelli}},\ }\bibfield  {title} {\enquote {\bibinfo {title}
  {Variable-density mixing in buoyancy-driven turbulence},}\ }\href {\doibase
  10.1017/S0022112008001481} {\bibfield  {journal} {\bibinfo  {journal} {J.
  Fluid Mech.}\ }\textbf {\bibinfo {volume} {605}},\ \bibinfo {pages}
  {145--180} (\bibinfo {year} {2008})}\BibitemShut {NoStop}%
\bibitem [{\citenamefont {Aslangil}, \citenamefont {Livescu},\ and\
  \citenamefont {Banerjee}(2020{\natexlab{b}})}]{aslangil_etal_2020b}%
  \BibitemOpen
  \bibfield  {author} {\bibinfo {author} {\bibfnamefont {D.}~\bibnamefont
  {Aslangil}}, \bibinfo {author} {\bibfnamefont {D.}~\bibnamefont {Livescu}}, \
  and\ \bibinfo {author} {\bibfnamefont {A.}~\bibnamefont {Banerjee}},\
  }\bibfield  {title} {\enquote {\bibinfo {title} {Variable-density
  buoyancy-driven turbulence with asymmetric initial density distribution},}\
  }\href {\doibase 10.1016/j.physd.2020.132444} {\bibfield  {journal} {\bibinfo
   {journal} {Physica D: Nonlinear Phenomena}\ }\textbf {\bibinfo {volume}
  {406}},\ \bibinfo {pages} {132444} (\bibinfo {year}
  {2020}{\natexlab{b}})}\BibitemShut {NoStop}%
\bibitem [{\citenamefont {Aslangil}, \citenamefont {Livescu},\ and\
  \citenamefont {Banerjee}(2019)}]{Aslangil_book_ch_2019}%
  \BibitemOpen
  \bibfield  {author} {\bibinfo {author} {\bibfnamefont {D.}~\bibnamefont
  {Aslangil}}, \bibinfo {author} {\bibfnamefont {D.}~\bibnamefont {Livescu}}, \
  and\ \bibinfo {author} {\bibfnamefont {A.}~\bibnamefont {Banerjee}},\
  }\bibfield  {title} {\enquote {\bibinfo {title} {Flow regimes in
  buoyancy-driven homogeneous variable-density turbulence},}\ }in\ \href@noop
  {} {\emph {\bibinfo {booktitle} {Progress in Turbulence VIII}}},\ \bibinfo
  {editor} {edited by\ \bibinfo {editor} {\bibfnamefont {R.}~\bibnamefont
  {{\"O}rl{\"u}}}, \bibinfo {editor} {\bibfnamefont {A.}~\bibnamefont
  {Talamelli}}, \bibinfo {editor} {\bibfnamefont {J.}~\bibnamefont {Peinke}}, \
  and\ \bibinfo {editor} {\bibfnamefont {M.}~\bibnamefont {Oberlack}}}\
  (\bibinfo  {publisher} {Springer International Publishing},\ \bibinfo
  {address} {Cham},\ \bibinfo {year} {2019})\ pp.\ \bibinfo {pages}
  {235--240}\BibitemShut {NoStop}%
\bibitem [{\citenamefont {Rayleigh}(1882)}]{rayleigh_1882}%
  \BibitemOpen
  \bibfield  {author} {\bibinfo {author} {\bibnamefont {Rayleigh}},\ }\bibfield
   {title} {\enquote {\bibinfo {title} {Investigation of the character of the
  equilibrium of an incompressible heavy fluid of variable density},}\ }\href
  {\doibase 10.1112/plms/s1-14.1.170} {\bibfield  {journal} {\bibinfo
  {journal} {Proceedings of the London Mathematical Society}\ }\textbf
  {\bibinfo {volume} {s1-14}},\ \bibinfo {pages} {170--177} (\bibinfo {year}
  {1882})},\ \Eprint
  {http://arxiv.org/abs/https://londmathsoc.onlinelibrary.wiley.com/doi/pdf/10.1112/plms/s1-14.1.170}
  {https://londmathsoc.onlinelibrary.wiley.com/doi/pdf/10.1112/plms/s1-14.1.170}
  \BibitemShut {NoStop}%
\bibitem [{\citenamefont {Taylor}(1950)}]{taylor_1949}%
  \BibitemOpen
  \bibfield  {author} {\bibinfo {author} {\bibfnamefont {G.~I.}\ \bibnamefont
  {Taylor}},\ }\bibfield  {title} {\enquote {\bibinfo {title} {The instability
  of liquid surfaces when accelerated in a direction perpendicular to their
  planes. i},}\ }\href {\doibase 10.1098/rspa.1950.0052} {\bibfield  {journal}
  {\bibinfo  {journal} {Proceedings of the Royal Society of London A:
  Mathematical, Physical and Engineering Sciences}\ }\textbf {\bibinfo {volume}
  {201}},\ \bibinfo {pages} {192--196} (\bibinfo {year} {1950})},\ \Eprint
  {http://arxiv.org/abs/http://rspa.royalsocietypublishing.org/content/201/1065/192.full.pdf}
  {http://rspa.royalsocietypublishing.org/content/201/1065/192.full.pdf}
  \BibitemShut {NoStop}%
\bibitem [{\citenamefont {Ramaprabhu}, \citenamefont {Karkhanis},\ and\
  \citenamefont {Lawrie}(2013)}]{ramaprabhu_etal_2013}%
  \BibitemOpen
  \bibfield  {author} {\bibinfo {author} {\bibfnamefont {P.}~\bibnamefont
  {Ramaprabhu}}, \bibinfo {author} {\bibfnamefont {V.}~\bibnamefont
  {Karkhanis}}, \ and\ \bibinfo {author} {\bibfnamefont {A.~G.~W.}\
  \bibnamefont {Lawrie}},\ }\bibfield  {title} {\enquote {\bibinfo {title}
  {{The Rayleigh-Taylor Instability driven by an accel-decel-accel profile}},}\
  }\href {\doibase 10.1063/1.4829765} {\bibfield  {journal} {\bibinfo
  {journal} {Physics of Fluids}\ }\textbf {\bibinfo {volume} {25}},\ \bibinfo
  {pages} {115104} (\bibinfo {year} {2013})}\BibitemShut {NoStop}%
\bibitem [{\citenamefont {Aslangil}, \citenamefont {Banerjee},\ and\
  \citenamefont {Lawrie}(2016)}]{aslangil_etal_2016}%
  \BibitemOpen
  \bibfield  {author} {\bibinfo {author} {\bibfnamefont {D.}~\bibnamefont
  {Aslangil}}, \bibinfo {author} {\bibfnamefont {A.}~\bibnamefont {Banerjee}},
  \ and\ \bibinfo {author} {\bibfnamefont {A.~G.~W.}\ \bibnamefont {Lawrie}},\
  }\bibfield  {title} {\enquote {\bibinfo {title} {Numerical investigation of
  initial condition effects on {Rayleigh-Taylor} instability with acceleration
  reversals},}\ }\href {\doibase 10.1103/PhysRevE.94.053114} {\bibfield
  {journal} {\bibinfo  {journal} {Phys. Rev. E}\ }\textbf {\bibinfo {volume}
  {94}},\ \bibinfo {pages} {053114} (\bibinfo {year} {2016})}\BibitemShut
  {NoStop}%
\bibitem [{\citenamefont {Aslangil}\ \emph {et~al.}(2020)\citenamefont
  {Aslangil}, \citenamefont {Farley}, \citenamefont {Lawrie},\ and\
  \citenamefont {Banerjee}}]{aslangil_etal_2020c}%
  \BibitemOpen
  \bibfield  {author} {\bibinfo {author} {\bibfnamefont {D.}~\bibnamefont
  {Aslangil}}, \bibinfo {author} {\bibfnamefont {Z.}~\bibnamefont {Farley}},
  \bibinfo {author} {\bibfnamefont {A.~G.}\ \bibnamefont {Lawrie}}, \ and\
  \bibinfo {author} {\bibfnamefont {A.}~\bibnamefont {Banerjee}},\ }\bibfield
  {title} {\enquote {\bibinfo {title} {{Rayleigh-Taylor Instability with
  Varying Periods of Zero Acceleration}},}\ }\href {\doibase 10.1115/1.4048348}
  {\bibfield  {journal} {\bibinfo  {journal} {Journal of Fluids Engineering}\ }
  (\bibinfo {year} {2020}),\ 10.1115/1.4048348},\ \Eprint
  {http://arxiv.org/abs/https://asmedigitalcollection.asme.org/fluidsengineering/article-pdf/doi/10.1115/1.4048348/6563325/fe-20-1362.pdf}
  {https://asmedigitalcollection.asme.org/fluidsengineering/article-pdf/doi/10.1115/1.4048348/6563325/fe-20-1362.pdf}
  \BibitemShut {NoStop}%
\bibitem [{\citenamefont {Livescu}(2020)}]{livescu_2020}%
  \BibitemOpen
  \bibfield  {author} {\bibinfo {author} {\bibfnamefont {D.}~\bibnamefont
  {Livescu}},\ }\bibfield  {title} {\enquote {\bibinfo {title} {Turbulence with
  large thermal and compositional density variations},}\ }\href {\doibase
  10.1146/annurev-fluid-010719-060114} {\bibfield  {journal} {\bibinfo
  {journal} {Annual Review of Fluid Mechanics}\ }\textbf {\bibinfo {volume}
  {52}},\ \bibinfo {pages} {309--341} (\bibinfo {year} {2020})},\ \Eprint
  {http://arxiv.org/abs/https://doi.org/10.1146/annurev-fluid-010719-060114}
  {https://doi.org/10.1146/annurev-fluid-010719-060114} \BibitemShut {NoStop}%
\bibitem [{\citenamefont {D.}(1960)}]{richtmyer_1960}%
  \BibitemOpen
  \bibfield  {author} {\bibinfo {author} {\bibfnamefont {R.~R.}\ \bibnamefont
  {D.}},\ }\bibfield  {title} {\enquote {\bibinfo {title} {Taylor instability
  in shock acceleration of compressible fluids},}\ }\href {\doibase
  10.1002/cpa.3160130207} {\bibfield  {journal} {\bibinfo  {journal}
  {Communications on Pure and Applied Mathematics}\ }\textbf {\bibinfo {volume}
  {13}},\ \bibinfo {pages} {297--319} (\bibinfo {year} {1960})},\ \Eprint
  {http://arxiv.org/abs/https://onlinelibrary.wiley.com/doi/pdf/10.1002/cpa.3160130207}
  {https://onlinelibrary.wiley.com/doi/pdf/10.1002/cpa.3160130207} \BibitemShut
  {NoStop}%
\bibitem [{\citenamefont {Meshkov}(1969)}]{meshkov_1969}%
  \BibitemOpen
  \bibfield  {author} {\bibinfo {author} {\bibfnamefont {E.~E.}\ \bibnamefont
  {Meshkov}},\ }\bibfield  {title} {\enquote {\bibinfo {title} {Instability of
  the interface of two gases accelerated by a shock wave},}\ }\href {\doibase
  10.1007/BF01015969} {\bibfield  {journal} {\bibinfo  {journal} {Fluid
  Dynamics}\ }\textbf {\bibinfo {volume} {4}},\ \bibinfo {pages} {101--104}
  (\bibinfo {year} {1969})}\BibitemShut {NoStop}%
\bibitem [{\citenamefont {Baltzer}\ and\ \citenamefont
  {Livescu}(2020)}]{baltzer_livescu_2020}%
  \BibitemOpen
  \bibfield  {author} {\bibinfo {author} {\bibfnamefont {J.~R.}\ \bibnamefont
  {Baltzer}}\ and\ \bibinfo {author} {\bibfnamefont {D.}~\bibnamefont
  {Livescu}},\ }\bibfield  {title} {\enquote {\bibinfo {title}
  {Variable-density effects in incompressible non-buoyant shear-driven
  turbulent mixing layers},}\ }\href {\doibase 10.1017/jfm.2020.466} {\bibfield
   {journal} {\bibinfo  {journal} {Journal of Fluid Mechanics}\ }\textbf
  {\bibinfo {volume} {900}},\ \bibinfo {pages} {A16} (\bibinfo {year}
  {2020})}\BibitemShut {NoStop}%
\bibitem [{\citenamefont {Charonko}\ and\ \citenamefont
  {Prestridge}(2017)}]{charonko_prestridge_2017}%
  \BibitemOpen
  \bibfield  {author} {\bibinfo {author} {\bibfnamefont {J.~J.}\ \bibnamefont
  {Charonko}}\ and\ \bibinfo {author} {\bibfnamefont {K.}~\bibnamefont
  {Prestridge}},\ }\bibfield  {title} {\enquote {\bibinfo {title}
  {Variable-density mixing in turbulent jets with coflow},}\ }\href {\doibase
  10.1017/jfm.2017.379} {\bibfield  {journal} {\bibinfo  {journal} {Journal of
  Fluid Mechanics}\ }\textbf {\bibinfo {volume} {825}},\ \bibinfo {pages}
  {887--921} (\bibinfo {year} {2017})}\BibitemShut {NoStop}%
\bibitem [{\citenamefont {Zhou}(2017{\natexlab{a}})}]{zhou_2017a}%
  \BibitemOpen
  \bibfield  {author} {\bibinfo {author} {\bibfnamefont {Y.}~\bibnamefont
  {Zhou}},\ }\bibfield  {title} {\enquote {\bibinfo {title} {Rayleigh--taylor
  and richtmyer--meshkov instability induced flow, turbulence, and mixing.
  i},}\ }\href {\doibase https://doi.org/10.1016/j.physrep.2017.07.005}
  {\bibfield  {journal} {\bibinfo  {journal} {Physics Reports}\ }\textbf
  {\bibinfo {volume} {720-722}},\ \bibinfo {pages} {1 -- 136} (\bibinfo {year}
  {2017}{\natexlab{a}})},\ \bibinfo {note} {rayleigh-Taylor and
  Richtmyer-Meshkov instability induced flow, turbulence, and mixing.
  I}\BibitemShut {NoStop}%
\bibitem [{\citenamefont {Zhou}(2017{\natexlab{b}})}]{zhou_2017b}%
  \BibitemOpen
  \bibfield  {author} {\bibinfo {author} {\bibfnamefont {Y.}~\bibnamefont
  {Zhou}},\ }\bibfield  {title} {\enquote {\bibinfo {title} {Rayleigh--taylor
  and richtmyer-meshkov instability induced flow, turbulence, and mixing.
  ii},}\ }\href {\doibase https://doi.org/10.1016/j.physrep.2017.07.008}
  {\bibfield  {journal} {\bibinfo  {journal} {Physics Reports}\ } (\bibinfo
  {year} {2017}{\natexlab{b}}),\
  https://doi.org/10.1016/j.physrep.2017.07.008}\BibitemShut {NoStop}%
\bibitem [{\citenamefont {Aluie}(2013)}]{aluie_2013}%
  \BibitemOpen
  \bibfield  {author} {\bibinfo {author} {\bibfnamefont {H.}~\bibnamefont
  {Aluie}},\ }\bibfield  {title} {\enquote {\bibinfo {title} {Scale
  decomposition in compressible turbulence},}\ }\href {\doibase
  https://doi.org/10.1016/j.physd.2012.12.009} {\bibfield  {journal} {\bibinfo
  {journal} {Physica D: Nonlinear Phenomena}\ }\textbf {\bibinfo {volume}
  {247}},\ \bibinfo {pages} {54 -- 65} (\bibinfo {year} {2013})}\BibitemShut
  {NoStop}%
\bibitem [{\citenamefont {Livescu}\ and\ \citenamefont
  {Li}(2017)}]{livescu_li_2017}%
  \BibitemOpen
  \bibfield  {author} {\bibinfo {author} {\bibfnamefont {D.}~\bibnamefont
  {Livescu}}\ and\ \bibinfo {author} {\bibfnamefont {Z.}~\bibnamefont {Li}},\
  }\bibfield  {title} {\enquote {\bibinfo {title} {Subgrid-scale backscatter
  after the shock-turbulence interaction},}\ }\href {\doibase
  10.1063/1.4971738} {\bibfield  {journal} {\bibinfo  {journal} {AIP Conference
  Proceedings}\ }\textbf {\bibinfo {volume} {1793}},\ \bibinfo {pages} {150009}
  (\bibinfo {year} {2017})},\ \Eprint
  {http://arxiv.org/abs/https://aip.scitation.org/doi/pdf/10.1063/1.4971738}
  {https://aip.scitation.org/doi/pdf/10.1063/1.4971738} \BibitemShut {NoStop}%
\bibitem [{\citenamefont {O'Brien}\ \emph {et~al.}(2014)\citenamefont
  {O'Brien}, \citenamefont {Urzay}, \citenamefont {Ihme}, \citenamefont
  {Moin},\ and\ \citenamefont {Saghafian}}]{Obrien_etal_2014}%
  \BibitemOpen
  \bibfield  {author} {\bibinfo {author} {\bibfnamefont {J.}~\bibnamefont
  {O'Brien}}, \bibinfo {author} {\bibfnamefont {J.}~\bibnamefont {Urzay}},
  \bibinfo {author} {\bibfnamefont {M.}~\bibnamefont {Ihme}}, \bibinfo {author}
  {\bibfnamefont {P.}~\bibnamefont {Moin}}, \ and\ \bibinfo {author}
  {\bibfnamefont {A.}~\bibnamefont {Saghafian}},\ }\bibfield  {title} {\enquote
  {\bibinfo {title} {Subgrid-scale backscatter in reacting and inert supersonic
  hydrogen--air turbulent mixing layers},}\ }\href {\doibase
  10.1017/jfm.2014.62} {\bibfield  {journal} {\bibinfo  {journal} {J. Fluid
  Mech.}\ }\textbf {\bibinfo {volume} {743}},\ \bibinfo {pages} {554--584}
  (\bibinfo {year} {2014})}\BibitemShut {NoStop}%
\end{thebibliography}%


\end{document}